\documentclass[aps,prd,preprint,superscriptaddress,amssymb,amsmath,nofootinbib]{revtex4}

\usepackage{color}
\usepackage{slashed}
\usepackage{graphicx}
\usepackage{dcolumn}
\usepackage{bm}
\usepackage{xcolor}
\DeclareMathAlphabet{\pazocal}{OMS}{zplm}{m}{n}

\begin{document}

\preprint{}

\title{Multi-Higgs boson signals of a modified muon Yukawa coupling at a muon collider}

\author{Radovan Dermisek}
\email[]{dermisek@iu.edu}
\affiliation{Department of Physics, Indiana University, Bloomington, Indiana, 47405, USA}

\author{Keith Hermanek}
\email[]{khermane@iu.edu}
\affiliation{Department of Physics, Indiana University, Bloomington, Indiana, 47405, USA}

\author{Taegyu~Lee}
\email[]{taeglee@iu.edu}
\affiliation{Department of Physics, Indiana University, Bloomington, Indiana, 47405, USA}

\author{Navin McGinnis}
\email[]{nmcginnis@arizona.edu}
\affiliation{Department of Physics, University of Arizona, Tucson, Arizona 85721, USA}
\affiliation{TRIUMF, 4004 Wesbrook Mall, Vancouver, British Columbia, Canada V6T 2A3}

\author{Sangsik Yoon}
\email[]{yoon12@iu.edu}
\affiliation{Department of Physics, Indiana University, Bloomington, Indiana, 47405, USA}


\date{March 2, 2024}

\begin{abstract}

We study di-Higgs and tri-Higgs boson productions at a muon collider as functions of the modification of the muon Yukawa coupling resulting from new physics parameterized by the dimension-six mass operator. We show that the di-Higgs signal can be used to observe a deviation in  the muon Yukawa coupling at the 10\% level for $\sqrt{s} = 10$ TeV and at the 3.5\% level for $\sqrt{s} = 30$ TeV. The tri-Higgs signal improves the sensitivity dramatically with increasing  $\sqrt{s}$, reaching 0.8\% at $\sqrt{s} = 30$ TeV. We also study all processes involving Goldstone bosons originating from the same operator, discuss possible model dependence resulting from other operators of dimension-six and higher, and identify $\mu^+ \mu^- \to hh$, $\mu^+ \mu^- \to hhh$, and $\mu^+ \mu^- \to hZ_LZ_L$ as golden channels. We further extend the study to an effective field theory including two Higgs doublets with type-II couplings and show that di-Higgs and tri-Higgs signals involving heavy Higgs bosons can be enhanced in the alignment limit by a factor of $(\tan \beta)^4$ and $(\tan \beta)^6$, respectively, which results in the potential sensitivity to a  modified muon Yukawa coupling at the $10^{-6}$ level already at  a $\sqrt{s} = 10 $ TeV muon collider. The results can easily be customized for other extensions of the Higgs sector.

\end{abstract}


\maketitle

\section{Introduction}

Among many exciting physics opportunities at a  muon collider  are more precise measurements of muon properties and possible discoveries of new physics through their deviations from predictions of the standard model (SM) \cite{Accettura:2023ked,Aime:2022flm,MuonCollider:2022xlm,Capdevilla:2020qel, Buttazzo:2020eyl,Yin:2020afe,Han:2021lnp,Dermisek:2021mhi}. The Large Hadron Collider (LHC) is expected to measure the muon Yukawa coupling with about 5\% precision through the decay $h \to \mu^+\mu^-$, while various options for a muon collider, characterized by the center of mass energy  and expected integrated luminosity, promise to increase the precision to a few percent or even 0.3\%~\cite{Accettura:2023ked}.

A  muon Yukawa coupling disagreeing with the SM prediction would clearly indicate new physics. However, confirming it by directly observing new particles responsible for a modification of the muon Yukawa coupling might be beyond the reach of a given collider. In addition, the sign of the Yukawa coupling, or a complex phase in general, is not determined by $h \to \mu^+\mu^-$ measurement, potentially leaving a large effect of new physics undetected. 

In this paper we study multi-Higgs  boson signals which in general accompany a modification of the muon Yukawa coupling independently of the scale and other details of new physics. As long as the dominant effect of new physics on the muon Yukawa coupling is captured by the dimension-six mass operator, $\bar{l}_{L}\mu_{R}H\left(H^{\dagger}H\right)$, where $l_L$ is the lepton doublet and $H$ is the Higgs doublet, the cross sections for $\mu^+ \mu^- \to hh$ and $\mu^+ \mu^- \to hhh$ are uniquely tied to the modification of  the muon Yukawa coupling. As a result of negligible SM backgrounds for these processes, these signals could provide the first evidence for new physics even before a deviation of the muon Yukawa coupling from the SM prediction is established by $h \to \mu^+\mu^-$. For example, the opposite sign of the muon Yukawa coupling leads to a very strong di-Higgs signal that can be seen even at a very low energy muon collider. Furthermore, if mass operators of higher dimensions also contribute significantly to the muon Yukawa coupling, signals with more Higgs bosons in final states are expected (and could be even stronger than $hh$ or $hhh$). By measuring all resulting multi-Higgs boson signals, the  Wilson coefficients of all contributing operators including the sizes of their complex phases can be determined.

Furthermore, we study possible signals of a modified muon Yukawa coupling in effective field theories with extended Higgs sectors where the effects of new physics are parametrized by additional operators. Focusing on a two Higgs doublet model (2HDM) with  type-II couplings to fermions, we find that di-Higgs and tri-Higgs signals involving heavy Higgs bosons can be enhanced in the alignment limit by a factor of $(\tan \beta)^4$ and $(\tan \beta)^6$, respectively, compared to $hh$ and $hhh$ for a given modification of the muon Yukawa coupling. As a result of the enhancement, these signals are potentially sensitive to even tiny modifications of the coupling that would not be observed by measuring $h \to \mu^+\mu^-$ at any currently considered future collider. For example, an observable $HHH$ signal, where $H$ is the heavy $CP$-even Higgs boson, is predicted at a $\sqrt{s} = 10 $ TeV muon collider from a modification of the muon Yukawa coupling at the $10^{-6}$ level. 

Di-Higgs and tri-Higgs boson productions at a muon collider were previously studied in connection with the Muon $g-2$ anomaly in Ref.~\cite{Dermisek:2021mhi}. It has long been known that  chirally enhanced contributions to the Muon $g-2$ can result from the mixing of the muon with new leptons, which in general modifies muon Yukawa  and gauge couplings~\cite{Kannike:2011ng, Dermisek:2013gta}. If the new leptons are heavy, their impact at low energies would be specified by the dimension-six mass operator and operators with covariant derivatives preserving chirality of the muon, in addition to the dipole operators contributing to the Muon $g-2$. Thus, through a modification of the muon Yukawa coupling, the Muon $g-2$ is also related to di-Higgs and tri-Higgs signals resulting from the same  operator. Such a  connection between the dipole operators and the mass operator is generally expected in models for new physics~\cite{Crivellin:2021rbq, Dermisek:2022aec, Dermisek:2023nhe}. Similarly, in models with an extended Higgs sector, such as the 2HDM type-II, similar connections between the operators can be made, and the $(\tan \beta)^6$ enhancement advertised above can be understood from related enhancements in contributions of heavy Higgses to Muon $g-2$~\cite{Dermisek:2020cod, Dermisek:2021ajd} or the electric dipole moment~\cite{Dermisek:2023tgq}. Additional studies of multi-Higgs boson signals at the LHC in connection with modified Yukawa couplings of quarks can be found in~\cite{Bauer:2017cov,Bar-Shalom:2018rjs,Alasfar:2019pmn,Egana-Ugrinovic:2021uew}.

Di-boson and tri-boson signals of a modified muon Yukawa coupling  with the focus on final states involving Goldstone bosons (longitudinal gauge bosons, $W_L$ and $Z_L$) were studied in detail in Ref.~\cite{Han:2021lnp}. Among the large number of possible processes, $\mu^+ \mu^- \to W_L^+W_L^-h$ and $Z_Lhh$ were identified as the optimal examples. While it is certainly the case that all the allowed combinations of di-bosons or tri-bosons result from a deviation in the muon Yukawa coupling, in addition to just di-Higgs and tri-Higgs, there are two main reasons that highly favor pure Higgs final states. Pure Higgs final states feature negligible SM backgrounds, and more importantly, the final states with $W$s and $Z$s, or mixtures of gauge and Higgs bosons can also originate from other dimension-six operators that are not related to the muon Yukawa coupling. We argue that besides $hh$ and $hhh$ there is only one additional \textit{golden mode} ($hZ_LZ_L$) not affected by any other dimension-six operators and thus directly related to contributions to the muon Yukawa coupling. We also discuss possible effects of other dimension-eight operators.

This paper is organized as follows. In Sec.~\ref{sec:SM} we consider a modification of the muon Yukawa coupling assuming the Higgs sector of the SM. We present detailed predictions for di-Higgs and tri-Higgs productions  resulting from the dimension-six mass operator and multi-Higgs productions from operators of higher dimensions. We also present results for di-boson and tri-boson productions involving Goldstone bosons, and discuss the relative sizes of the signals, SM backgrounds, and model dependence of predictions for various final states. In Sec.~\ref{sec:2HDM} we assume the Higgs sector is that of the 2HDM type-II and extend possible signatures of a modified Yukawa coupling to di-boson and tri-boson signals involving heavy Higgs bosons. We follow with a similar discussion of various final states as in the SM case. We summarize results and conclude in Sec.~\ref{sec:conclusions}. Unitarity constraints on possible modifications of the muon Yukawa coupling, used in the main text, are discussed in Appendix~\ref{sec:23scat}. Results for other dimension-six mass operators in the 2HDM case, besides the one discussed in the main text, are summarized in Appendix~\ref{sec:c1c2c3}.

\section{Modified muon Yukawa coupling -- SM Higgs sector }
\label{sec:SM}

Consider the effective Lagrangian for the second generation of leptons assuming the SM Higgs sector:
\begin{equation}
\mathcal{L} = -y_{\mu}\,\bar{l}_{L}\mu_{R}H \;-\; C_{\mu H}\,\bar{l}_{L}\mu_{R}H\left(H^{\dagger}H\right) + {\rm H.c.},
\label{eq:eff_lagrangian_SM}
\end{equation}
%
where  the components of the lepton doublet are $l_{L}=(\nu_{\mu}, \mu_{L})^{T}$. The first term  is the  usual muon Yukawa coupling in the SM. When the Higgs field develops a vacuum expectation value (VEV), $H=(G^+,v+(h+iG)/\sqrt{2})^T$ with $v=174$ GeV, the dimension-six operator in the second term generates an additional contribution to the muon mass 
\begin{equation}
m_\mu = y_\mu \,v + C_{\mu H} \,v^3.
\end{equation} 
This  operator  is the only dimension-six operator that contributes to the muon mass and Yukawa coupling in the Warsaw basis~\cite{Grzadkowski:2010es}.

  \begin{figure}[t!]
\includegraphics[scale=0.8]{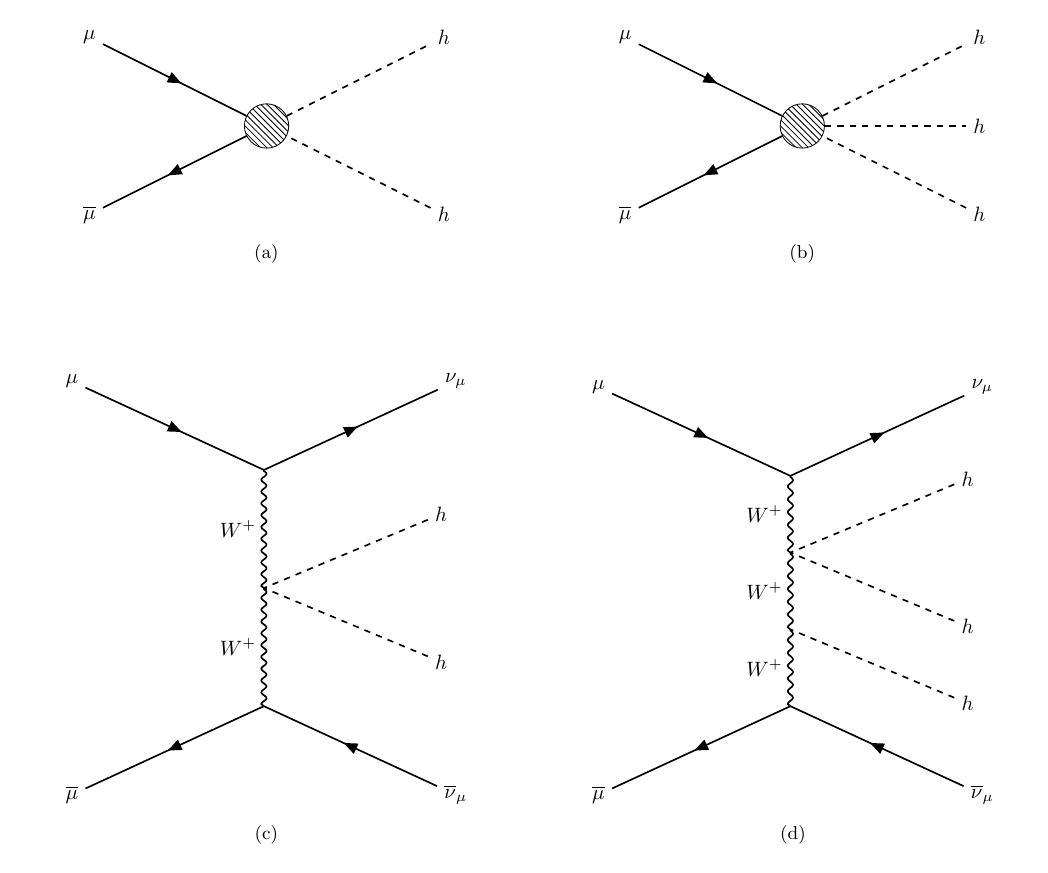}
\caption{Feynman diagrams for the annihilation processes (a) $\mu^+\mu^-\rightarrow hh$ and (b) $\mu^+\mu^-\rightarrow hhh$ resulting from the dimension-six contact term. Examples of contributing diagrams to VBF processes in the SM, (c) $\mu^+\mu^-\rightarrow \nu_\mu \bar{\nu}_\mu hh$ and (d) $\mu^+\mu^-\rightarrow \nu_\mu \bar{\nu}_\mu hhh$, which are the main background for di-Higgs and tri-Higgs signals in (a) and (b).}
\label{fig:VBF}
\end{figure}
In the basis where the muon mass is real and positive, the resulting interactions with the SM Higgs boson are described by
\begin{equation}
\mathcal{L}\supset - \frac{1}{\sqrt{2}} \, \lambda^h_{\mu\mu}\, \bar{\mu}_L\mu_R h - \frac{1}{2} \, \lambda^{hh}_{\mu\mu}\, \bar{\mu}_L\mu_R h^2 - \frac{1}{3!}\, \lambda^{hhh}_{\mu\mu}\, \bar{\mu}_L\mu_R h^3 + {\rm H.c.},
\label{eq:lagrangian_h}	
\end{equation}
where the couplings are given by
\begin{eqnarray}
	\lambda^h_{\mu\mu} &=&  \frac{m_\mu}{v} +2 \,C_{\mu H}\, v^2,\\
	\lambda^{hh}_{\mu\mu} &=& 3\, C_{\mu H}\, v ,\\
	\lambda^{hhh}_{\mu\mu} &=& \frac{3}{\sqrt{2}}\,C_{\mu H},
\end{eqnarray}
and are in general complex.

It is convenient to parametrize the departure of the Yukawa coupling from the SM prediction by:
\begin{equation}
\lambda^h_{\mu\mu} = \frac{m_\mu}{v} \, \kappa_\mu 
\label{eq:kappa}
\end{equation}
and also define
\begin{equation}
\kappa_\mu = 1 + \Delta \kappa_\mu ,
\label{eq:Delta_kappa}
\end{equation}
where $\Delta \kappa_\mu = 2C_{\mu H}v^3/m_\mu$. We can rewrite the couplings above as
\begin{eqnarray}
	\lambda^{hh}_{\mu\mu} &=& \frac{3m_\mu}{2v^2}\,\Delta\kappa_\mu , \\
	\lambda^{hhh}_{\mu\mu} &=& \frac{3m_\mu}{2\sqrt{2}v^3}\,\Delta\kappa_\mu.
\end{eqnarray}

\begin{figure}[t!]
\includegraphics[scale=0.35]{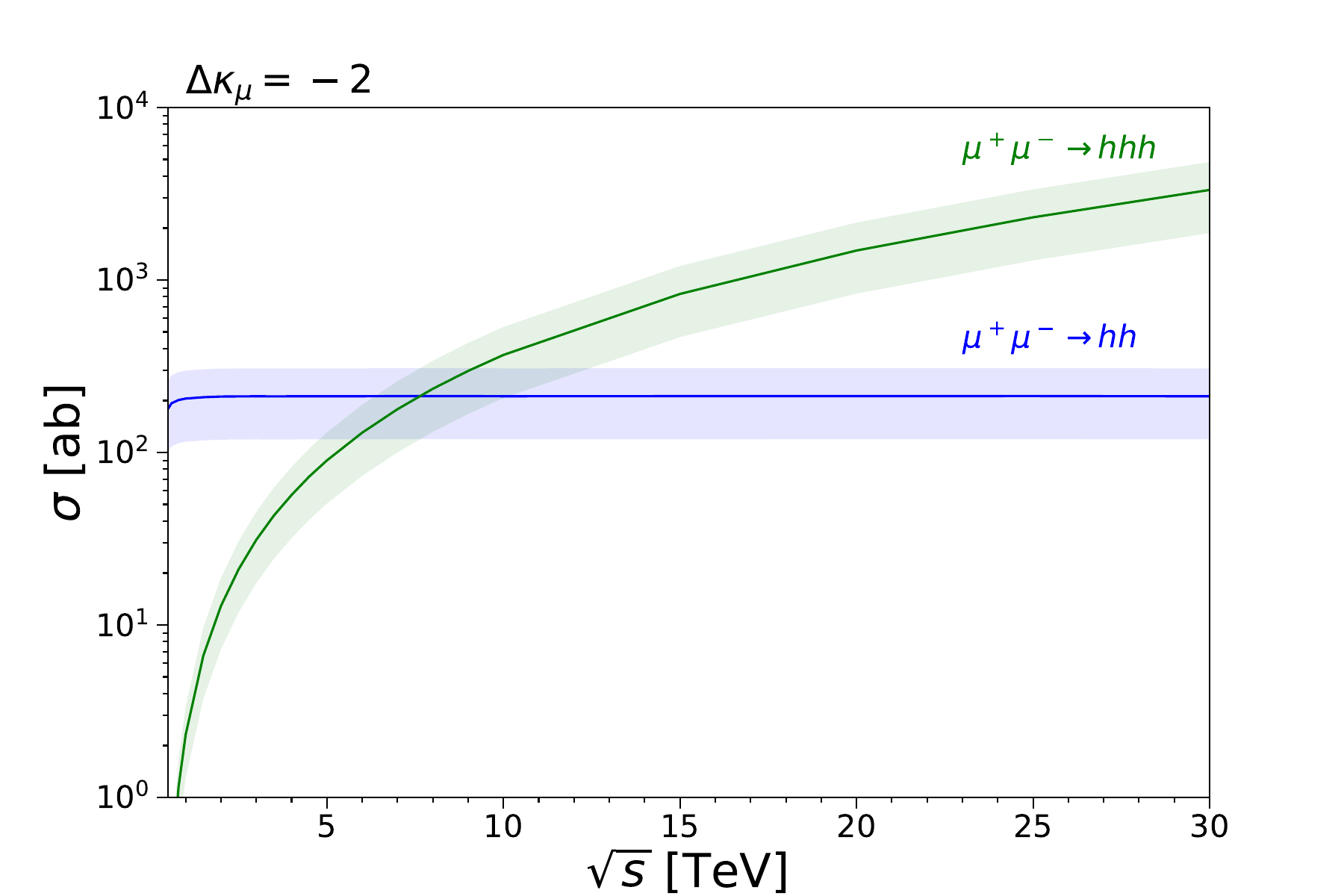}
\caption{Total cross sections for $\mu^+\mu^-\rightarrow h h$ and  $\mu^+\mu^-\rightarrow h h h$ as functions of $\sqrt{s}$ corresponding to $\Delta \kappa_\mu=-2$ (solid lines) and $95\%$ C.L. range for the opposite sign muon Yukawa coupling (shaded regions). }
\label{fig:energy_SM}
\end{figure}

The cross sections for  $\mu^+\mu^- \to hh$ and $\mu^+\mu^- \to  hhh$, resulting from the new contact interactions [see Fig.~\ref{fig:VBF} (a) and (b)] were calculated in Ref.~\cite{Dermisek:2021mhi} and, neglecting the muon mass and the Higgs mass, are given by 
\begin{eqnarray}
	\sigma_{\mu^+\mu^-\rightarrow hh} &=& \frac{|\lambda^{hh}_{\mu\mu} |^2 }{64\pi}  \; = \; \frac{9}{256\pi}\left(\frac{m_\mu}{v^2}\right)^2|\Delta\kappa_\mu|^2  \label{eq:cross_section_dihiggs},\\
	\sigma_{\mu^+\mu^-\rightarrow hhh} &=&  \frac{|\lambda^{hhh}_{\mu\mu} |^2 }{6144\pi^3} \, s \; = \; \frac{3s}{2^{14}\pi^3}\left(\frac{m_\mu}{v^3}\right)^2| \Delta\kappa_\mu |^2. \label{eq:cross_section_trihiggs}
\end{eqnarray}
Note that the contributions to the cross sections come from couplings which scale as $v/\Lambda^2$ for di-Higgs and $1/\Lambda^2$ for tri-Higgs. Thus, the cross sections must scale with energy as $s^0$ for di-Higgs and $s$ for tri-Higgs, by dimensional analysis.
These formulas are excellent approximations well above the production thresholds. Exact formulas for cross sections including threshold effects resulting from nonzero Higgs mass can be found in Ref.~\cite{Dermisek:2021mhi}. The cross sections, calculated from the effective Lagrangian implemented in {\tt FeynRules}~\cite{Degrande:2011ua} using {\tt MadGraph5}~\cite{Alwall:2014hca}, are plotted in Fig.~\ref{fig:energy_SM} as functions of $\sqrt{s}$ for $\Delta \kappa_\mu = -2$ which corresponds to the opposite sign muon Yukawa coupling compared to its SM value ($\kappa_\mu = -1$). Shaded regions  indicate the current $95\%$ C.L. range for the opposite sign muon Yukawa coupling~\cite{CMS:2020xwi}.
The total cross section for  $\mu^+ \mu^- \to hh$, away from the threshold, is about 210 ab 
 independently of the center of mass energy, while the cross section for $\mu^+ \mu^- \to hhh$ grows quadratically with $\sqrt{s}$ and becomes larger than $\sigma_{\mu^+\mu^-\rightarrow hh}$ above $\sqrt{s} \simeq 7.6 $  TeV. These di-Higgs and tri-Higgs cross sections are by many orders of magnitude larger than the corresponding SM backgrounds, $\sigma(\mu^+ \mu^- \to hh)_{\rm SM} = 1.6\times10^{-4}\;{\rm ab}$ and $\sigma(\mu^+ \mu^- \to hhh)_{\rm SM} = 2.9\times10^{-5}\;{\rm ab}$ at $\sqrt{s} = 3 $ TeV, and fall as $1/s^2$ and $1/s$ at larger energies, respectively.\footnote {Note that the di-Higgs or tri-Higgs boson productions at a muon collider are dominated by vector boson fusion. We will comment on these and other backgrounds later in this subsection. }
Even for the Higgs quartic coupling  as large as the current upper limit~\cite{CMS:2022kdx,ATLAS:2019qdc}, the contributions from the SM diagrams to $\mu^+ \mu^- \to hh$ and $hhh$ are negligible at $\sqrt{s}> 3\;{\rm TeV}$. Thus, we have neglected standard model and interference terms in Eqs.~(\ref{eq:cross_section_dihiggs}) and (\ref{eq:cross_section_trihiggs}).

For the target  integrated luminosity  at a muon collider depending on its center of mass energy~\cite{Accettura:2023ked},
 \begin{equation}
\mathcal{L}_{\rm int} = 10\, {\rm ab}^{-1}\, \left(\frac{E_{cm}}{10 \,{\rm TeV}}\right)^2,
\label{eq:lum}	
\end{equation}
we see that even a low energy muon collider would easily see the di-Higgs signal associated with the opposite sign muon Yukawa coupling. For example, 191 di-Higgs and 30 tri-Higgs events are expected already at $\sqrt{s} = 3$ TeV. 
 
\begin{figure}[h!]
\includegraphics[scale=0.75]{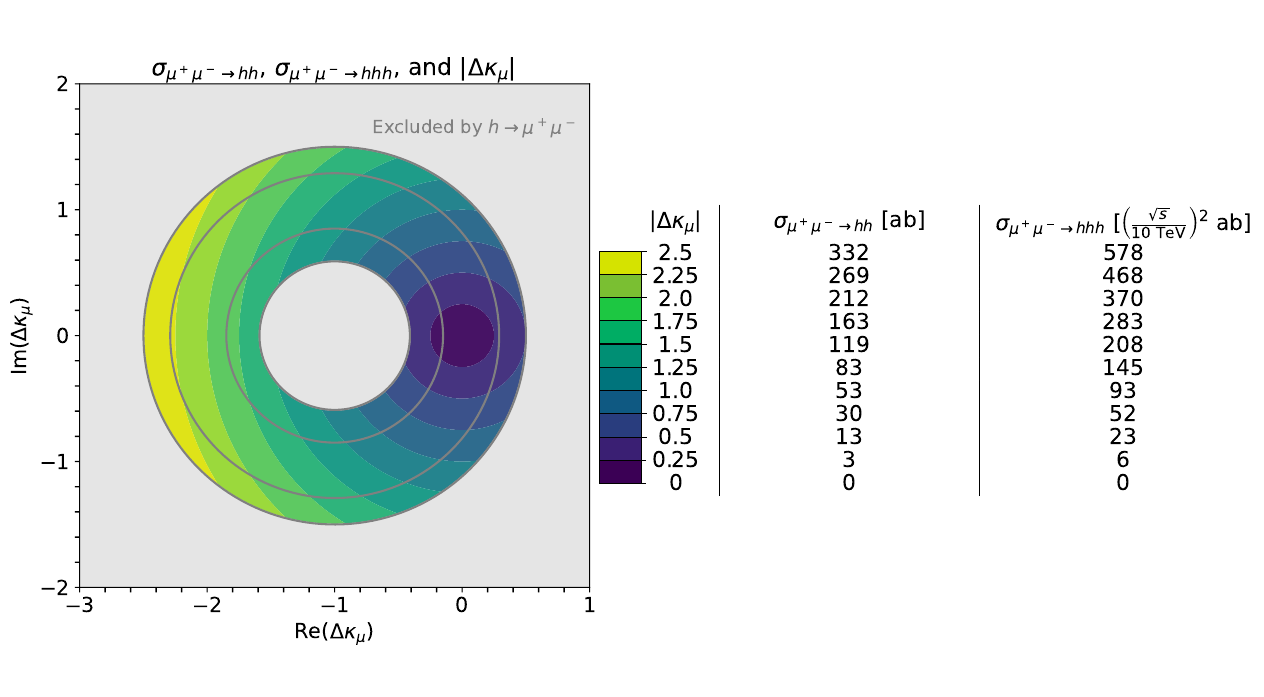}
\caption{Contours of constant di-Higgs and tri-Higgs production cross sections and the corresponding $|\Delta \kappa_\mu |$ in the plane of real and imaginary parts of $\Delta\kappa_\mu$. Gray circles and shaded regions correspond to 68\% C.L. and 95\% C.L. exclusion limits from  the CMS search for $h\to \mu^+\mu^-$~\cite{CMS:2020xwi}.}
\label{fig:complex-kappa_SM}
\end{figure}
 
 The cross sections for $\mu^+ \mu^- \to hh$ and for $\mu^+ \mu^- \to hhh$ at $\sqrt{s} = 10$ TeV  are also plotted in the plane of real and imaginary parts of  $\Delta \kappa_\mu$ in Fig.~\ref{fig:complex-kappa_SM}. The cross sections for tri-Higgs production at different $\sqrt{s}$ can be obtained by simple rescaling as indicated in the legend. Note that although the cross sections depend only on the $|\Delta \kappa_\mu| $, the constraints from $h\to \mu^+\mu^-$ cannot be written in terms of  $|\Delta \kappa_\mu |$ only. Therefore, the combined measurement of $h \to \mu^+ \mu^-$ and $\mu^+ \mu^- \to h h$ or $\mu^+ \mu^- \to h h h$ determine also the size of the complex phase of the muon Yukawa coupling.

\begin{figure}[t!]
  \includegraphics[scale=0.24]{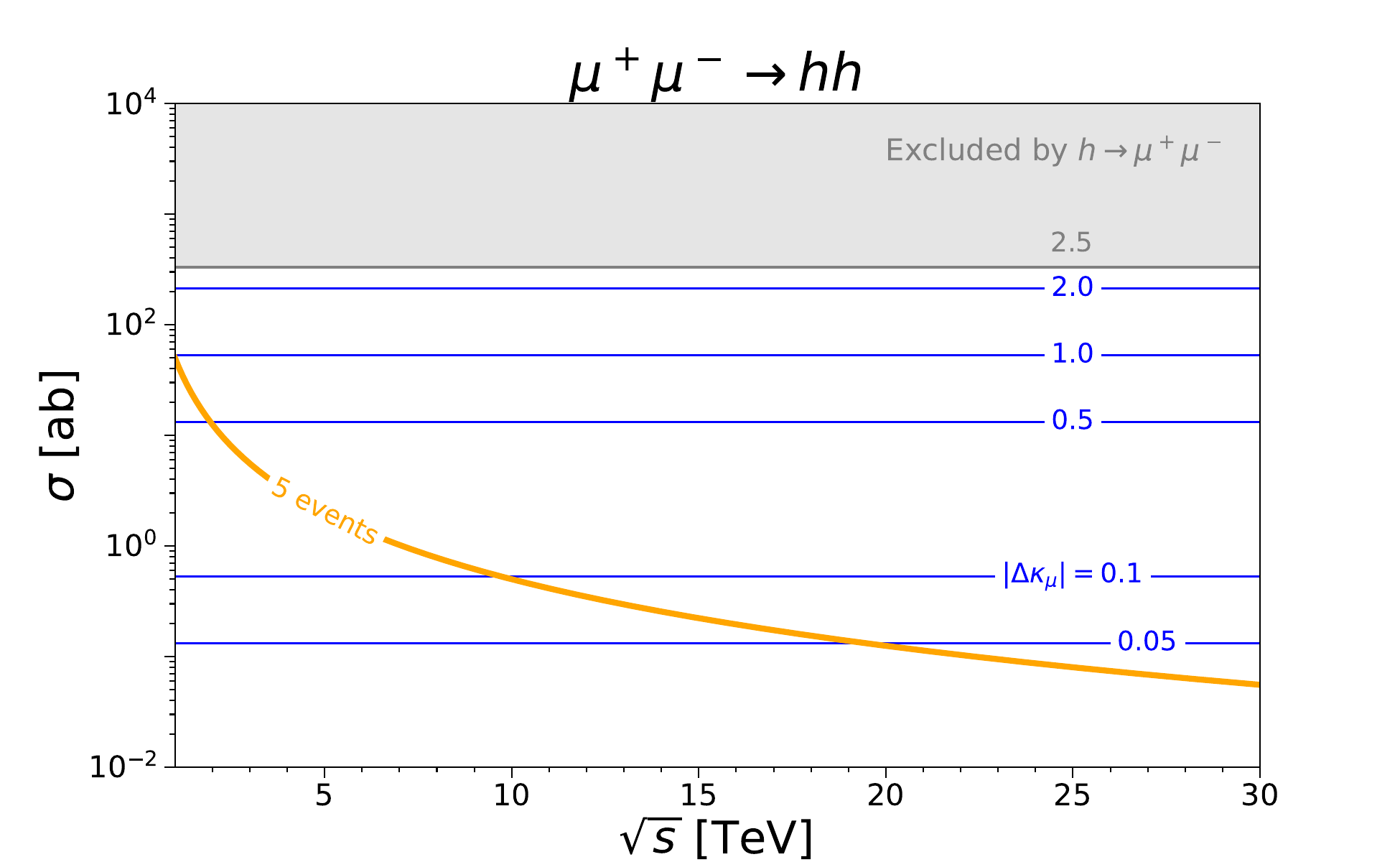}
  \includegraphics[scale=0.24]{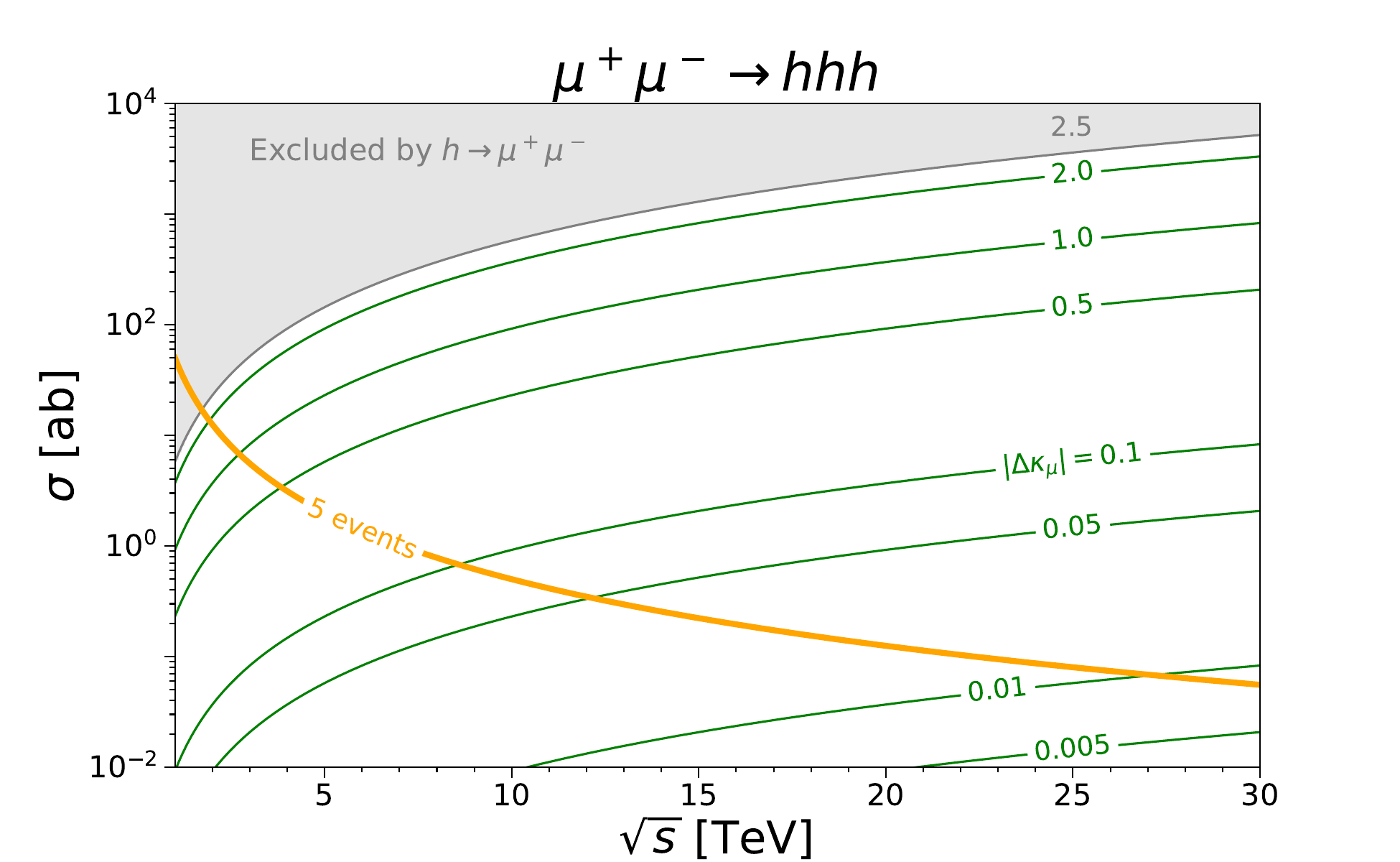}
\caption{Di-Higgs and tri-Higgs production cross sections as functions of $\sqrt{s}$ for various $|\Delta\kappa_\mu|$. Gray shaded regions are excluded at 95\% C.L. by the CMS search for $h\to \mu^+\mu^-$~\cite{CMS:2020xwi}. Orange lines indicate the cross sections corresponding to five signal events.}
\label{fig:sqrts_kappa_SM}
\end{figure}

 To extend the range of $|\Delta \kappa_\mu| $ to small values, to which the current searches are not yet sensitive, we plot the di-Higgs and tri-Higgs production cross sections as functions of $\sqrt{s}$ for various $|\Delta\kappa_\mu|$ in Fig.~\ref{fig:sqrts_kappa_SM}. Orange lines indicate the cross sections corresponding to five signal events for integrated luminosity given in Eq.~(\ref{eq:lum}). Taking this as an estimate of the sensitivity of a muon collider to the modification of the muon Yukawa coupling, we see that the di-Higgs signal can be used to observe a deviation in  the muon Yukawa coupling at the 10\% level for $\sqrt{s} = 10$ TeV and at the 3.5\% level for $\sqrt{s} = 30$ TeV. The tri-Higgs signal leads to only a slightly better sensitivity at $\sqrt{s} = 10$ TeV, namely 7\%,  but would improve dramatically with increasing  $\sqrt{s}$, reaching 0.8\% at $\sqrt{s} = 30$ TeV (and 0.07\% at $\sqrt{s} = 100$ TeV).

  The quoted sensitivities should be viewed as an estimate of the ultimate sensitivities of a muon collider that assume close to perfect signal reconstruction and background rejection, in addition to combining signals resulting from different decay modes of the Higgs boson. The SM di-Higgs or tri-Higgs boson productions at a muon collider are dominated by vector boson fusion (VBF)~\cite{Costantini:2020stv,Ruiz:2021tdt,AlAli:2021let}. Examples of contributing diagrams are  given in Figs.~\ref{fig:VBF} 1(c) and 1(d). This production mechanism leads to other particles in final states, and thus is, in principle, distinguishable from pure $hh$ or $hhh$ signals. For example, the dominant $W^+W^-$ mediated process with neutrinos in final states can be distinguished via a cut on the total invariant mass of the visible particles. For $\sqrt{s}=10\;{\rm TeV}$, we find that requiring the total invariant mass of the visible particles to be larger than 9.6 TeV for di-Higgs and 6 TeV for tri-Higgs is sufficient to reduce this source of background below one event. 
 
 For specific decay modes of the Higgs boson there are additional backgrounds resulting from SM processes with final states that cannot be completely distinguished from the Higgs boson. For the dominant decay mode, $h\to b \bar b$, there is $ZZ$ or $ZZZ$ background with $Z \to b\bar b$, a fraction of which would be reconstructed as the di-Higgs or tri-Higgs signal. For quoted sensitivities at $\sqrt{s}=10\;{\rm TeV}$, we find that in order for the backgrounds from $ZZ$ and $ZZZ$ to be smaller than the signal,  we should be able to distinguish $Z \to b\bar b$ from $h\to b\bar b$ with 95\% efficiency for the di-Higgs and about 30\% efficiency for the tri-Higgs.  
We see that the  backgrounds for  the di-Higgs signal are more challenging and might result in lower sensitivities than the ones quoted above. However, the tri-Higgs signal is stronger for $\sqrt{s} > 7.6$ TeV and the backgrounds do not seem to pose a big challenge.

Interactions involving Goldstone bosons resulting from the $\mathcal{O}_{\mu H}$  operator  are described by 
\begin{eqnarray}
    \pazocal{L}  \supset  &-&\frac{1}{\sqrt{2}}\,\lambda^G_{\mu\mu}\,\bar\mu_L\mu_R G-\lambda_{\mu\mu}^{hG}\,\bar{\mu}_L\mu_R hG -\frac{1}{2!}\,\lambda_{\mu\mu}^{GG}\,\bar{\mu}_L\mu_R GG -\lambda_{\mu\mu}^{G^+G^-}\bar{\mu}_L\mu_R G^+G^- \nonumber \\
    & -&\frac{1}{2!}\,\lambda_{\mu\mu}^{hhG}\,\bar{\mu}_L\mu_R h^2G  - \frac{1}{2!}\,\lambda_{\mu\mu}^{hGG}\,\bar{\mu}_L\mu_R hGG - \frac{1}{3!}\,\lambda_{\mu\mu}^{GGG}\,\bar{\mu}_L\mu_R GGG \nonumber \\
    & -&\lambda_{\mu\mu}^{hG^+G^-}\bar{\mu}_L\mu_R hG^+G^- - \lambda_{\mu\mu}^{GG^+G^-}\bar{\mu}_L\mu_R GG^+G^- + {\rm H.c.},
\label{eqn:SM_couplings_Gauge}
\end{eqnarray}
where the couplings are summarized in Table~\ref{table:SM_G_couplings} in terms of the Wilson coefficient and also in terms of $\Delta\kappa_\mu$. The cross sections for corresponding di-boson and tri-boson productions involving longitudinal gauge bosons are summarized in Tables~\ref{table:SM_G_di-boson} and \ref{table:SM_G_tri-boson}. These processes, studied in detail in Ref.~{\cite{Han:2021lnp}}, suffer from large SM backgrounds (falling with energy) and thus the rates in Tables~\ref{table:SM_G_di-boson} and \ref{table:SM_G_tri-boson} are good approximations only at large $\sqrt{s}$.  In addition, we will see that all the final states involving longitudinal gauge bosons except for $h Z_L Z_L$ also result from other dimension-six operators that do not affect muon mass and Yukawa coupling, and thus are not necessarily correlated with the modification of the muon Yukawa coupling. 

\begin{table*}[h!]
\caption{Coupling constants involving goldstone bosons defined in Eq.~(\ref{eqn:SM_couplings_Gauge}).}
\begin{ruledtabular}
\begin{tabular}{ccc}
  &  In terms of $C_{\mu H}$ & In terms of $\Delta\kappa_\mu$\\
 \hline
 $\lambda_{\mu\mu}^{G}$   & $i\frac{m_\mu}{v}$ & $i\frac{m_\mu}{v}$ \\
 \hline
 $\lambda_{\mu\mu}^{hG}$   & $iv\;C_{\mu H}$  & $\frac{im_\mu}{2v^2}\Delta\kappa_\mu$ \\
\hline  
 $\lambda_{\mu\mu}^{GG}$   & $v\;C_{\mu H}$  & $\frac{m_\mu}{2v^2}\Delta\kappa_\mu$\\
\hline
 $\lambda_{\mu\mu}^{G^+G^-}$   & $v\;C_{\mu H}$   & $\frac{m_\mu}{2v^2}\Delta\kappa_\mu$ \\
\hline
 $\lambda_{\mu\mu}^{hhG}$   & $\frac{i}{\sqrt{2}}\;C_{\mu H}$ & $\frac{im_\mu}{2\sqrt{2}v^3}\Delta\kappa_\mu$ \\
\hline
 $\lambda_{\mu\mu}^{hGG}$ & $\frac{1}{\sqrt{2}}\;C_{\mu H}$  & $\frac{m_\mu}{2\sqrt{2}v^3}\Delta\kappa_\mu$ \\
\hline
 $\lambda_{\mu\mu}^{GGG}$  &  $\frac{3i}{\sqrt{2}}\;C_{\mu H}$ & $\frac{3im_\mu}{2\sqrt{2}v^3}\Delta\kappa_\mu$ \\
 \hline
  $\lambda_{\mu\mu}^{hG^+G^-}$   & $\frac{1}{\sqrt{2}}\;C_{\mu H}$ & $\frac{m_\mu}{2\sqrt{2}v^3}\Delta\kappa_\mu$\\
\hline
 $\lambda_{\mu\mu}^{GG^+G^-}$   & $\frac{i}{\sqrt{2}}\;C_{\mu H}$  & $\frac{im_\mu}{2\sqrt{2}v^3}\Delta\kappa_\mu$ \\
\end{tabular}
\end{ruledtabular}
\label{table:SM_G_couplings}
\end{table*}

\begin{table*}[h!]
\caption{Cross sections for di-boson productions involving longitudinal gauge bosons.}
\begin{ruledtabular}
\begin{tabular}{ccc}
  & In terms of $\Delta\kappa_\mu$ & \;\;\;\;\;\;In units of $\sigma_{\mu^+\mu^-\rightarrow hh}$ \;\;\;\;\;\; \\
 \hline
 $\sigma_{\mu^+\mu^-\rightarrow hZ_L}$ & $\frac{1}{128\pi}\left(\frac{m_\mu}{v^2}\right)^2|\Delta\kappa_\mu|^2$ &  $\frac{2}{9}$  \\
\hline  
 $\sigma_{\mu^+\mu^-\rightarrow Z_LZ_L}$ & $\frac{1}{256\pi}\left(\frac{m_\mu}{v^2}\right)^2|\Delta\kappa_\mu|^2$  & $\frac{1}{9}$\\
\hline
 $\sigma_{\mu^+\mu^-\rightarrow W_L^+W_L^-}$  & $\frac{1}{128\pi}\left(\frac{m_\mu}{v^2}\right)^2|\Delta\kappa_\mu|^2$  &  $\frac{2}{9}$\\
 \end{tabular}
\end{ruledtabular}
\label{table:SM_G_di-boson}
\end{table*}

\begin{table*}[h!]
\caption{Cross sections for tri-boson productions involving longitudinal gauge bosons.}
\begin{ruledtabular}
\begin{tabular}{ccc}
  & In terms of $\Delta\kappa_\mu$ & \;\;\;\;\;\;In units of $\sigma_{\mu^+\mu^-\rightarrow hhh}$ \;\;\;\;\;\; \\
 \hline
 $\sigma_{\mu^+\mu^-\rightarrow hhZ_L}$  & $\frac{s}{2^{14}\pi^3}\left(\frac{m_\mu}{v^3}\right)^2|\Delta\kappa_\mu|^2$  &  $\frac{1}{3}$\\
\hline
 $\sigma_{\mu^+\mu^-\rightarrow hZ_LZ_L}$ & $\frac{s}{2^{14}\pi^3}\left(\frac{m_\mu}{v^3}\right)^2|\Delta\kappa_\mu|^2$ & $\frac{1}{3}$\\
\hline
 $\sigma_{\mu^+\mu^-\rightarrow Z_LZ_LZ_L}$  &  $\frac{3s}{2^{14}\pi^3}\left(\frac{m_\mu}{v^3}\right)^2|\Delta\kappa_\mu|^2$  & $1$\\
\hline
  $\sigma_{\mu^+\mu^-\rightarrow hW_L^+W_L^-}$  &  $\frac{s}{2^{13}\pi^3}\left(\frac{m_\mu}{v^3}\right)^2|\Delta\kappa_\mu|^2$  & $\frac{2}{3}$\\
\hline
 $\sigma_{\mu^+\mu^-\rightarrow Z_LW_L^+W_L^-}$  & $\frac{s}{2^{13}\pi^3}\left(\frac{m_\mu}{v^3}\right)^2|\Delta\kappa_\mu|^2$  & $\frac{2}{3}$\\
 \end{tabular}
\end{ruledtabular}
\label{table:SM_G_tri-boson}
\end{table*}
\subsection{Other operators and the golden channels}

As already mentioned, the $\mathcal{O}_{\mu H}$ operator in Eq.~(\ref{eq:eff_lagrangian_SM})  is the only dimension-six operator that contributes to the muon mass and Yukawa coupling in the Warsaw basis. However, in specific models, it is generally expected that the mass operator is accompanied by other dimension-six operators, most importantly the dipole operators, $(\bar{l}_L\sigma^{\mu\nu}\mu_R)\tau^IHW^I_{\mu\nu}$ and $(\bar{l}_L\sigma^{\mu\nu}\mu_R)HB_{\mu\nu}$, already mentioned in the Introduction, and operators that involve covariant derivatives acting on the  lepton fields or the Higgs fields, such as $C_R (\overline{\mu}_R H^{\dagger}) i \slashed{D} (\mu_R H)$ or $C_L (\overline{l}_L H) i \slashed{D} (l_L H^{\dagger})$, see Ref.~\cite{Dermisek:2021mhi}.  Parts of these operators, where derivatives act on the lepton fields, can be reduced to $\mathcal{O}_{\mu H}$ via equations of motion for the muon fields and thus are included in our discussion. The remaining pieces with derivatives acting on the Higgs doublet can be decomposed into symmetric and antisymmetric combinations, $ H^{\dagger}(i D_{\mu} H) = ( H^{\dagger}(i D_{\mu} H) + (H^{\dagger} i \overleftarrow{D}_{\mu}) H )/2 + ( H^{\dagger}(i D_{\mu} H) - (H^{\dagger} i \overleftarrow{D}_{\mu}) H )/2$. Integrating by parts on the symmetric combination again leads to  $\mathcal{O}_{\mu H}$ (proportional to $y_{\mu}$; up to total derivatives), while the antisymmetric part results in independent  $(LL)$ and $(RR)$ operators in the Warsaw basis~\cite{Grzadkowski:2010es}: $C_{H l}^{(1)} (H^{\dagger}i \overleftrightarrow{D}_{\mu} H) \left(\bar{l}_L\gamma^\mu l_L\right), C_{H l}^{(3)} (H^{\dagger}i \overleftrightarrow{D}_{\mu}^a H) \left(\bar{l}_L \tau^a \gamma^\mu l_L\right),$ and $C_{H \mu} (H^{\dagger}i \overleftrightarrow{D}_{\mu} H) \left(\bar{\mu}_R\gamma^\mu \mu_R\right)$. While these operators do not affect the muon mass or Yukawa coupling, they modify muon gauge couplings to $Z$ and $W$, and they do lead to di-boson and tri-boson signals. 

Covariant derivative operators lead to $\mu^+ \mu^- \rightarrow W_L^+W_L^-$, $Z_L h$, $W^\pm W_L^\mp$, and $Zh$ di-boson processes and $\mu^+\mu^-\rightarrow W^\pm W_L^\mp h$, $W^\pm W_L^\mp Z_L$, $W_L^+W_L^-Z$, $Zhh$, and $ZZ_LZ_L$ tri-boson processes. 
The dipole operators lead to $hZ$, $Z_L Z$, $W_L^\pm W^\mp$, $W^+W^-$, $Z_LW^+W^-$, $ZW^\pm_LW^\mp$, and $hW^+W^-$.
Thus, among the di-boson processes resulting from the dimension-six mass operator, only  $\mu^+ \mu^- \rightarrow hh$ is not affected by other operators. Similarly, among the tri-boson processes resulting from the dimension-six mass operator, only  $\mu^+ \mu^- \rightarrow hhh$ and $h Z_LZ_L$  are not affected. These conclusions can be independently verified by the Feynman rules resulting from effective operators~\cite{Dedes:2017zog}. In addition, due to the $(LL)$ or $(RR)$ nature of derivative operators, their contributions interfere with SM backgrounds. Moreover, the resulting cross sections grow faster at large energies compared to cross sections for processes resulting from the dimension-six mass operator: di-boson processes grow as $s$ and tri-boson processes as $s^2$.

Thus we find that only $\mu^+ \mu^- \rightarrow hh, \, hhh$ and $ h Z_L Z_L$ processes are unique signals of a modified muon Yukawa coupling. From Table~\ref{table:SM_G_tri-boson} we see that the signal  cross section for $h Z_L Z_L$ is 3 times smaller than for the tri-Higgs production. Besides larger cross sections, di-Higgs and tri-Higgs final states also benefit from negligible SM backgrounds, and thus are more sensitive probes of a modified muon Yukawa coupling.

So far our discussion was limited to dimension-six operators. If contributions of  mass operators of higher dimensions, $\mathcal{O}^{(n)}_{\mu H} = \bar{l}_L \mu_R H \left(H^\dagger H\right)^n$, where $n= 2,3, \dots$ correspond to dimension $8,10,\dots$ operators (with $n=1$ representing the dimension-six operator) are  not negligible, it would reflect in a different ratio of di-Higgs and tri-Higgs events, and in new signals with up to $2n+1$ Higgs bosons in final states. The previous equations corresponding to a dimension-six mass operator can be straightforwardly generalized.  The muon Yukawa coupling is then given by
\begin{equation}
	\lambda_{\mu\mu}^h = \frac{m_\mu}{v} +  \sum_n 2n \,C_{\mu H}^{(n)}\,v^{2n},
\end{equation}
where $C_{\mu H}^{(n)}$ is the Wilson coefficient of $\mathcal{O}_{\mu H}^{(n)}$, and thus  
\begin{equation}
	\Delta\kappa_\mu = \sum_n 2n \, C_{\mu H}^{(n)} \,\frac{v^{2n+1}}{m_\mu} .
\label{eq:general_dkappa}
\end{equation}
Similarly, the effective di-Higgs and tri-Higgs couplings are 
\begin{eqnarray}
	\lambda_{\mu\mu}^{hh} &=& \sum_n \left(2n+1\right)n \,C_{\mu H}^{(n)}\,v^{2n-1} \\
	\lambda_{\mu\mu}^{hhh} &=& \sum_n \frac{\left(2n+1\right)n\left(2n-1\right)}{\sqrt{2}}\,C_{\mu H}^{(n)}\,v^{2n-2} \, .
\end{eqnarray}
The contributions resulting from an individual $\mathcal{O}^{(n)}_{\mu H}$  operator, including the contributions to couplings involving  Goldstone bosons, are summarized in Table~\ref{table:SM_G_couplings_n-operator}.

\begin{table*}[t!]
\caption{Contributions to  $\Delta\kappa_\mu$ and muon couplings to two and three Higgs or Goldstone bosons resulting from $\mathcal{O}^{(n)}_{\mu H}$  in units of $C_{\mu H}^{(n)}v^{2n-2}$. }
\begin{ruledtabular}
\begin{tabular}{ccccccccccc}
 $\Delta\kappa_\mu$& $\lambda_{\mu\mu}^{hh}$& $\lambda_{\mu\mu}^{hG}$ & $\lambda_{\mu\mu}^{GG}$ & $\lambda_{\mu\mu}^{G^+G^-}$ & $\lambda_{\mu\mu}^{hhh}$ & $\lambda_{\mu\mu}^{hhG}$ & $\lambda_{\mu\mu}^{hGG}$ & $\lambda_{\mu\mu}^{GGG}$ & $\lambda_{\mu\mu}^{hG^+G^-}$ & $\lambda_{\mu\mu}^{GG^+G^-}$   \\    
 \hline
$2n\frac{v^3}{m_\mu}$ & $(2n+1)nv$& $inv$  & $nv$  & $nv$ & $\frac{(2n+1)n(2n-1)}{\sqrt{2}}$ &$\frac{in(2n-1)}{\sqrt{2}}$ & $\frac{n(2n-1)}{\sqrt{2}}$  & $\frac{3in}{\sqrt{2}}$ & $\frac{n(2n-1)}{\sqrt{2}}$  & $\frac{in}{\sqrt{2}}$ \\
\end{tabular}
\end{ruledtabular}
\label{table:SM_G_couplings_n-operator}
\end{table*}

In general,  the effective muon  coupling to $k$ Higgs bosons, defined by extending the Lagrangian in Eq.~(\ref{eq:lagrangian_h}) to 
\begin{equation}
	\mathcal{L} \supset - \frac{1}{\sqrt{2}} \, \lambda^h_{\mu\mu} \, \bar{\mu}_L\mu_R h -\sum_{k\geq 2}\, \frac{1}{k!} \, \lambda_{\mu\mu}^{h^k} \, \bar{\mu}_L\mu_R h^k \, ,
\end{equation}
 can be written as 
 \begin{equation}
	\lambda_{\mu\mu}^{h^k} = \sum_n \frac{(2n+1)!}{2^{k/2}(2n+1-k)!} \, C_{\mu H}^{(n)}\,v^{2n+1-k} \,  ,
\label{eq:coupling_k}
 \end{equation}
and the total cross section for $\mu^+\mu^- \rightarrow h^k$, neglecting the Higgs mass, is given by
 \begin{equation}
 	\sigma_{\mu^+\mu^- \rightarrow h^k} = \frac{s^{k-2}}{2^{4k-3}\pi^{2k-3}k!(k-1)!(k-2)!} \,\left| \lambda_{\mu\mu}^{h^k}\right|^2 \, .
\label{eq:xsection_k}
\end{equation}
From these results we can make several interesting observations. If the contribution of one operator to $\Delta\kappa_\mu$ dominates, or for simplicity only one of the operators is present, then the cross sections for $k$-Higgs productions are all proportional to $|\Delta\kappa_\mu |$. Cross sections for di-Higgs and tri-Higgs productions are also proportional to $(2n+1)^2$ and $(4n^2-1)^2$ respectively, and thus the rates resulting from operators of dimensions higher than six are dramatically larger than those presented in the main results. The ratio of these cross sections in such a case is given by 
\begin{equation}
	\frac{\sigma_{\mu^+\mu^-\rightarrow hhh} }{\sigma_{\mu^+\mu^-\rightarrow hh}} = \frac{(2n-1)^2}{192\pi^2}\frac{s}{v^2},
\end{equation}
and thus with increasing $n$ the tri-Higgs signal starts dominating at lower  $\sqrt{s}$. In addition, for $n>1$, large signals with more Higgs bosons in final states are expected and   could provide further sensitive probes of a modified muon Yukawa coupling.  For example,  assuming that only the dimension-eight mass operator is present ($n=2$), we get
 \begin{eqnarray}
 	\lambda_{\mu\mu}^{hhhh} &=& \frac{15}{2}\,\frac{m_\mu}{v^4}\, \Delta\kappa_\mu ,\\
	\lambda_{\mu\mu}^{hhhhh} &=& \frac{15}{2\sqrt{2}}\,\frac{m_\mu}{v^5}\,\Delta\kappa_\mu, \end{eqnarray} 
and 
   \begin{eqnarray}
	\sigma_{\mu^+\mu^-\rightarrow hhhh} &=&\frac{5^2s^2}{2^{20}\pi^5}\,\left(\frac{m_\mu}{v^4}\right)^2\,\left| \Delta\kappa_\mu\right |^2,\\
	\sigma_{\mu^+\mu^-\rightarrow hhhhh} &=&\frac{5\,s^3}{3\cdot 2^{27}\pi^7}\,\left(\frac{m_\mu}{v^5}\right)^2 \, \left| \Delta\kappa_\mu\right |^2.
 \end{eqnarray}
 
 \begin{figure}[h!]
  \includegraphics[scale=0.24]{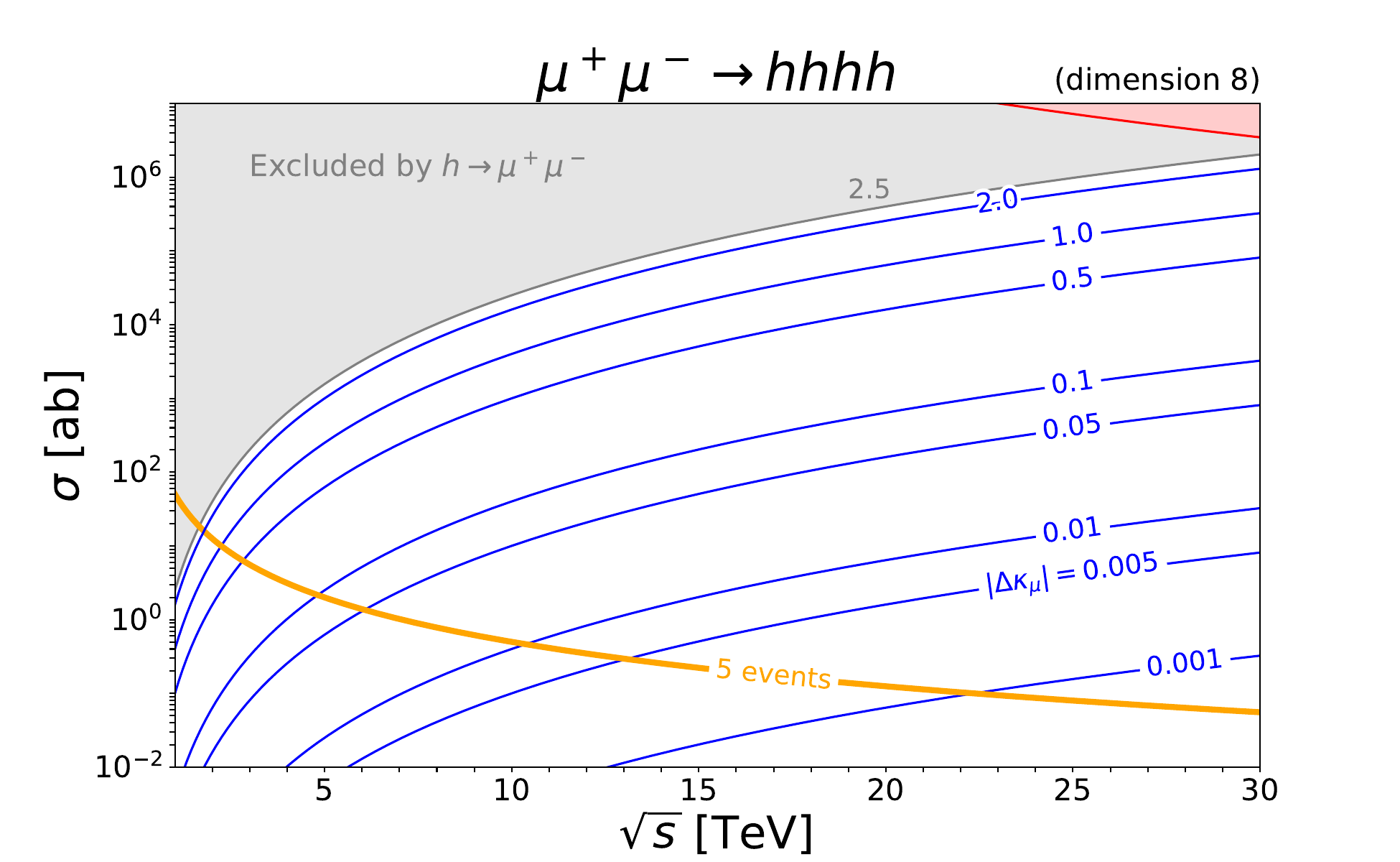}
  \includegraphics[scale=0.24]{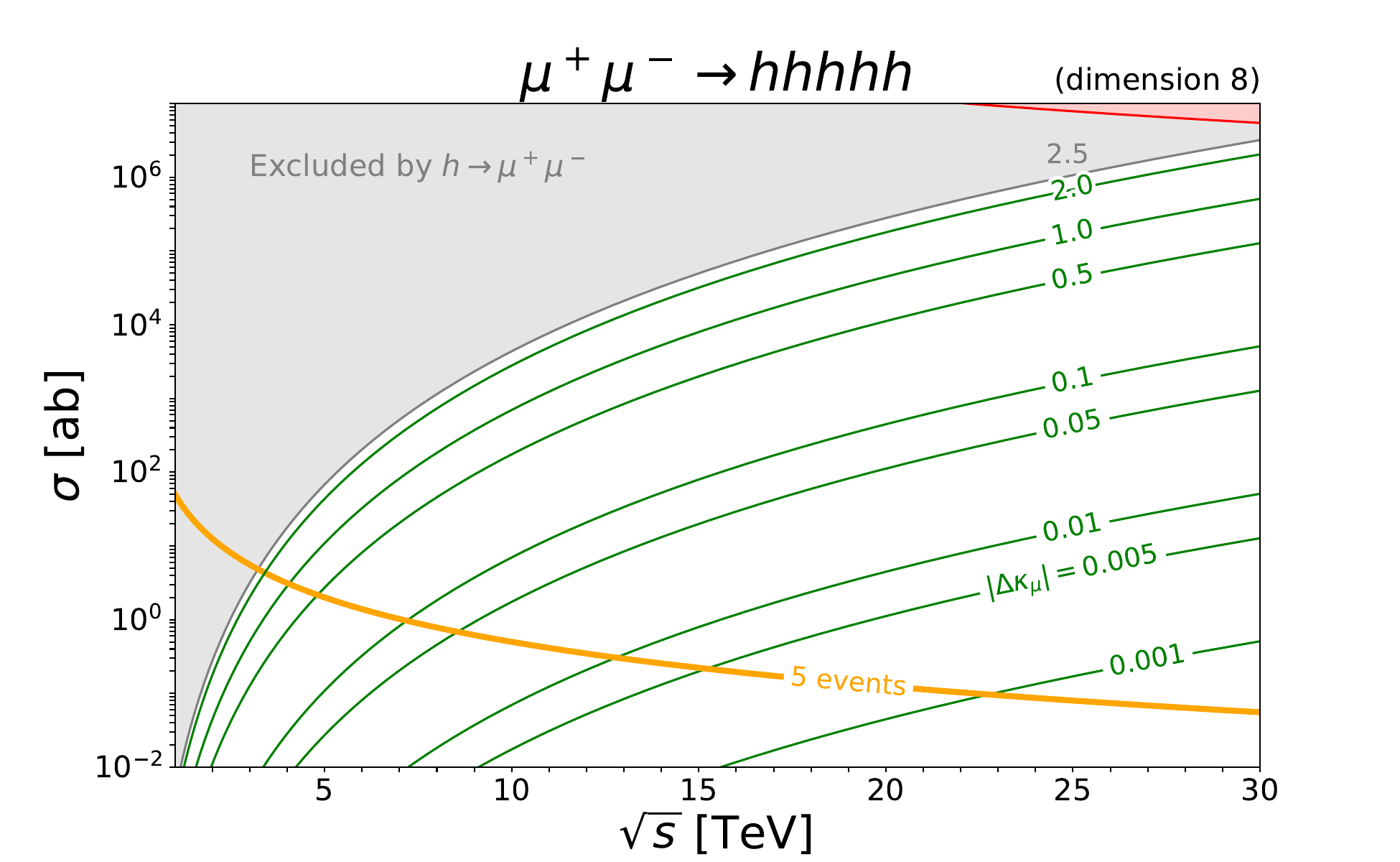}
\caption{Production cross sections for four and five Higgs bosons as functions of $\sqrt{s}$ for various $|\Delta\kappa_\mu|$ assuming that only the dimension-eight mass operator is present.  Gray shaded regions are excluded at 95\% C.L. by the CMS search for $h\to \mu^+\mu^-$~\cite{CMS:2020xwi}. The red shaded regions are excluded by unitarity constraints. Orange lines indicate the cross sections corresponding to five signal events.}
\label{fig:hhhh-hhhhh-dim8}
\end{figure}
These cross sections are plotted as functions of $\sqrt{s}$ for various $|\Delta\kappa_\mu|$ in Fig.~\ref{fig:hhhh-hhhhh-dim8}. 
Indicated unitarity constraints on the largest possible $|\Delta\kappa_\mu|$ resulting from the dimension-eight mass operator, given by Eqs.~(\ref{eq:unitarity_4h}) and (\ref{eq:unitarity_5h}), will be discussed in detail in the following section and  Appendix~\ref{sec:23scat}. We see that compared to the tri-Higgs signal resulting from the dimension-six operator, the four or five Higgs final states are potentially sensitive to an order of magnitude smaller $|\Delta\kappa_\mu|$.

However, if two or more operators of various dimensions contribute to $\Delta\kappa_\mu$ it is  possible that, due to accidental cancellations, some of the $k$-Higgs final states are highly suppressed. Although this is not expected, we note that, by measuring cross sections for all $\mu^+\mu^-\rightarrow h^k$ together with $h\to \mu^+ \mu^-$, the magnitudes of all Wilson coefficients  can be determined, and further constraints on their phases can be obtained. Specifically, the $n$th-Wilson coefficient contributes to $k \leq 2n + 1$ Higgs final states, whereas the $(n-1)$th coefficient contributes to $k \leq 2n - 1$, meaning the magnitude of $C_{\mu H}^{(n)}$ can only be probed by measurements involving $k = 2n + 1$ or $k = 2n$ Higgses. The lower multiplicity $2 \leq k \leq 2n - 1$ processes will contain information about the magnitudes and interferences between the remaining $C_{\mu H}^{(n-1)}$ terms. Lastly, by using $h \rightarrow \mu^+ \mu^-$ as the remaining observable, a total of $2n$ constraints are found, determining the magnitudes and phases of all $n$-contributing Wilson coefficients, up to a possible sign of each phase.

It is interesting to note that there are several effective couplings involving Goldstone bosons, that are directly related to $\Delta\kappa_\mu$ regardless of its origin. From Table~\ref{table:SM_G_couplings_n-operator} we see that, for example,
\begin{equation}
	\lambda_{\mu\mu}^{GG} = \sum_{n} n\,C_{\mu H}^{(n)}\,v^{2n-1} = \frac{m_\mu}{2v^2}\,\Delta\kappa_\mu \, ,
\end{equation}
and thus the contribution of mass operators to $\mu^+ \mu^- \rightarrow Z_L Z_L$ is always directly related to a modification of the muon Yukawa coupling. The contributions to other di-boson processes, $\mu^+ \mu^- \rightarrow  W_L^\pm W_L^\mp$ and $Z_L h$ are also directly related to a modification of the muon Yukawa coupling. Among all possible tri-boson processes this is the case for  only $\mu^+ \mu^- \rightarrow  W_L^+ W_L^- Z_L$ and $Z_L Z_LZ_L$, see Table~\ref{table:SM_G_couplings_n-operator}. Thus one might be tempted to conclude that these processes are more direct probes of a modified Yukawa coupling. However, as discussed above, all these final states are also affected by other dimension-six operators not related to the Yukawa coupling at all. Considering other dimension-eight operators that do not contribute to the muon Yukawa coupling, we find that also $hh$, $hhh$, and $hZ_L Z_L$ are affected at this level~\cite{Murphy:2020rsh}.

\section{Modified muon Yukawa coupling --  2HDM type-II}
\label{sec:2HDM}
In the low energy theory described by 2HDM with type-II couplings of Higgs doublets to SM leptons, there are four independent dimension-six operators parametrizing the effect of new physics that modify the muon mass and Yukawa coupling. The effective Lagrangian of mass operators up to dimension-six is given by~\cite{Dermisek:2023tgq}
\begin{equation}
\begin{split}
    \mathcal{L} &= -y_{\mu}\bar{l}_{L}\mu_{R}H_d \;-\; C_{\mu H_d}\bar{l}_{L}\mu_{R}H_d\left(H^{\dagger}_d H_d\right) \\
    & - C_{\mu H_u}^{(1)} \bar{l}_{L}\mu_{R}H_d\left(H^{\dagger}_u H_u\right) - C_{\mu H_u}^{(2)} \bar{l}_{L}\mu_{R} \cdot H_u^{\dagger }\left(H_d \cdot H_u\right) - C_{\mu H_u}^{(3)} \bar{l}_{L}\mu_{R} \cdot H_u^{\dagger }\left(H_d^{\dagger} \cdot H_u^{\dagger} \right) + {\rm H.c.},
\end{split}
\label{eq:eff_lagrangian_2HDM}
\end{equation}
where $\mathcal{O}_{\mu H_d}$ closely resembles the dimension-six mass operator discussed in the previous section, while $\mathcal{O}_{\mu H_u}^{(1)}$, $\mathcal{O}_{\mu H_u}^{(2)}$, and $\mathcal{O}_{\mu H_u}^{(3)}$ contain both Higgs doublets simultaneously. The components of the two Higgs doublets are $H_d=(H_d^+, H_d^0)^{T}$ and $H_u=(H_u^0, H_u^-)^{T}$. For example, the $\mathcal{O}_{\mu H_d}$ operator can be written in component notation as $-C_{\mu H_d}\bar{\mu}_L\mu_R(H_d^0H_d^+H_d^- + H_d^0{H_d^{0}}^*H_d^0)$.  The $SU(2)\times U(1)_Y$  quantum numbers and the $Z_2$ charges of the fields resulting in type-II couplings are $l_L\;({\bf 2}_{-1/2},+)$, $\mu_R\;({\bf 1}_{-1},-)$, $H_u \;({\bf 2}_{-1/2},+)$, and $H_d \;({\bf 2}_{1/2},-)$. The explicit $``\cdot"$ represents contraction of $SU(2)$ doublets via the antisymmetric $\epsilon_{ij}$, where $\epsilon_{12} = - \epsilon_{21} = +1$. Note that there is yet another operator, $-C_{\mu H_u}^{\, '}\bar l_L\mu_RH_u(H_u^\dagger H_d)$, allowed by all symmetries, which, however, can be written as a linear combination of $\mathcal{O}_{\mu H_u}^{(1)}$ and $\mathcal{O}_{\mu H_u}^{(2)}$ operators:
\begin{equation}
\mathcal{O}_{\mu H_u}^{\, '} =  \mathcal{O}_{\mu H_u}^{(1)} - \mathcal{O}_{\mu H_u}^{(2)}.
\label{eq:O'}
\end{equation}
Furthermore, this operator after electroweak symmetry breaking (EWSB) does not contribute to the muon mass. 

\begin{table}[t!]
    \caption{$SU(2)_L \times U(1)_Y \times Z_2$ quantum numbers of 22 possible UV completions involving vectorlike leptons $L$ and $E$. $(\pm, \pm)$ represents $Z_2$ charge assignments of $L$ and $E$, respectively, required to generate at tree level the dimension-six mass operators labeled at the top of each column. When two operators are generated simultaneously in the same model, the corresponding Wilson coefficients are correlated  by a factor indicated in front of $(\pm, \pm)$. }
\centering
 \begin{tabular}{||c c c c c||} 
 \hline
 $L \oplus E$ & \ \ \ $\mathcal{O}_{\mu H_d}$ & \ \ \ $\mathcal{O}_{\mu H_u}^{(1)} $ & \ \ \ $\mathcal{O}_{\mu H_u}^{(2)} $ & \ \ \ $\mathcal{O}_{\mu H_u}^{(3)}$ \\ [1.0ex] 
 \hline
 $\mathbf{2}_{-1/2} \oplus \mathbf{1}_{-1}$ & \ \ \ $(+,-)$ \ \ \ & \ \ \ $(-,-)$ & \ \ \ $(+,+)$ & \ \ \ $(-,+)$ \\
 $\mathbf{2}_{-1/2} \oplus \mathbf{3}_{-1}$ & \ \ \ $(+,-)$ \ \ \ & \ \ \ $(+,+); (-,-)$ & \ \ \ $-\frac{1}{2} (+,+); -2 (-,-)$ & \ \ \ $(-,+)$ \\
 $\mathbf{2}_{-3/2} \oplus \mathbf{1}_{-1}$ & \ \ \ $(+,-)$ \ \ \ & \ \ \ $(-,-)$ & \ \ \ $(-,+)$ & \ \ \ $(+,+)$ \\
 $\mathbf{2}_{-3/2} \oplus \mathbf{3}_{-1}$ & \ \ \ $(+,-)$ \ \ \ & \ \ \ $(-,-);(-,+)$ & \ \ \ $-2 (-,-); - \frac{1}{2} (-,+)$ & \ \ \ $(+,+)$ \\
 $\mathbf{2}_{-1/2} \oplus \mathbf{1}_{0}$ & \ \ \ \ \ \ & \ \ \ $(+,+); (-,+)$ & \ \ \ $-1 (+,+); -1 (-,+)$ & \ \ \ \\
 $\mathbf{2}_{-1/2} \oplus \mathbf{3}_{0}$ & \ \ \ $(+,-)$ \ \ \ & \ \ \ $(+,+); (-,+)$ & \ \ \ $+1(+,+); +1(-,+)$ & \ \ \ $(-,-)$ \\ [1ex] 
 \hline
 \end{tabular}
    \label{tab:cmuhu_reps}
\end{table}

Although in general all the dimension-six mass operators can be present simultaneously, specific UV completions typically generate just one or two of them as dominant ones. There are 22 possible UV completions that generate these operators at tree level, categorized  in  Table~\ref{tab:cmuhu_reps} according to gauge and $Z_2$ charge assignments of new leptons, $L \oplus E$, that are assumed to come in vectorlike pairs. Note that the models which generate $C_{\mu H_d}$ at tree level \textit{do not} generate any operator involving $H_u$, and that only representations containing an $SU(2)$ triplet or a SM singlet can generate two operators ($\mathcal{O}_{\mu H_u}^{(1)}$ and $\mathcal{O}_{\mu H_u}^{(2)}$) simultaneously at tree level.\footnote{However, in general all the operators are expected to be generated at loop level. For example, the model with $\textbf{2}_{-1/2} \oplus \textbf{1}_{-1}$, and $(+,-) \ Z_2$ charges generates only $C_{\mu H_d}$ at tree level, but generates all other operators at loop level which was studied in detail in Ref.~\cite{Dermisek:2023tgq}.}  When these operators are simultaneously generated, the corresponding Wilson coefficients are correlated, $C_{\mu H_u}^{(2)} = a \times C_{\mu H_u}^{(1)}$, where $a$ is shown in the fourth column of Table~\ref{tab:cmuhu_reps} by the multiplying factor in front of $(\pm, \pm)$. Note that for the models with SM singlets  (fifth line in the table), the generated $C_{\mu H_u}^{(1)}$ and $C_{\mu H_u}^{(2)}$ can be replaced by just $-C_{\mu H_u}^{\, '}$ through Eq.~(\ref{eq:O'}), and thus these models do not contribute to the muon mass for any $Z_2$ charges.
Finally, besides the models in Table~\ref{tab:cmuhu_reps}, there is an infinite number of possible UV completions that generate the mass operators at loop level \cite{Dermisek:2023nhe}. 

In the main text we will present detailed results for the $\mathcal{O}_{\mu H_d}$ operator only. The results for other operators sufficient for obtaining predictions in any specific model will be summarized in Appendix~\ref{sec:c1c2c3}. We focus on the $\mathcal{O}_{\mu H_d}$ operator because it is generated by the new leptons with the same quantum numbers as SM leptons (including the $Z_2$ charges). Furthermore, as we will see, the di-Higgs and tri-Higgs signals resulting from this operator feature the largest possible $\tan \beta$ enhancements.

When the neutral components of the two Higgs doublets develop vacuum expectation values, $\left<H_d^0\right> = v_d = v\cos\beta$ and $\left<H_u^0\right> = v_u = v\sin\beta$, the neutral and charged components of the doublets can be written in terms of mass eigenstates and Goldstone bosons as
\begin{eqnarray}
	H_d^0 &=& v_d + \frac{1}{\sqrt{2}}(-h\sin\alpha+H\cos\alpha)+\frac{i}{\sqrt{2}}(G\cos\beta-A\sin\beta), \\
	H_u^0 &=& v_u + \frac{1}{\sqrt{2}}(h\cos\alpha+H\sin\alpha)-\frac{i}{\sqrt{2}}(G\sin\beta+A\cos\beta),
\end{eqnarray}
and 
\begin{eqnarray}
	H_d^{\pm} &=& \cos\beta G^{\pm}- \sin\beta H^{\pm}, \\
	H_u^{\pm} &=& -\sin\beta G^{\pm} - \cos\beta H^{\pm},
\end{eqnarray}
where $h$ and $H$ are the light and heavy $CP$-even Higgs bosons, $A$ is the $CP$-odd Higgs boson, $H^\pm$ are the charged Higgs bosons, and $G$ and $G^{\pm}$ are the neutral and charged  Goldstone bosons, respectively. The angle $\alpha$ is a rotation angle that diagonalizes the $CP$-even Higgs boson mass matrix. For more details on the notation, see Ref.~\cite{Dermisek:2023tgq}. 
 
 The $\mathcal{O}_{\mu H_d}$  operator generates an additional contribution to the muon mass 
\begin{equation}
m_\mu = y_\mu v_d + C_{\mu H_d} v_d^3,
\label{eq:mmu_CmuHd}
\end{equation} 
and, in the basis where the muon mass is real and positive, the resulting interactions with the Higgs bosons in the 2HDM are described by
\begin{eqnarray}
    \pazocal{L}  \supset 
     &-&\frac{1}{\sqrt{2}}\lambda^h_{\mu\mu}\bar{\mu}_L\mu_R h -\frac{1}{\sqrt{2}}\lambda^H_{\mu\mu}\bar{\mu}_L\mu_R H -\frac{1}{\sqrt{2}}\lambda^A_{\mu\mu}\bar{\mu}_L\mu_R A \nonumber \\
     &-& \frac{1}{2!}\lambda_{\mu\mu}^{hh}\bar{\mu}_L\mu_R h^2- \frac{1}{2!}\lambda_{\mu\mu}^{AA}\bar{\mu}_L\mu_R A^2 - \frac{1}{2!} \lambda_{\mu\mu}^{HH}\bar{\mu}_L\mu_R H^2  - \lambda_{\mu\mu}^{hH}\bar{\mu}_L\mu_R hH \nonumber \\  
     &-& \lambda_{\mu\mu}^{hA}\bar{\mu}_L\mu_R hA - \lambda_{\mu\mu}^{HA}\bar{\mu}_L\mu_R HA 
    - \lambda_{\mu\mu}^{H^+H^-}\bar{\mu}_L\mu_R H^+H^- - \frac{1}{3!} \lambda_{\mu\mu}^{hhh}\bar{\mu}_L\mu_R h^3 \nonumber \\  
    &-& \frac{1}{3!} \lambda_{\mu\mu}^{AAA}\bar{\mu}_L\mu_R A^3 - \frac{1}{3!}\lambda_{\mu\mu}^{HHH}\bar{\mu}_L\mu_R H^3 -\frac{1}{2!} \lambda_{\mu\mu}^{hhH}\bar{\mu}_L\mu_R h^2H - \frac{1}{2!} \lambda_{\mu\mu}^{hhA}\bar{\mu}_L\mu_R h^2A    \nonumber \\ 
    & -&\frac{1}{2!}\lambda_{\mu\mu}^{hAA}\bar{\mu}_L\mu_R hA^2 - \frac{1}{2!} \lambda_{\mu\mu}^{hHH}\bar{\mu}_L\mu_R hH^2   - \frac{1}{2!} \lambda_{\mu\mu}^{AHH}\bar{\mu}_L\mu_R AH^2  \nonumber \\
        & -& \frac{1}{2!} \lambda_{\mu\mu}^{HAA}\bar{\mu}_L\mu_R HA^2 
    - \lambda_{\mu\mu}^{hH^+H^-}\bar{\mu}_L\mu_R hH^+H^- 
    -  \lambda_{\mu\mu}^{HH^+H^-}\bar{\mu}_L\mu_R HH^+H^-  \nonumber \\ 
     & -& \lambda_{\mu\mu}^{AH^+H^-}\bar{\mu}_L\mu_R AH^+H^-  
    - \lambda_{\mu\mu}^{hHA}\bar{\mu}_L\mu_R hHA +{\rm H.c.},
\label{eqn:couplings_2HDM}
\end{eqnarray}
where the couplings are summarized in Table~\ref{table:couplings} in terms of the Wilson coefficient, the VEV, $\alpha$ and $\beta$,  and also in the alignment limit, $\alpha = \beta - \frac{\pi}{2}$, where $h$ is SM-like.
The last column contains couplings in the alignment limit written in terms of $\Delta\kappa_\mu$ that, with the same definitions of $\kappa_\mu$ and $\Delta \kappa_\mu$ as in Eqs.~(\ref{eq:kappa}) and (\ref{eq:Delta_kappa}), is given by  
\begin{equation}
\Delta \kappa_\mu = 2\, C_{\mu H_d}\, \frac{v_d^3}{m_\mu}.
\label{eq:Delta_kappa_CmuHd}
\end{equation}

\begin{table*}[h!]
\caption{Coupling constants describing interactions with the 2HDM Higgs bosons in Eq.~(\ref{eqn:couplings_2HDM}).}
\begin{ruledtabular}
\begin{tabular}{cccc}
    & In general & Alignment limit ($\alpha = \beta - \frac{\pi}{2})$ & In terms of $\Delta\kappa_\mu$\\
 \hline
  $\lambda_{\mu\mu}^{h}$  &  $\frac{m_\mu+2v^3\cos^3{\beta}\;C_{\mu H_d}}{v}\left(\frac{-\sin\alpha}{\cos\beta}\right)$ & $\frac{m_\mu+2v^3\cos^3{\beta}\;C_{\mu H_d}}{v}$ & $\frac{m_\mu}{v}(1+\Delta\kappa_\mu)$\\
 \hline
   $\lambda_{\mu\mu}^{H}$  &  $\frac{m_\mu+2v^3\cos^3{\beta}\;C_{\mu H_d}}{v}\left(\frac{\cos\alpha}{\cos\beta}\right)$ &$\frac{m_\mu+2v^3\cos^3{\beta}\;C_{\mu H_d}}{v}\tan\beta$& $\frac{m_\mu}{v}(1+\Delta\kappa_\mu)\tan\beta$\\
 \hline
 $\lambda_{\mu\mu}^{A}$  &  $-i\frac{m_\mu}{v}\tan\beta$  & $-i\frac{m_\mu}{v}\tan\beta$ & $-i\frac{m_\mu}{v}\tan\beta$\\
\hline  
 $\lambda_{\mu\mu}^{hh}$  &  $3v\cos{\beta}\sin^2{\alpha}\;C_{\mu H_d}$ & $3v\cos^3{\beta}\;C_{\mu H_d}$ & $\frac{3m_\mu}{2v^2}\Delta\kappa_\mu$\\
\hline  
 $\lambda_{\mu\mu}^{AA}$  & $v\cos{\beta}\sin^2{\beta}\;C_{\mu H_d}$ & $v\sin^2{\beta}\cos{\beta}\;C_{\mu H_d}$ & $\frac{m_\mu}{2v^2}\Delta\kappa_\mu\tan^2\beta$ \\
\hline 
 $\lambda_{\mu\mu}^{HH}$  & $3v\cos{\beta}\cos^2{\alpha}\;C_{\mu H_d}$  & $3v\sin^2{\beta}\cos{\beta}\;C_{\mu H_d}$ & $\frac{3m_\mu}{2v^2}\Delta\kappa_\mu\tan^2\beta$\\
\hline
 $\lambda_{\mu\mu}^{hH}$  & $-3v\cos{\beta}\sin{\alpha}\cos{\alpha}\;C_{\mu H_d}$ & $3v\sin\beta\cos^2{\beta}\;C_{\mu H_d}$ & $\frac{3m_\mu}{2v^2}\Delta\kappa_\mu\tan\beta$\\
\hline
 $\lambda_{\mu\mu}^{hA}$  & $iv\cos{\beta}\sin\beta\sin\alpha\;C_{\mu H_d}$ & $-iv\sin\beta\cos^2{\beta}\;C_{\mu H_d}$ & $-\frac{i m_\mu}{2v^2}\Delta\kappa_\mu\tan\beta$\\
\hline
 $\lambda_{\mu\mu}^{HA}$  &  $-iv\cos\beta\sin\beta\cos\alpha\;C_{\mu H_d}$ & $-iv\sin^2\beta\cos\beta\;C_{\mu H_d}$  & $-\frac{im_\mu}{2v^2}\Delta\kappa_\mu\tan^2\beta$\\
\hline
 $\lambda_{\mu\mu}^{H^+H^-}$  &  $v\cos\beta\sin^2{\beta}\;C_{\mu H_d}$ & $v\sin^2{\beta}\cos\beta\;C_{\mu H_d}$  & $\frac{m_\mu}{2v^2}\Delta\kappa_\mu\tan^2\beta$\\
\hline
 $\lambda_{\mu\mu}^{hhh}$  & $-\frac{3}{\sqrt{2}}\sin^3{\alpha}\;C_{\mu H_d}$  & $\frac{3}{\sqrt{2}}\cos^3{\beta}\;C_{\mu H_d}$  & $\frac{3m_\mu}{2\sqrt{2}v^3}\Delta\kappa_\mu$\\
\hline
 $\lambda_{\mu\mu}^{AAA}$  &  $-\frac{3}{\sqrt{2}}i\sin^3{\beta}\;C_{\mu H_d}$ & $-\frac{3}{\sqrt{2}}i\sin^3{\beta}\;C_{\mu H_d}$  & $-\frac{3im_\mu}{2\sqrt{2}v^3}\Delta\kappa_\mu\tan^3\beta$\\
\hline
 $\lambda_{\mu\mu}^{HHH}$  & $\frac{3}{\sqrt{2}}\cos^3{\alpha}\;C_{\mu H_d}$ & $\frac{3}{\sqrt{2}}\sin^3{\beta}\;C_{\mu H_d}$ & $\frac{3m_\mu}{2\sqrt{2}v^3}\Delta\kappa_\mu\tan^3\beta$\\
\hline
 $\lambda_{\mu\mu}^{hhH}$  &  $\frac{3}{\sqrt{2}}\sin^2{\alpha}\cos\alpha\;C_{\mu H_d}$  & $\frac{3}{\sqrt{2}}\sin\beta\cos^2{\beta}\;C_{\mu H_d}$  & $\frac{3m_\mu}{2\sqrt{2}v^3}\Delta\kappa_\mu\tan\beta$\\
\hline
 $\lambda_{\mu\mu}^{hhA}$  & $-\frac{i}{\sqrt{2}}\sin\beta\sin^2{\alpha}\;C_{\mu H_d}$ & $-\frac{i}{\sqrt{2}}\sin\beta\cos^2{\beta}\;C_{\mu H_d}$ & $-\frac{im_\mu}{2\sqrt{2}v^3}\Delta\kappa_\mu\tan\beta$\\
\hline
 $\lambda_{\mu\mu}^{hAA}$  & $-\frac{1}{\sqrt{2}}\sin^2{\beta}\sin{\alpha}\;C_{\mu H_d}$ & $\frac{1}{\sqrt{2}}\sin^2{\beta}\cos{\beta}\;C_{\mu H_d}$ & $\frac{m_\mu}{2\sqrt{2}v^3}\Delta\kappa_\mu\tan^2\beta$ \\
\hline
 $\lambda_{\mu\mu}^{hHH}$  & $-\frac{3}{\sqrt{2}}\sin\alpha\cos^2{\alpha}\;C_{\mu H_d}$ & $\frac{3}{\sqrt{2}}\sin^2{\beta}\cos\beta\;C_{\mu H_d}$ & $\frac{3m_\mu}{2\sqrt{2}v^3}\Delta\kappa_\mu\tan^2\beta$ \\
\hline
 $\lambda_{\mu\mu}^{AHH}$  & $-\frac{i}{\sqrt{2}}\sin\beta\cos^2{\alpha}\;C_{\mu H_d}$ & $-\frac{i}{\sqrt{2}}\sin^3\beta\;C_{\mu H_d}$  & $-\frac{im_\mu}{2\sqrt{2}v^3}\Delta\kappa_\mu\tan^3\beta$\\
\hline
 $\lambda_{\mu\mu}^{HAA}$  & $\frac{1}{\sqrt{2}}\sin^2{\beta}\cos\alpha\;C_{\mu H_d}$ &  $\frac{1}{\sqrt{2}}\sin^3{\beta}\;C_{\mu H_d}$  & $\frac{m_\mu}{2\sqrt{2}v^3}\Delta\kappa_\mu\tan^3\beta$\\
\hline
 $\lambda_{\mu\mu}^{hH^+H^-}$  &  $-\frac{1}{\sqrt{2}}\sin^2{\beta}\sin\alpha\;C_{\mu H_d}$ & $\frac{1}{\sqrt{2}}\sin^2{\beta}\cos\beta\;C_{\mu H_d}$ & $\frac{m_\mu}{2\sqrt{2}v^3}\Delta\kappa_\mu\tan^2\beta$\\
\hline
 $\lambda_{\mu\mu}^{HH^+H^-}$  &  $\frac{1}{\sqrt{2}}\sin^2{\beta}\cos{\alpha}\;C_{\mu H_d}$ & $\frac{1}{\sqrt{2}}\sin^3{\beta}\;C_{\mu H_d}$ & $\frac{m_\mu}{2\sqrt{2}v^3}\Delta\kappa_\mu\tan^3\beta$\\
\hline
 $\lambda_{\mu\mu}^{AH^+H^-}$  & $-\frac{i}{\sqrt{2}}\sin^3{\beta}\;C_{\mu H_d}$ & $-\frac{i}{\sqrt{2}}\sin^3{\beta}\;C_{\mu H_d}$ & $-\frac{im_\mu}{2\sqrt{2}v^3}\Delta\kappa_\mu\tan^3\beta$\\
\hline
 $\lambda_{\mu\mu}^{hHA}$  & $\frac{i}{\sqrt{2}}\sin\beta\sin\alpha\cos\alpha\;C_{\mu H_d}$ & $-\frac{i}{\sqrt{2}}\sin^2\beta\cos\beta\;C_{\mu H_d}$  & $-\frac{im_\mu}{2\sqrt{2}v^3}\Delta\kappa_\mu\tan^2\beta$\\
\end{tabular}
\end{ruledtabular}
\label{table:couplings}
\end{table*}

\begin{figure}[t!]
 \includegraphics[scale=0.3]{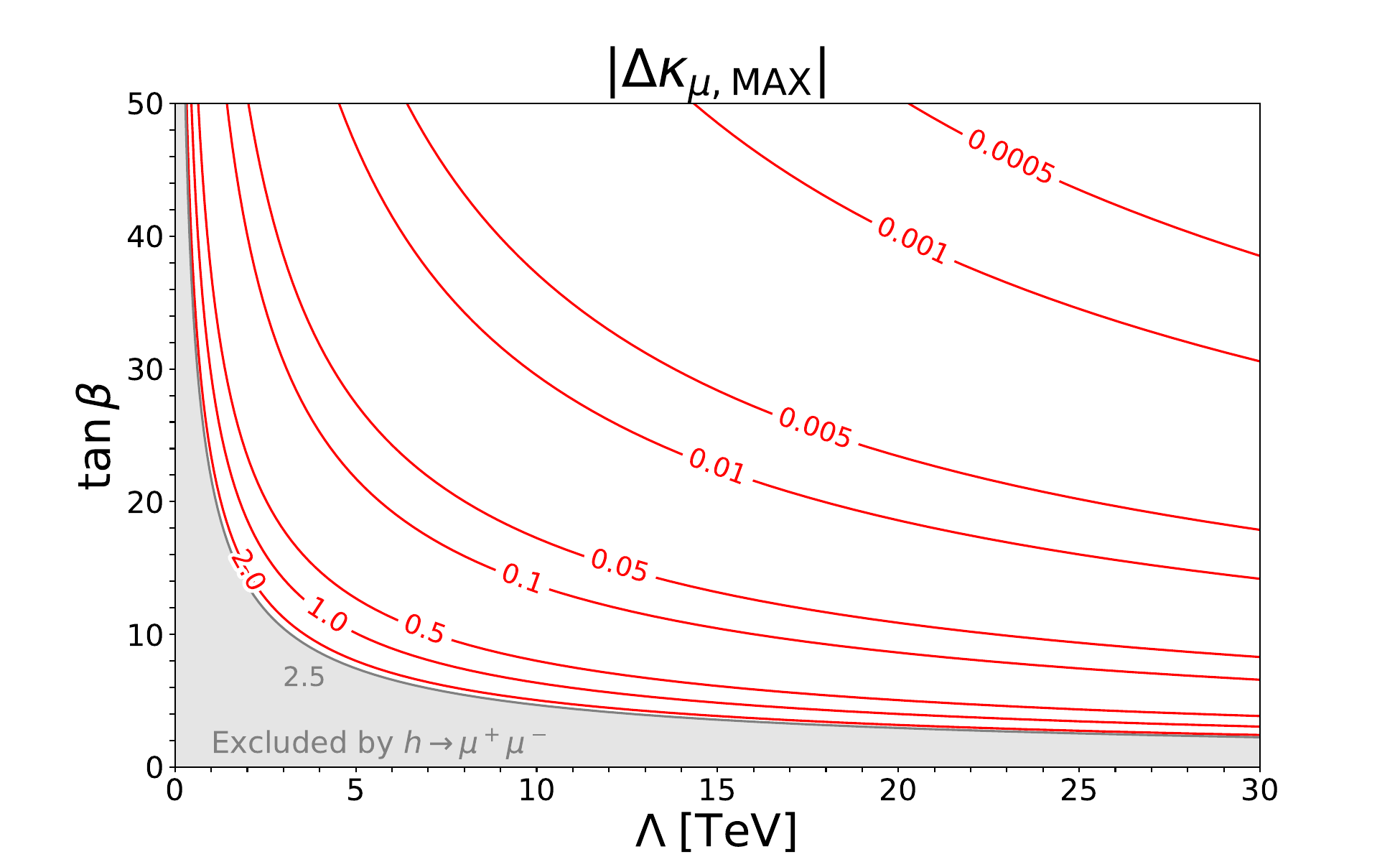}
\caption{Contours of constant $|\Delta\kappa_{\mu, {\rm MAX}}|$ in $\tan\beta$ -- $\Lambda$ plane. The gray  shaded region is excluded at 95\% C.L. by the CMS search for $h\to \mu^+\mu^-$~\cite{CMS:2020xwi}.}
\label{fig:complex-kappa_contour}
\end{figure}

In specific models, after integrating out heavy degrees of freedom at a given scale $\Lambda$, the maximum size of generated $C_{\mu H_d}$ can be limited by perturbativity of the couplings. Without specifying the model, the size of higher-dimensional operators can be constrained by the requirement of preserving $S$-matrix unitarity via partial wave analysis for scattering processes involving the operators themselves in the high energy limit~\cite{DiLuzio:2016sur,Allwicher:2021jkr}. Because of the dimensionality of the operators, $C_i \propto \Lambda^{4-d}$ for $d > 4$, high energy scattering processes are limited by powers of $\sqrt{s} \leq \Lambda$ rather than $\sqrt{s} \rightarrow \infty$ when on-shell resonances of new particles occur at the cutoff scale $\Lambda$ in the theory. Specifically, for the operator of our interest, a unitarity limit on $C_{\mu H_d}$ can be found from the $2 \rightarrow 2$ and $2 \rightarrow 3$ scattering amplitudes for physical states. Details of the calculation can be found in Appendix~\ref{sec:23scat}. We find that
\begin{equation}
    |C_{\mu H_d}| 
    \leq \left(\frac{16\pi}{3 \cos \beta \sin^2 \beta }\right) \frac{1}{v \Lambda} \ \ \ \textrm{and} \ \ \ |C_{\mu H_d}| 
    \leq \left(\frac{128 \pi^2}{\sqrt{3} \sin^3 \beta} \right) \frac{1}{\Lambda^2}
\label{eq:ulim}
\end{equation}
for  $\mu^+\mu^-\rightarrow HH$ and $\mu^+\mu^-\rightarrow HHH$, respectively, where we have neglected the masses of heavy Higgses. In the range of the parameter space with  $\Lambda \geq1\;{\rm TeV}$, $1\leq \tan\beta \leq 50$, and $|\Delta\kappa_\mu|\leq 2.50$ the constraint from $\mu^+\mu^-\rightarrow HHH$ is always stronger. It can be rewritten as a bound on the largest possible $\Delta\kappa_\mu$ through Eq.~(\ref{eq:Delta_kappa_CmuHd}),
\begin{equation}
|\Delta\kappa_\mu| \leq  \frac{256\pi^2}{\sqrt{3}\tan^3\beta}\frac{v^3}{m_\mu \Lambda^2},
\label{eq:unitarity-Delta_kappa}
\end{equation}
or, for fixed $\Delta\kappa_\mu$, as a unitarity bound on  $\tan\beta$,
\begin{equation}
\tan\beta \leq  \left(\frac{256\pi^2}{\sqrt{3}|\Delta\kappa_\mu|}\frac{v^3}{m_\mu \Lambda^2}\right)^{1/3} .
\label{eq:unitarity-tanb}
\end{equation}
The maximum $|\Delta\kappa_{\mu}|$ that can be obtained for a given $\tan\beta$  and a scale of new physics $\Lambda$ is plotted in Fig.~\ref{fig:complex-kappa_contour}.

\begin{figure}[h!]
  \includegraphics[scale=0.24]{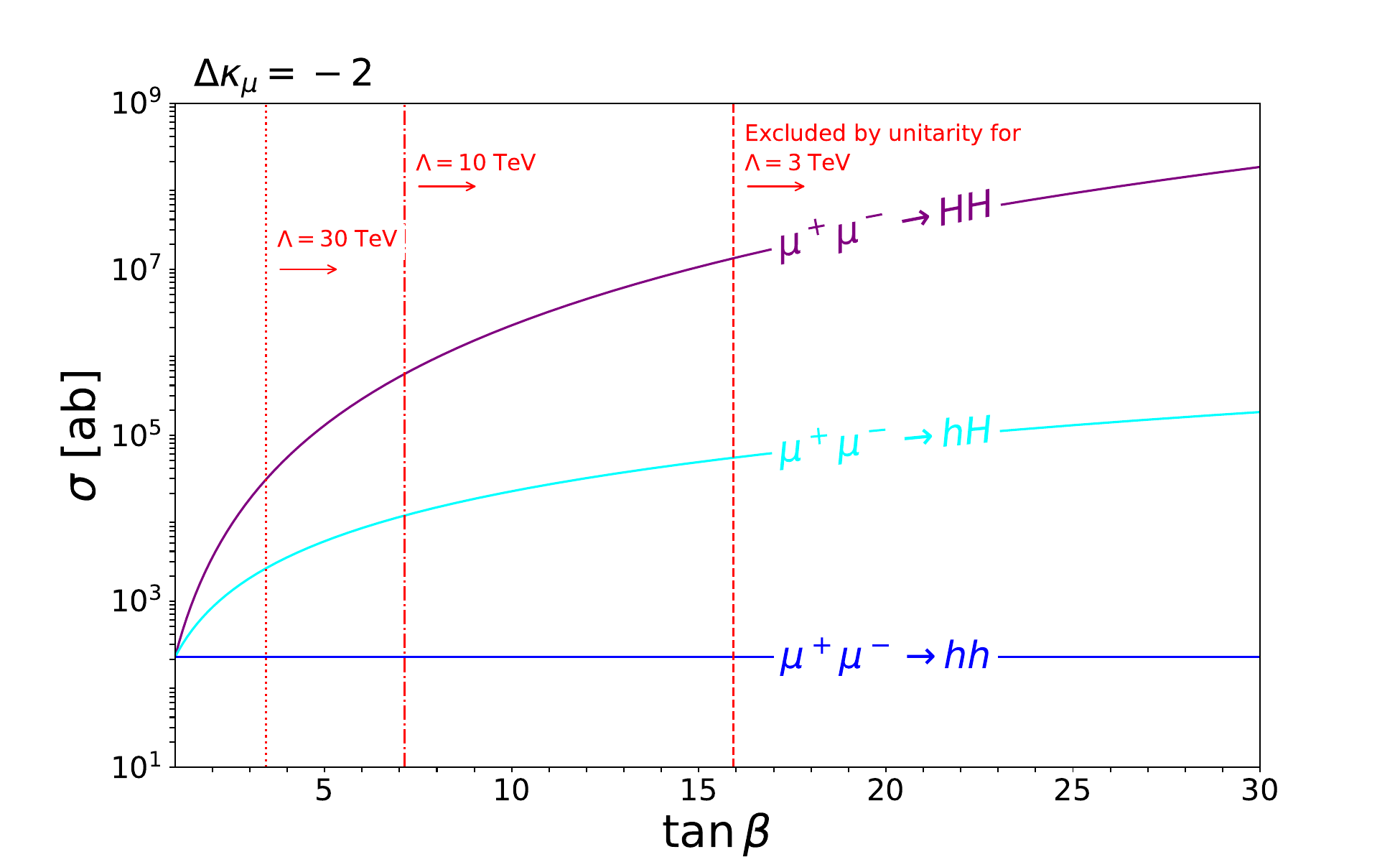}
  \includegraphics[scale=0.24]{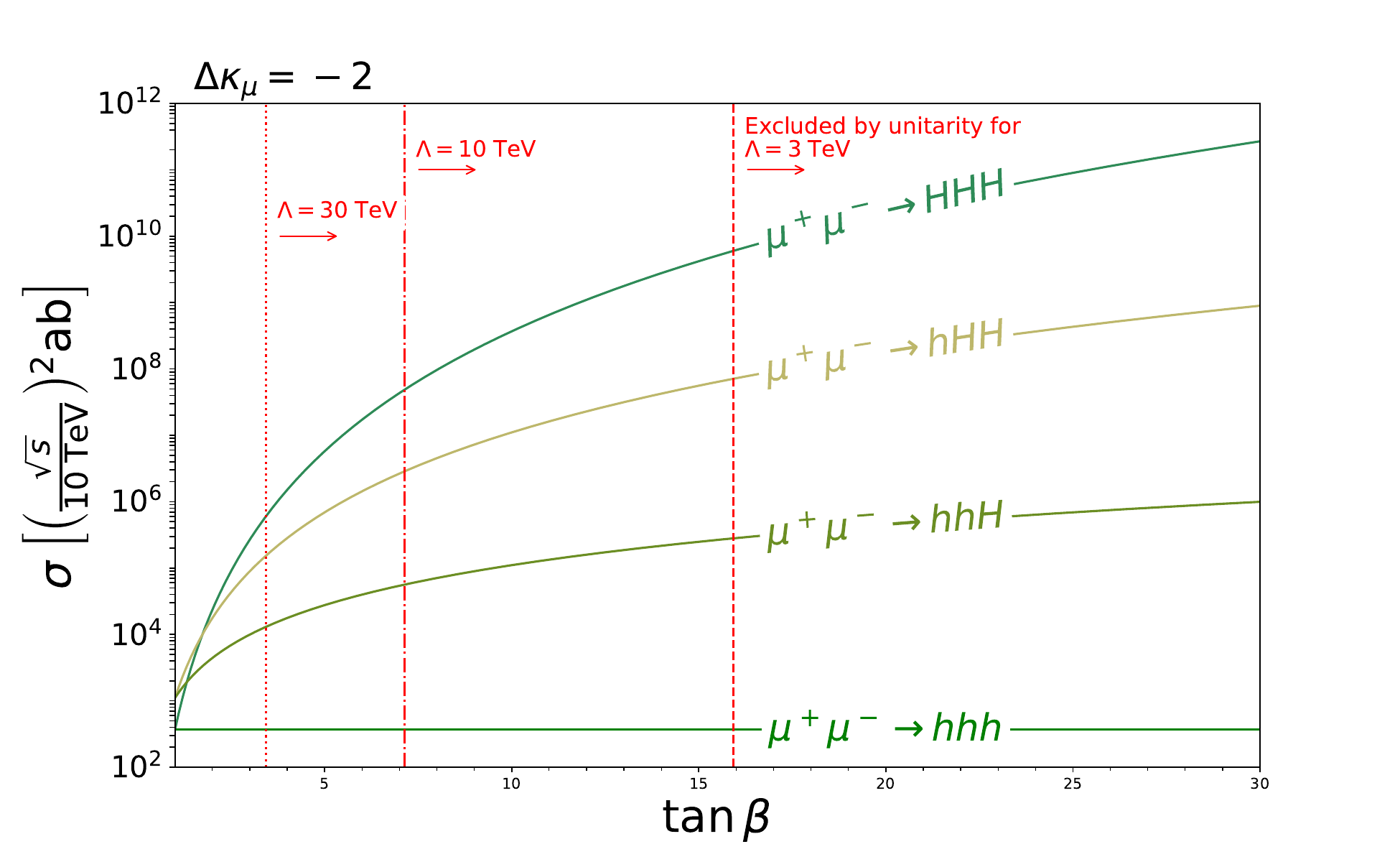}
\caption{Representative di-Higgs and tri-Higgs production cross sections in 2HDM as functions of $\tan\beta$ corresponding to $\Delta \kappa_\mu = -2$.  The unitarity bounds on $\tan \beta$ are indicated by vertical red lines for $\Lambda=30,\, 10$, and $3\;{\rm TeV}$ from left to right.}
\label{fig:2HDM_tanb}
\end{figure}

The cross sections for all combinations of di-Higgs final states and tri-Higgs final states are summarized in Tables~\ref{table:di-Higgs} and \ref{table:tri-Higgs}. Representative cross sections for $\Delta \kappa_\mu = -2$  as functions of $\tan \beta$ are shown in Fig.~\ref{fig:2HDM_tanb}. As in the SM case,  di-Higgs production cross sections do not depend on $\sqrt{s}$, while  tri-Higgs  cross sections grow quadratically with $\sqrt{s}$. The plotted cross sections are excellent approximations when the combined masses of given final states are much smaller than $\sqrt{s}$. The unitarity bound on $\tan \beta$ given in Eq.~(\ref{eq:unitarity-tanb}) for this choice of $\Delta\kappa_\mu$ is indicated for various $\Lambda$.  For tri-Higgs production cross sections only $\sqrt{s} < \Lambda$ should be considered. 
We see that models with 2HDM type-II Higgs sector can generate $\Delta \kappa_\mu = -2$  only for $\tan \beta \lesssim 16$ due to unitarity constraints for $\Lambda \gtrsim 3\;{\rm TeV}$. However, if heavy Higgses are kinematically accessible, the di-Higgs and tri-Higgs final states containing heavy Higgses  provide signals of the opposite sign muon Yukawa coupling which potentially exceed the SM di-Higgs and tri-Higgs signals by orders of magnitude. For example, for $\sqrt{s} = 10$ TeV and $\tan \beta = 5$, the cross section for $\mu^+ \mu^- \to HHH$ is 4 orders of magnitude larger than the cross section for $\mu^+ \mu^- \to hhh$.

\begin{table}[t!]
\caption{Cross sections for all combinations of di-Higgs final states.}
\begin{ruledtabular}
\begin{tabular}{ccc}
 & In terms of $\Delta \kappa_\mu$ & \;\;\;\;\;\;In units of $\sigma_{\mu^+\mu^-\rightarrow hh}$ \;\;\;\;\;\;\\
 \hline
 $\sigma_{\mu^+\mu^-\rightarrow hh}$  &  $\frac{9}{256\pi}\left(\frac{m_\mu}{v^2}\right)^2|\Delta\kappa_\mu|^2$ & 1\\
\hline  
 $\sigma_{\mu^+\mu^-\rightarrow AA}$  & $\frac{1}{256\pi}\left(\frac{m_\mu}{v^2}\right)^2|\Delta\kappa_\mu|^2\tan^4\beta$ & $\frac{1}{9}\tan^4\beta$ \\
\hline 
 $\sigma_{\mu^+\mu^-\rightarrow HH}$  & $\frac{9}{256\pi}\left(\frac{m_\mu}{v^2}\right)^2|\Delta\kappa_\mu|^2\tan^4\beta$  & $\tan^4\beta$ \\
\hline
 $\sigma_{\mu^+\mu^-\rightarrow hH}$  & $\frac{9}{128\pi}\left(\frac{m_\mu}{v^2}\right)^2|\Delta\kappa_\mu|^2\tan^2\beta$  & $2\tan^2\beta$ \\
\hline
 $\sigma_{\mu^+\mu^-\rightarrow hA}$ & $\frac{1}{128\pi}\left(\frac{m_\mu}{v^2}\right)^2|\Delta\kappa_\mu|^2\tan^2\beta$ & $\frac{2}{9}\tan^2\beta$\\
\hline
 $\sigma_{\mu^+\mu^-\rightarrow HA}$ & $\frac{1}{128\pi}\left(\frac{m_\mu}{v^2}\right)^2|\Delta\kappa_\mu|^2\tan^4\beta$  &  $\frac{2}{9}\tan^4\beta$ \\
\hline
 $\sigma_{\mu^+\mu^-\rightarrow H^+H^-}$ & $\frac{1}{128\pi}\left(\frac{m_\mu}{v^2}\right)^2|\Delta\kappa_\mu|^2\tan^4\beta$   &  $\frac{2}{9}\tan^4\beta$  \\
\end{tabular}
\end{ruledtabular}
\label{table:di-Higgs}
\end{table}

\begin{table}[h!]
\caption{Cross sections for all combinations of tri-Higgs final states.}
\begin{ruledtabular}
\begin{tabular}{ccc}
 & In terms of $\Delta \kappa_\mu$ & \;\;\;\;\;\;In units of $\sigma_{\mu^+\mu^-\rightarrow hhh}$ \;\;\;\;\;\;\\\hline
 $\sigma_{\mu^+\mu^-\rightarrow hhh}$  & $\frac{3s}{2^{14}\pi^3}\left(\frac{m_\mu}{v^3}\right)^2|\Delta\kappa_\mu|^2$  &   $1$  \\
\hline
 $\sigma_{\mu^+\mu^-\rightarrow AAA}$  & $\frac{3s}{2^{14}\pi^3}\left(\frac{m_\mu}{v^3}\right)^2|\Delta\kappa_\mu|^2\tan^6\beta$  &  $\tan^6\beta$  \\
\hline
 $\sigma_{\mu^+\mu^-\rightarrow HHH}$  & $\frac{3s}{2^{14}\pi^3}\left(\frac{m_\mu}{v^3}\right)^2|\Delta\kappa_\mu|^2\tan^6\beta$  & $\tan^6\beta$ \\
\hline
 $\sigma_{\mu^+\mu^-\rightarrow hhH}$ & $\frac{9s}{2^{14}\pi^3}\left(\frac{m_\mu}{v^3}\right)^2|\Delta\kappa_\mu|^2\tan^2\beta$  &  $3\tan^2\beta$ \\
\hline
 $\sigma_{\mu^+\mu^-\rightarrow hhA}$ & $\frac{s}{2^{14}\pi^3}\left(\frac{m_\mu}{v^3}\right)^2|\Delta\kappa_\mu|^2\tan^2\beta$  & $\frac{1}{3}\tan^2\beta$\\
\hline
 $\sigma_{\mu^+\mu^-\rightarrow hAA}$ & $\frac{s}{2^{14}\pi^3}\left(\frac{m_\mu}{v^3}\right)^2|\Delta\kappa_\mu|^2\tan^4\beta$  & $\frac{1}{3}\tan^4\beta$  \\
\hline
 $\sigma_{\mu^+\mu^-\rightarrow hHH}$ & $\frac{9s}{2^{14}\pi^3}\left(\frac{m_\mu}{v^3}\right)^2|\Delta\kappa_\mu|^2\tan^4\beta$  & $3\tan^4\beta$ \\
\hline
 $\sigma_{\mu^+\mu^-\rightarrow AHH}$ & $\frac{s}{2^{14}\pi^3}\left(\frac{m_\mu}{v^3}\right)^2|\Delta\kappa_\mu|^2\tan^6\beta$  & $\frac{1}{3}\tan^6\beta$ \\
\hline
 $\sigma_{\mu^+\mu^-\rightarrow HAA}$  &  $\frac{s}{2^{14}\pi^3}\left(\frac{m_\mu}{v^3}\right)^2|\Delta\kappa_\mu|^2\tan^6\beta$  & $\frac{1}{3}\tan^6\beta$\\
\hline
 $\sigma_{\mu^+\mu^-\rightarrow hH^+H^-}$  & $\frac{s}{2^{13}\pi^3}\left(\frac{m_\mu}{v^3}\right)^2|\Delta\kappa_\mu|^2\tan^4\beta$  & $\frac{2}{3}\tan^4\beta$\\
\hline
 $\sigma_{\mu^+\mu^-\rightarrow HH^+H^-}$  & $\frac{s}{2^{13}\pi^3}\left(\frac{m_\mu}{v^3}\right)^2|\Delta\kappa_\mu|^2\tan^6\beta$  & $\frac{2}{3}\tan^6\beta$\\
\hline
 $\sigma_{\mu^+\mu^-\rightarrow AH^+H^-}$ & $\frac{s}{2^{13}\pi^3}\left(\frac{m_\mu}{v^3}\right)^2|\Delta\kappa_\mu|^2\tan^6\beta$  & $\frac{2}{3}\tan^6\beta$\\
\hline
 $\sigma_{\mu^+\mu^-\rightarrow hHA}$ & $\frac{s}{2^{13}\pi^3}\left(\frac{m_\mu}{v^3}\right)^2|\Delta\kappa_\mu|^2\tan^4\beta$  & $\frac{2}{3}\tan^4\beta$\\
\end{tabular}
\end{ruledtabular}
\label{table:tri-Higgs}
\end{table}

\begin{figure}[h!]
  \includegraphics[scale=0.3]{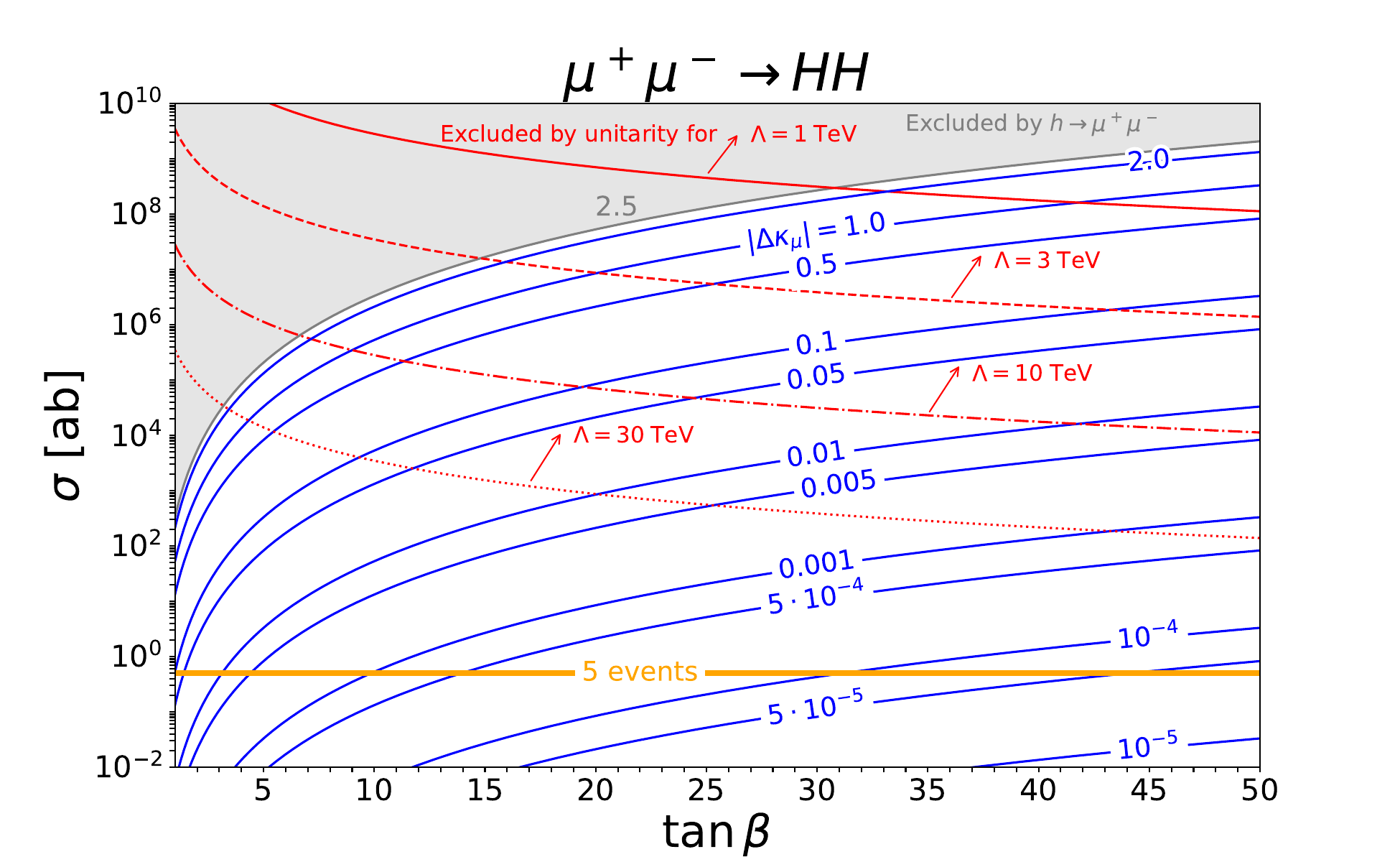}
\caption{The total cross section for $\mu^+\mu^-\rightarrow HH$ as a function of $\tan \beta=1$ for various $|\Delta \kappa_\mu|$. The gray  shaded region is excluded at 95\% C.L. by the CMS search for $h\to \mu^+\mu^-$~\cite{CMS:2020xwi}. The red lines with arrows indicate regions excluded by unitarity constraints for $\Lambda=30,\, 10,\, 3,$ and $1\;{\rm TeV}$. The orange line indicates the cross section corresponding to five signal events assuming the luminosity expected at a $\sqrt{s} = 10$ TeV muon collider.}
\label{fig:HH_tanb}
\end{figure}

The cross section for $\mu^+\mu^-\rightarrow HH$  is also plotted as a function of $\tan \beta$ for various $|\Delta \kappa_\mu|$ in Fig.~\ref{fig:HH_tanb}. The orange line indicates the cross section corresponding to five signal events assuming the luminosity expected at a $\sqrt{s} = 10$ TeV muon collider, see Eq.~(\ref{eq:lum}). We see that, as a result of $\tan^4 \beta$ enhancement, the $\mu^+\mu^-\rightarrow HH$ signal, if kinematically open, can be used to observe a deviation in  the muon Yukawa coupling at the 2\% level for $\tan \beta = 1$ and at the 0.004\% level for $\tan \beta = 50$.

\begin{figure}[t!]
  \includegraphics[scale=0.24]{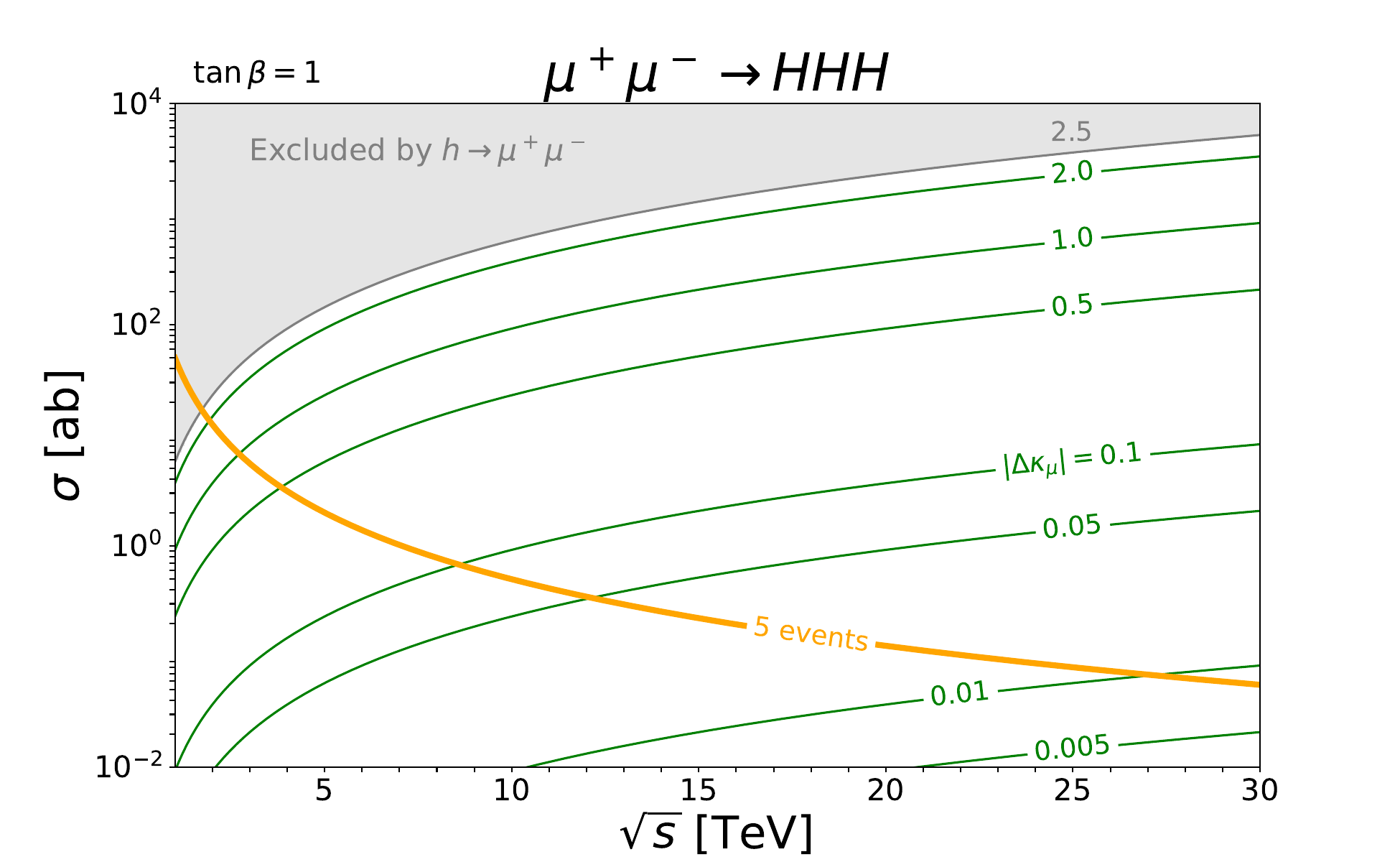}
  \includegraphics[scale=0.24]{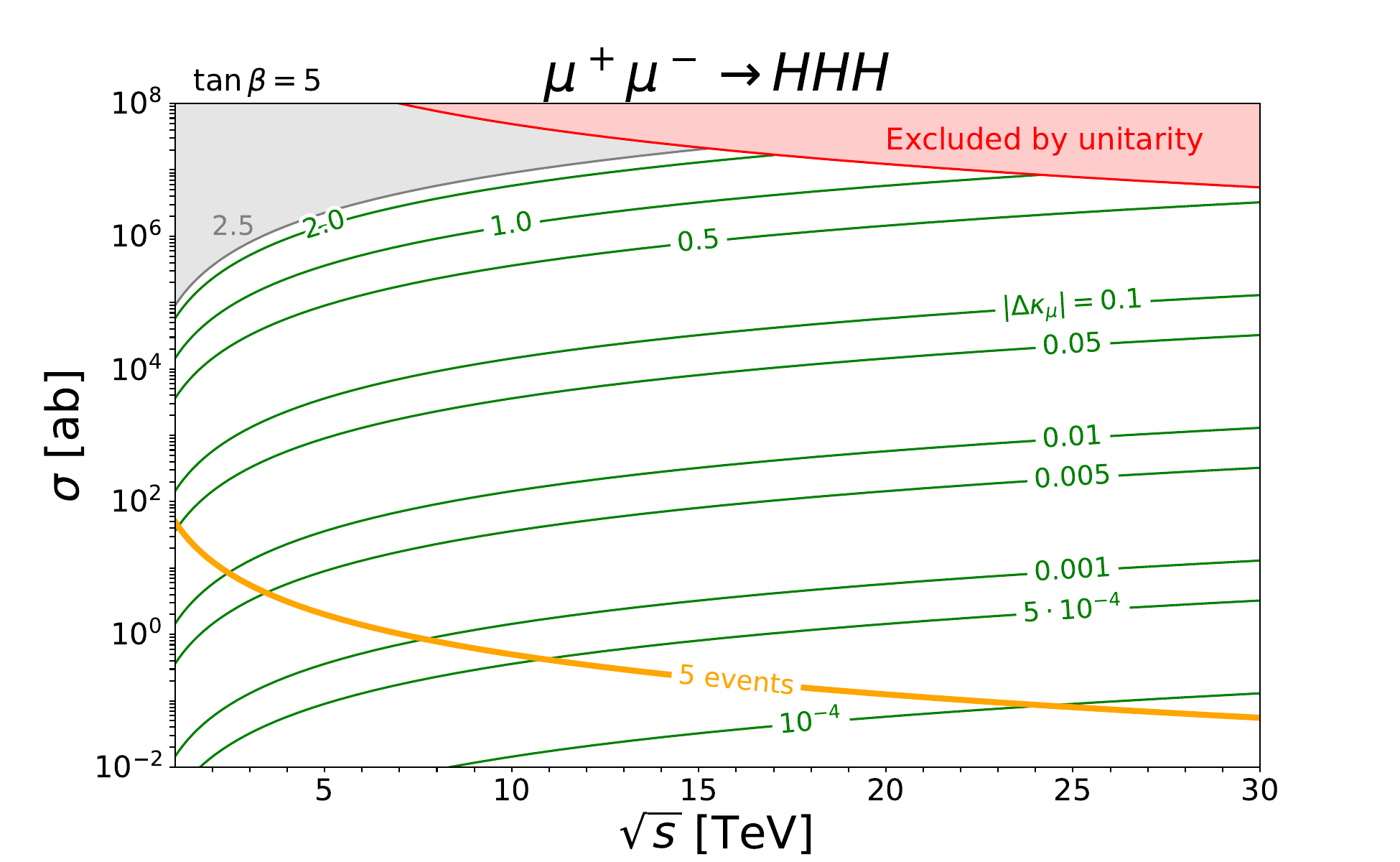}
    \includegraphics[scale=0.24]{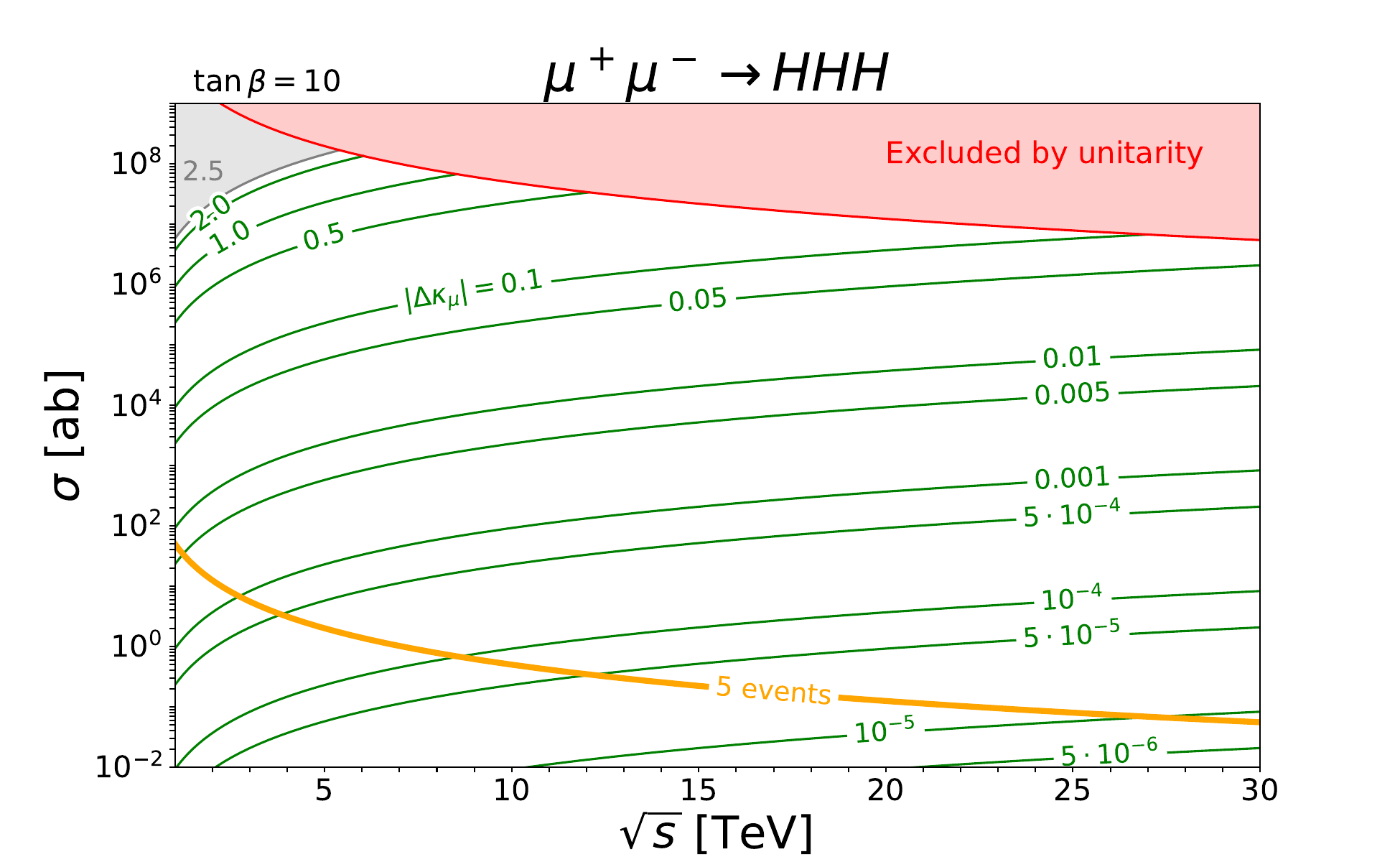}
  \includegraphics[scale=0.24]{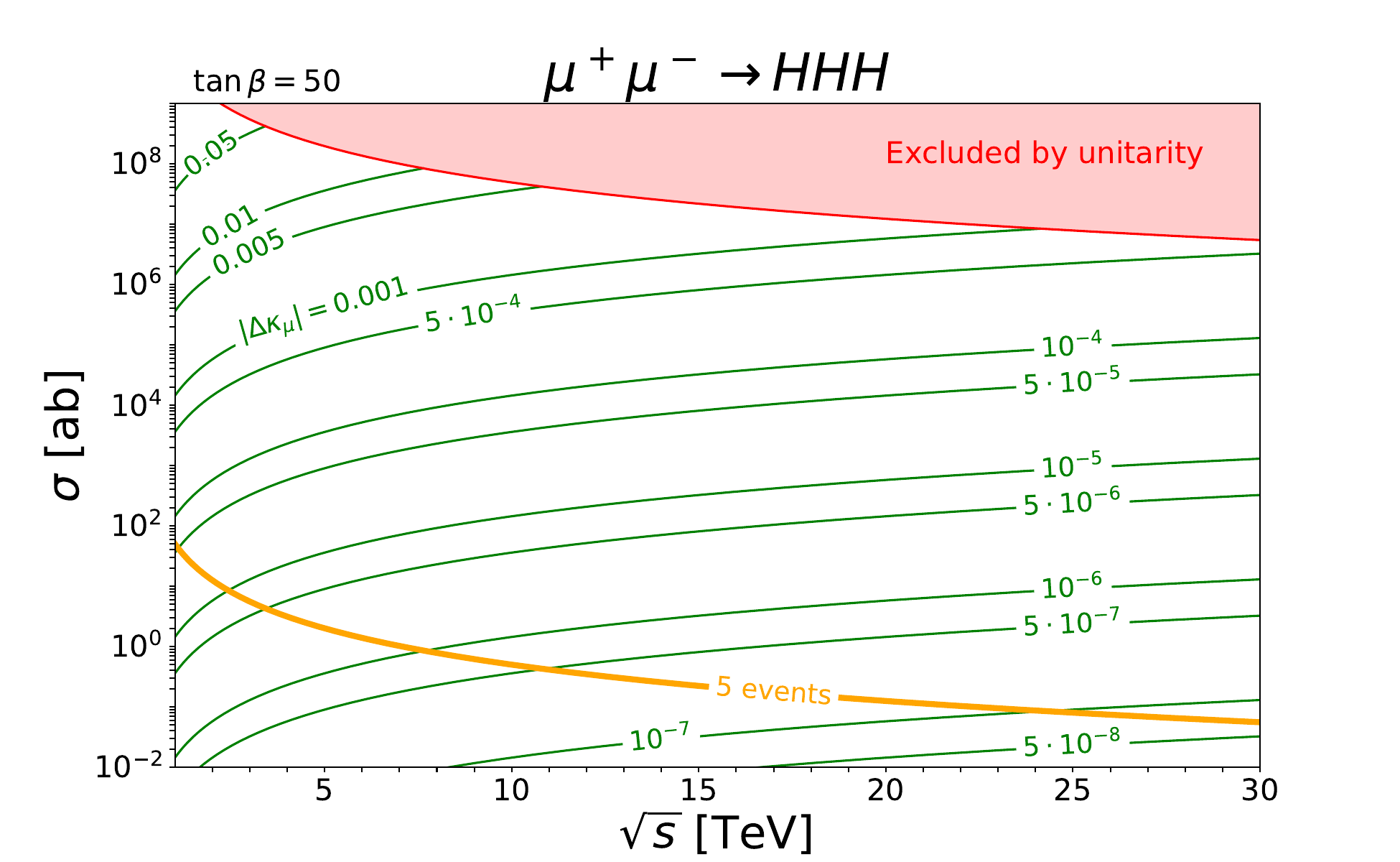}
\caption{The total cross section for $\mu^+\mu^-\rightarrow HHH$ as a function of $\sqrt{s}$ for various $|\Delta \kappa_\mu|$ and  $\tan \beta=1,5,10$ and 50. The gray  shaded regions are excluded at 95\% C.L. by the CMS search for $h\to \mu^+\mu^-$~\cite{CMS:2020xwi}. The red shaded regions are excluded by unitarity constraints. Orange lines indicate the cross sections corresponding to five signal events.}
\label{fig:HHH_sqrts}
\end{figure}

The cross section for $\mu^+\mu^-\rightarrow HHH$  is plotted as a function of $\sqrt{s}$ for various $|\Delta \kappa_\mu|$ and several choices of $\tan \beta$ in Fig.~\ref{fig:HHH_sqrts}. The red shaded regions are excluded by the unitarity limits on the cross section obtained from Eq.~(\ref{eq:unitarity-Delta_kappa}) and the formula for the cross section in terms of $|\Delta \kappa_\mu|$ in Table~\ref{table:tri-Higgs}. The boundary of the excluded region corresponds to the maximum $\Delta\kappa_\mu$ allowed by unitarity obtained from Eq.~(\ref{eq:unitarity-Delta_kappa})

\begin{equation}
\sigma_{\mu^+\mu^-\rightarrow HHH} = \frac{4\pi}{s}
\end{equation} 

which is independent of $\tan \beta$ and thus the same in all plots.  As orange lines indicate, as a result of $\tan^6 \beta$ enhancement,  the $\mu^+\mu^-\rightarrow HHH$ signal, if kinematically open, could lead to dramatically stronger sensitivity to muon Yukawa coupling. For example, at a $\sqrt{s} = 10$ TeV muon collider, deviations in the muon Yukawa coupling at the $7\times10^{-5}$ level could be tested for $\tan \beta = 10$ and at the $6\times10^{-7}$ level for $\tan \beta = 50$.

\begin{figure}[t!]
  \includegraphics[scale=0.24]{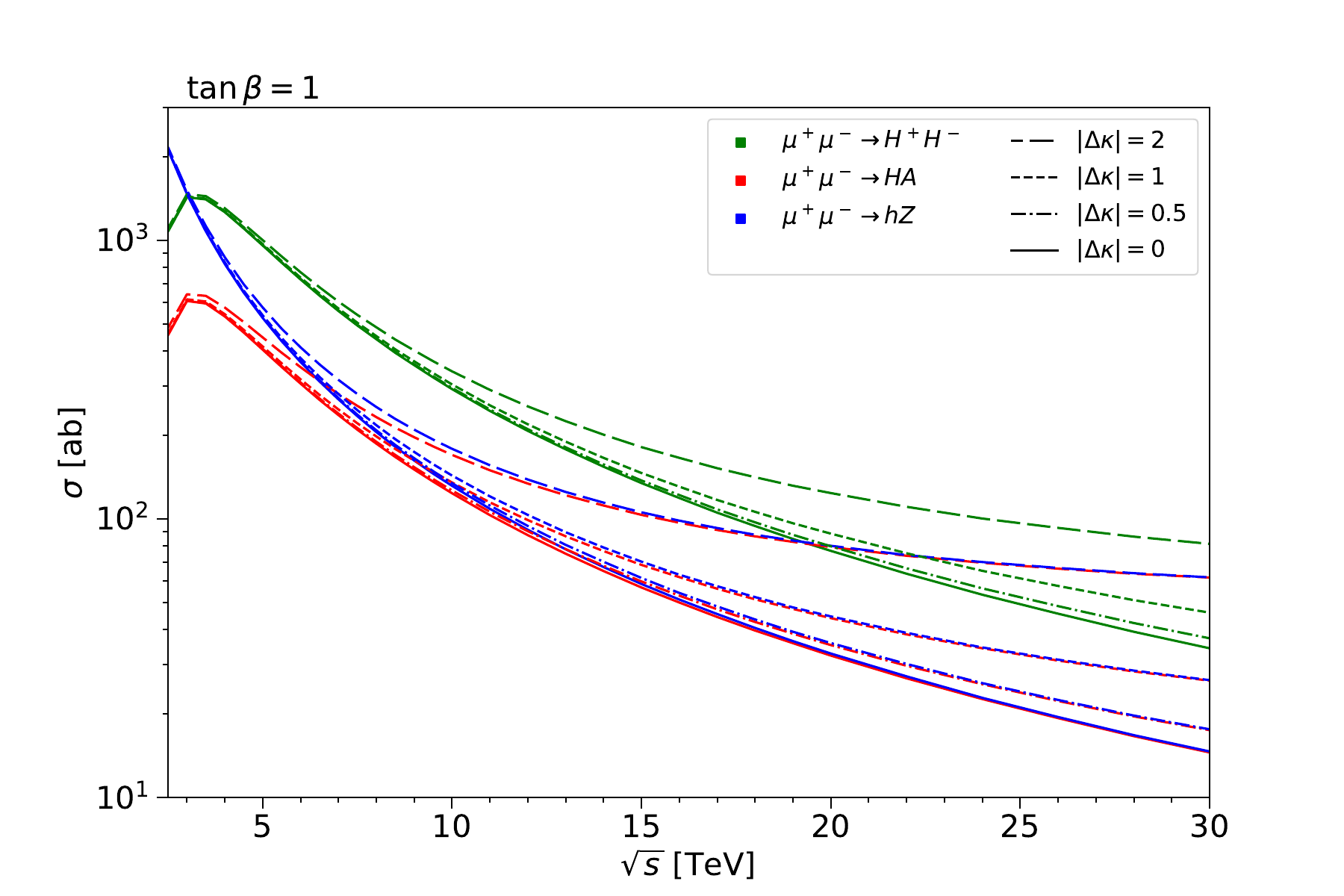}
  \includegraphics[scale=0.24]{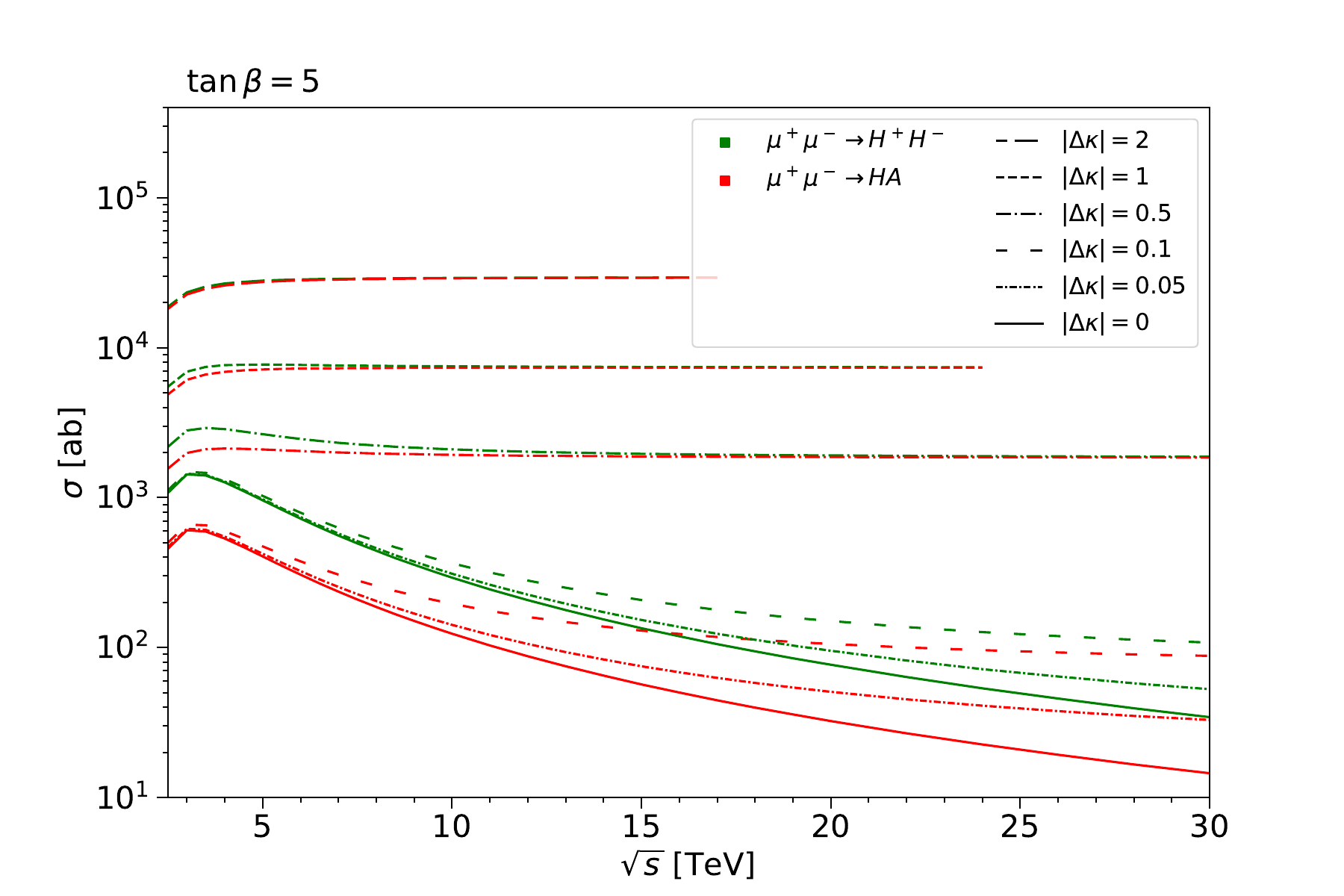}
    \includegraphics[scale=0.24]{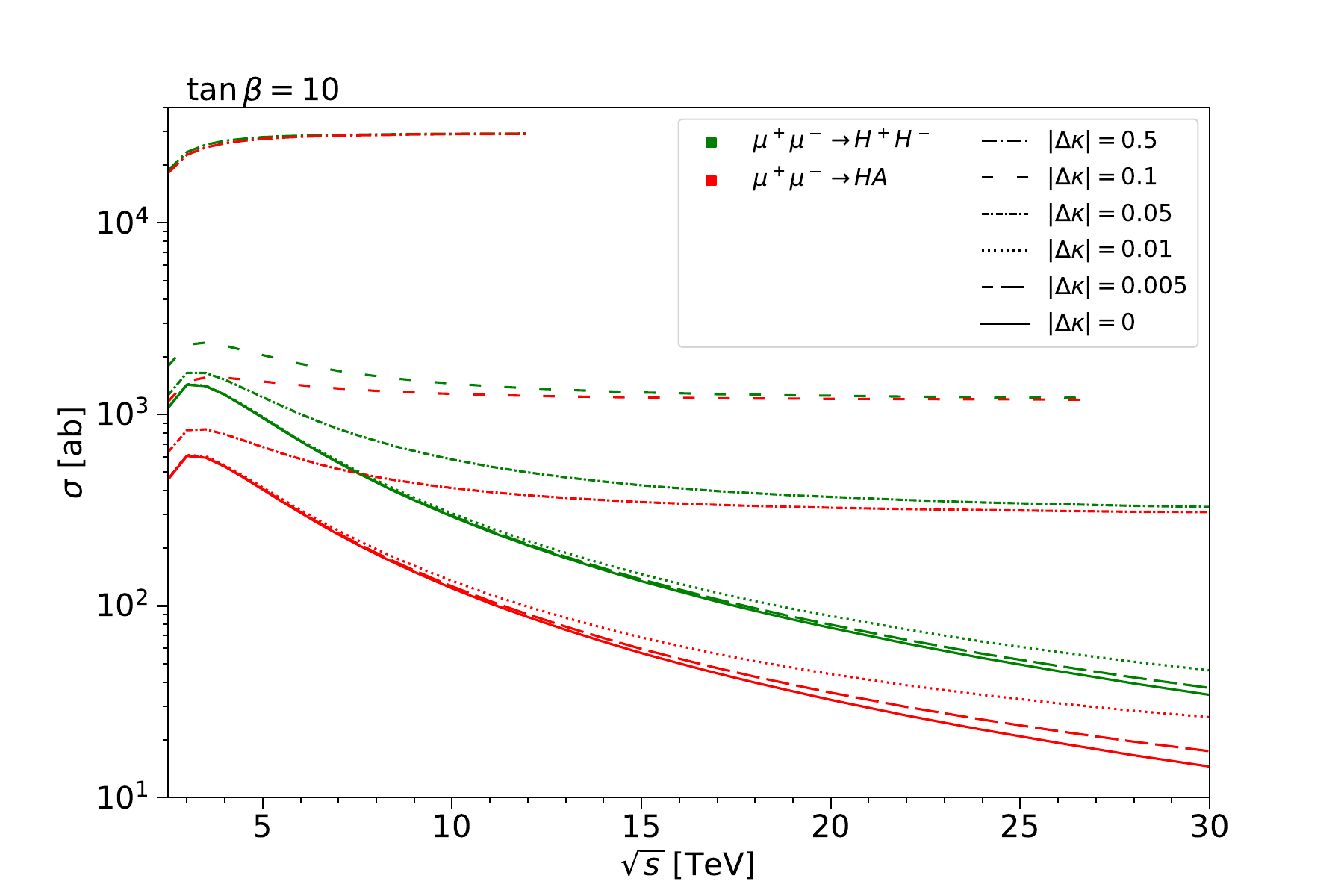}
  \includegraphics[scale=0.24]{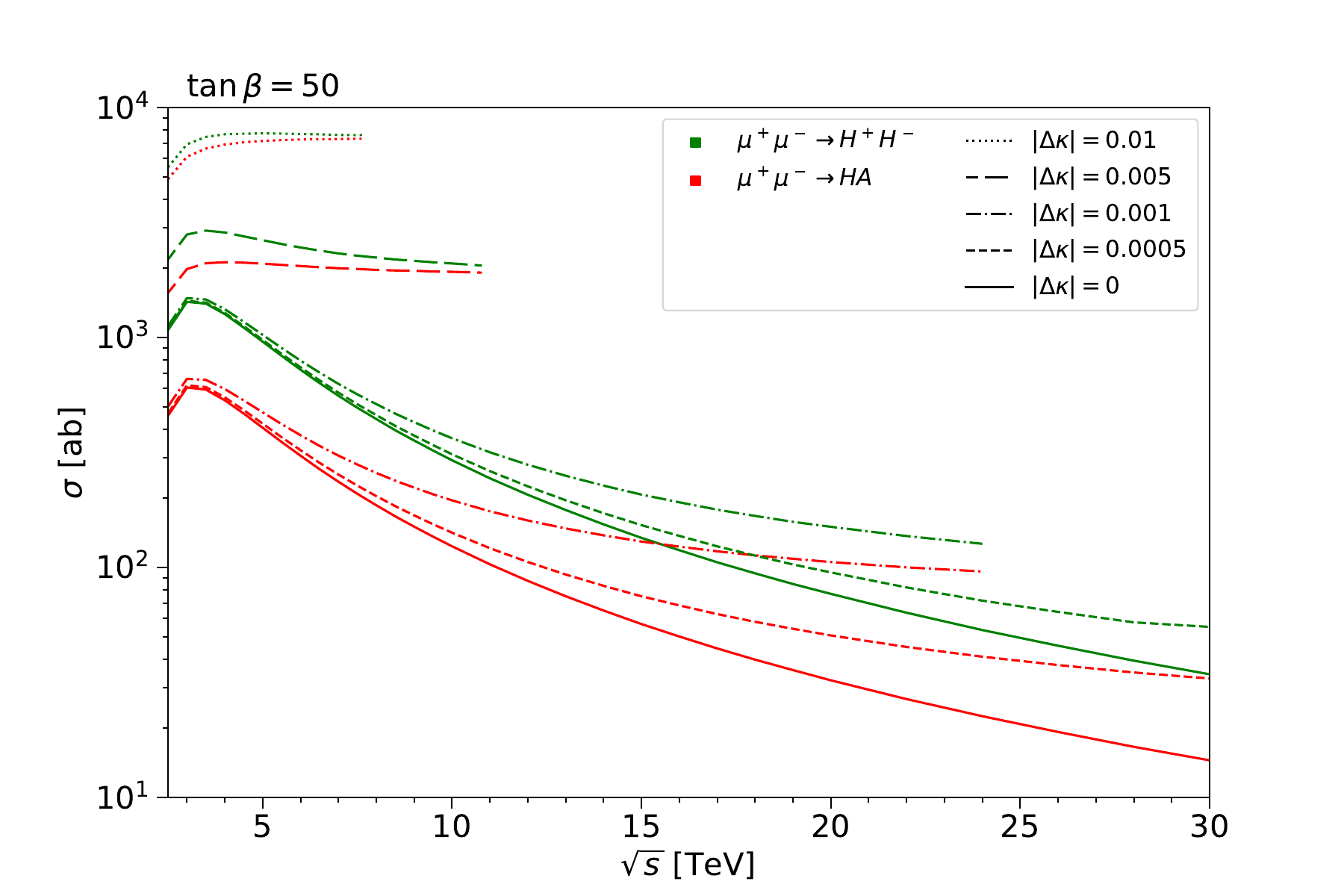}
\caption{The total cross sections for $\mu^+\mu^-\rightarrow HA$, $\mu^+\mu^-\rightarrow H^+H^-$ and $\mu^+\mu^-\rightarrow hZ$ as  functions of $\sqrt{s}$ for various $|\Delta \kappa_\mu|$ and  $\tan \beta=1,5,10$ and 50. Each line ends where the unitarity constraint is saturated for given $|\Delta \kappa_\mu|$ and  $\tan \beta$.}
\label{fig:HA_and_HpHm}
\end{figure}

Cross sections for other Higgs final states can be obtained by simple rescaling according to the last column of Tables~\ref{table:di-Higgs} and \ref{table:tri-Higgs}. However, for the final states that have sizable contributions from the 2HDM type-II without the dimension-six mass operator, the cross sections listed in these tables are good approximations only for large $\sqrt{s}$. For example, among the di-Higgs final states, $HA$ and $H^+H^-$ are also produced by $\mu^+\mu^-\rightarrow Z^*\rightarrow HA$ and $\mu^+\mu^-\rightarrow Z^*, \gamma^* \rightarrow H^+H^-$. Note that the cross sections for these processes behave as $1/s$ and that the interference with the processes for the same final states originating from the dimension-six mass operator is negligible because of the different chiralities of muons required. These cross sections, calculated from the effective Lagrangian implemented in {\tt FeynRules}~\cite{Degrande:2011ua} using {\tt MadGraph5}~\cite{Alwall:2014hca}, assuming 1 TeV masses of heavy Higgs bosons, are plotted
 in Fig.~\ref{fig:HA_and_HpHm}. The line for $|\Delta \kappa_\mu| = 0$ corresponds to the production cross section in the 2HDM type-II. The same comments apply to other tri-Higgs final states. In addition, for some of the other tri-Higgs final states there are contributions from Higgs cubic and quartic couplings that depend on the exact form of the potential.

Interactions involving Goldstone bosons resulting from the $\mathcal{O}_{\mu H_d}$  operator  are described by 
\begin{eqnarray}
    \pazocal{L}  \supset  &-&\frac{1}{\sqrt{2}}\lambda^G_{\mu\mu}\bar\mu_L\mu_R G-\lambda_{\mu\mu}^{hG}\bar{\mu}_L\mu_R hG -\lambda_{\mu\mu}^{HG}\bar{\mu}_L\mu_R HG -\lambda_{\mu\mu}^{AG}\bar{\mu}_L\mu_R AG -\frac{1}{2!}\lambda_{\mu\mu}^{GG}\bar{\mu}_L\mu_R GG \nonumber \\
    & -& \lambda_{\mu\mu}^{H^+G^-}\bar{\mu}_L\mu_R H^+G^- - \lambda_{\mu\mu}^{H^-G^+}\bar{\mu}_L\mu_R H^-G^+ - \lambda_{\mu\mu}^{G^+G^-}\bar{\mu}_L\mu_R G^+G^- \nonumber \\
    & -&\frac{1}{2!}\lambda_{\mu\mu}^{hhG}\bar{\mu}_L\mu_R h^2G -\lambda_{\mu\mu}^{hHG}\bar{\mu}_L\mu_R hHG -\lambda_{\mu\mu}^{hAG}\bar{\mu}_L\mu_R hAG \nonumber \\
    & -&\frac{1}{2!}\lambda_{\mu\mu}^{HHG}\bar{\mu}_L\mu_R HHG -\lambda_{\mu\mu}^{HAG}\bar{\mu}_L\mu_R HAG -\frac{1}{2!}\lambda_{\mu\mu}^{AAG}\bar{\mu}_L\mu_R AAG \nonumber \\
    & -&\frac{1}{2!}\lambda_{\mu\mu}^{hGG}\bar{\mu}_L\mu_R hGG -\frac{1}{2!}\lambda_{\mu\mu}^{HGG}\bar{\mu}_L\mu_R HGG -\frac{1}{2!}\lambda_{\mu\mu}^{AGG}\bar{\mu}_L\mu_R AGG \nonumber \\
    & -&\frac{1}{3!}\lambda_{\mu\mu}^{GGG}\bar{\mu}_L\mu_R GGG -\lambda_{\mu\mu}^{hG^+G^-}\bar{\mu}_L\mu_R hG^+G^- -\lambda_{\mu\mu}^{HG^+G^-}\bar{\mu}_L\mu_R HG^+G^-\nonumber \\
    & -&\lambda_{\mu\mu}^{AG^+G^-}\bar{\mu}_L\mu_R AG^+G^- -\lambda_{\mu\mu}^{GG^+G^-}\bar{\mu}_L\mu_R GG^+G^- -\lambda_{\mu\mu}^{hH^+G^-}\bar{\mu}_L\mu_R hH^+G^-\nonumber \\ 
    & -&\lambda_{\mu\mu}^{hG^+H^-}\bar{\mu}_L\mu_R hG^+H^- -\lambda_{\mu\mu}^{HH^+G^-}\bar{\mu}_L\mu_R HH^+G^- -\lambda_{\mu\mu}^{HG^+H^-}\bar{\mu}_L\mu_R HG^+H^- \nonumber \\ 
    & -&\lambda_{\mu\mu}^{AH^+G^-}\bar{\mu}_L\mu_R AH^+G^- -\lambda_{\mu\mu}^{AG^+H^-}\bar{\mu}_L\mu_R AG^+H^- -\lambda_{\mu\mu}^{GH^+H^-}\bar{\mu}_L\mu_R GH^+H^- \nonumber \\
    & -&\lambda_{\mu\mu}^{GH^+G^-}\bar{\mu}_L\mu_R GH^+G^- -\lambda_{\mu\mu}^{GG^+H^-}\bar{\mu}_L\mu_R GG^+H^- + {\rm H.c.},
\label{eqn:couplings_Gauge}
\end{eqnarray}
where the couplings are summarized in Table~\ref{table:G_couplings} in terms of the Wilson coefficient, $v$, $\alpha$ and $\beta$,  and also in the alignment limit, $\alpha = \beta - \frac{\pi}{2}$. The last column contains couplings in the alignment limit written in terms of $\Delta\kappa_\mu$. The cross sections for corresponding di-boson and tri-boson productions involving longitudinal gauge bosons are summarized in Tables~\ref{table:G_di-boson} and \ref{table:G_tri-boson}. For comparison, we plot $Zh$ in Fig.~\ref{fig:HA_and_HpHm} (only for $\tan \beta = 1$ since this process does not depend on $\tan \beta $). 

\begin{table*}[h!]
\caption{Coupling constants involving goldstone bosons defined in Eq.~(\ref{eqn:couplings_Gauge}).}
\begin{ruledtabular}
\begin{tabular}{cccc}
  & In general & Alignment limit ($\alpha = \beta - \frac{\pi}{2}$) & In terms of $\Delta\kappa_\mu$\\
 \hline
 $\lambda_{\mu\mu}^{G}$  &  $i\frac{m_\mu}{v}$ & $i\frac{m_\mu}{v}$ & $i\frac{m_\mu}{v}$\\
 \hline
 $\lambda_{\mu\mu}^{hG}$  &  $-iv\cos^2\beta\sin\alpha\;C_{\mu H_d}$ & $iv\cos^3\beta\;C_{\mu H_d}$  & $\frac{im_\mu}{2v^2}\Delta\kappa_\mu$\\
\hline  
 $\lambda_{\mu\mu}^{HG}$  & $iv\cos^2\beta\cos\alpha\;C_{\mu H_d}$ & $iv\cos^2\beta\sin\beta\;C_{\mu H_d}$  & $\frac{im_\mu}{2v^2}\Delta\kappa_\mu\tan\beta$\\
\hline 
 $\lambda_{\mu\mu}^{AG}$  & $-v\cos^2\beta\sin\beta\;C_{\mu H_d}$  & $-v\cos^2\beta\sin\beta\;C_{\mu H_d}$  & $-\frac{m_\mu}{2v^2}\Delta\kappa_\mu\tan\beta$\\
\hline
 $\lambda_{\mu\mu}^{GG}$  & $v\cos^3\beta\;C_{\mu H_d}$ & $v\cos^3\beta\;C_{\mu H_d}$  & $\frac{m_\mu}{2v^2}\Delta\kappa_\mu$\\
\hline
 $\lambda_{\mu\mu}^{H^+G^-}$, $\lambda_{\mu\mu}^{H^-G^+}$  & $-v\cos^2\beta\sin\beta\;C_{\mu H_d}$ &  $-v\cos^2\beta\sin\beta\;C_{\mu H_d}$  & $-\frac{m_\mu}{2v^2}\Delta\kappa_\mu\tan\beta$\\
\hline
 $\lambda_{\mu\mu}^{G^+G^-}$  &  $v\cos^3\beta\;C_{\mu H_d}$ & $v\cos^3\beta\;C_{\mu H_d}$   & $\frac{m_\mu}{2v^2}\Delta\kappa_\mu$\\
\hline
 $\lambda_{\mu\mu}^{hhG}$  &  $\frac{i}{\sqrt{2}}\cos\beta\sin^2\alpha\;C_{\mu H_d}$ & $\frac{i}{\sqrt{2}}\cos^3\beta\;C_{\mu H_d}$ & $\frac{im_\mu}{2\sqrt{2}v^3}\Delta\kappa_\mu$\\
\hline
 $\lambda_{\mu\mu}^{hHG}$  & $-\frac{i}{\sqrt{2}}\cos\beta\cos\alpha\sin\alpha\;C_{\mu H_d}$  & $\frac{i}{\sqrt{2}}\cos^2\beta\sin\beta\;C_{\mu H_d}$  & $\frac{im_\mu}{2\sqrt{2}v^3}\Delta\kappa_\mu\tan\beta$\\
\hline
 $\lambda_{\mu\mu}^{hAG}$  &  $\frac{1}{\sqrt{2}}\cos\beta\sin\beta\sin\alpha\;C_{\mu H_d}$ & $-\frac{1}{\sqrt{2}}\cos^2\beta\sin\beta\;C_{\mu H_d}$  & $-\frac{im_\mu}{2\sqrt{2}v^3}\Delta\kappa_\mu\tan\beta$\\
\hline
 $\lambda_{\mu\mu}^{HHG}$  & $\frac{i}{\sqrt{2}}\cos\beta\cos^2\alpha\;C_{\mu H_d}$ & $\frac{i}{\sqrt{2}}\cos\beta\sin^2\beta\;C_{\mu H_d}$ & $\frac{im_\mu}{2\sqrt{2}v^3}\Delta\kappa_\mu\tan^2\beta$\\
\hline
 $\lambda_{\mu\mu}^{HAG}$  &  $-\frac{1}{\sqrt{2}}\cos\beta\cos\alpha\sin\beta\;C_{\mu H_d}$  & $-\frac{1}{\sqrt{2}}\cos\beta\sin^2\beta\;C_{\mu H_d}$  & $-\frac{m_\mu}{2\sqrt{2}v^3}\Delta\kappa_\mu\tan^2\beta$\\
\hline
 $\lambda_{\mu\mu}^{AAG}$  & $\frac{3i}{\sqrt{2}}\cos\beta\sin^2{\beta}\;C_{\mu H_d}$ & $\frac{3i}{\sqrt{2}}\cos\beta\sin^2{\beta}\;C_{\mu H_d}$ & $\frac{3im_\mu}{2\sqrt{2}v^3}\Delta\kappa_\mu\tan^2\beta$\\
\hline
 $\lambda_{\mu\mu}^{hGG}$  & $-\frac{1}{\sqrt{2}}\cos^2{\beta}\sin{\alpha}\;C_{\mu H_d}$ & $\frac{1}{\sqrt{2}}\cos^3{\beta}\;C_{\mu H_d}$  & $\frac{m_\mu}{2\sqrt{2}v^3}\Delta\kappa_\mu$\\
\hline
 $\lambda_{\mu\mu}^{HGG}$  & $\frac{1}{\sqrt{2}}\cos^2\beta\cos\alpha\;C_{\mu H_d}$ & $\frac{1}{\sqrt{2}}\cos^2\beta\sin\beta\;C_{\mu H_d}$ & $\frac{m_\mu}{2\sqrt{2}v^3}\Delta\kappa_\mu\tan\beta$ \\
\hline
 $\lambda_{\mu\mu}^{AGG}$  & $-\frac{3i}{\sqrt{2}}\cos^2\beta\sin\beta\;C_{\mu H_d}$ & $-\frac{3i}{\sqrt{2}}\cos^2\beta\sin\beta\;C_{\mu H_d}$ & $-\frac{3im_\mu}{2\sqrt{2}v^3}\Delta\kappa_\mu\tan\beta$\\
\hline
 $\lambda_{\mu\mu}^{GGG}$  & $\frac{3i}{\sqrt{2}}\cos^3\beta\;C_{\mu H_d}$ &  $\frac{3i}{\sqrt{2}}\cos^3\beta\;C_{\mu H_d}$ & $\frac{3im_\mu}{2\sqrt{2}v^3}\Delta\kappa_\mu$\\
 \hline
  $\lambda_{\mu\mu}^{hG^+G^-}$  &  $-\frac{1}{\sqrt{2}}\cos^2\beta\sin\alpha\;C_{\mu H_d}$ & $\frac{1}{\sqrt{2}}\cos^3\beta\;C_{\mu H_d}$ & $\frac{m_\mu}{2\sqrt{2}v^3}\Delta\kappa_\mu$\\
\hline
 $\lambda_{\mu\mu}^{HG^+G^-}$  &  $\frac{1}{\sqrt{2}}\cos^2\beta\cos\alpha\;C_{\mu H_d}$ & $\frac{1}{\sqrt{2}}\cos^2\beta\sin\beta\;C_{\mu H_d}$  & $\frac{m_\mu}{2\sqrt{2}v^3}\Delta\kappa_\mu\tan\beta$\\
\hline
 $\lambda_{\mu\mu}^{AG^+G^-}$  & $-\frac{i}{\sqrt{2}}\cos^2\beta\sin\beta\;C_{\mu H_d}$ & $-\frac{i}{\sqrt{2}}\cos^2\beta\sin\beta\;C_{\mu H_d}$  & $-\frac{im_\mu}{2\sqrt{2}v^3}\Delta\kappa_\mu\tan\beta$
 \\
\hline
 $\lambda_{\mu\mu}^{GG^+G^-}$  & $\frac{i}{\sqrt{2}}\cos^3\beta\;C_{\mu H_d}$ & $\frac{i}{\sqrt{2}}\cos^3\beta\;C_{\mu H_d}$  & $\frac{im_\mu}{2\sqrt{2}v^3}\Delta\kappa_\mu$\\
\hline
 $\lambda_{\mu\mu}^{hH^+G^-}$, $\lambda_{\mu\mu}^{hG^+H^-}$  & $\frac{1}{\sqrt{2}}\cos\beta\sin\beta\sin\alpha\;C_{\mu H_d}$ & $-\frac{1}{\sqrt{2}}\cos^2\beta\sin\beta\;C_{\mu H_d}$  & $-\frac{m_\mu}{2\sqrt{2}v^3}\Delta\kappa_\mu\tan\beta$ \\
\hline
 $\lambda_{\mu\mu}^{HH^+G^-}$, $\lambda_{\mu\mu}^{HG^+H^-}$  & $-\frac{1}{\sqrt{2}}\cos\beta\cos\alpha\sin\beta\;C_{\mu H_d}$ & $-\frac{1}{\sqrt{2}}\cos\beta\sin^2\beta\;C_{\mu H_d}$  & $-\frac{m_\mu}{2\sqrt{2}v^3}\Delta\kappa_\mu\tan^2\beta$\\
\hline
 $\lambda_{\mu\mu}^{AH^+G^-}$, $\lambda_{\mu\mu}^{AG^+H^-}$ & $\frac{i}{\sqrt{2}}\cos\beta\sin^2\beta\;C_{\mu H_d}$ & $\frac{i}{\sqrt{2}}\cos\beta\sin^2\beta\;C_{\mu H_d}$  & $\frac{im_\mu}{2\sqrt{2}v^3}\Delta\kappa_\mu\tan^2\beta$\\
\hline
 $\lambda_{\mu\mu}^{GH^+H^-}$  & $\frac{i}{\sqrt{2}}\cos\beta\sin^2\beta\;C_{\mu H_d}$ & $\frac{i}{\sqrt{2}}\cos\beta\sin^2\beta\;C_{\mu H_d}$  & $\frac{im_\mu}{2\sqrt{2}v^3}\Delta\kappa_\mu\tan^2\beta$\\
\hline
 $\lambda_{\mu\mu}^{GH^+G^-}$, $\lambda_{\mu\mu}^{GG^+H^-}$  & $-\frac{i}{\sqrt{2}}\cos^2\beta\sin\beta\;C_{\mu H_d}$ & $-\frac{i}{\sqrt{2}}\cos^2\beta\sin\beta\;C_{\mu H_d}$  & $-\frac{im_\mu}{2\sqrt{2}v^3}\Delta\kappa_\mu\tan\beta$\\
\end{tabular}
\end{ruledtabular}
\label{table:G_couplings}
\end{table*}

\begin{table*}[h!]
\caption{Cross sections for di-boson productions involving longitudinal gauge bosons.}
\begin{ruledtabular}
\begin{tabular}{ccc}
  & In terms of $\Delta\kappa_\mu$ & \;\;\;\;\;\;In units of $\sigma_{\mu^+\mu^-\rightarrow hh}$ \;\;\;\;\;\; \\
 \hline
 $\sigma_{\mu^+\mu^-\rightarrow hZ_L}$ & $\frac{1}{128\pi}\left(\frac{m_\mu}{v^2}\right)^2|\Delta\kappa_\mu|^2$ &  $\frac{2}{9}$  \\
\hline  
 $\sigma_{\mu^+\mu^-\rightarrow HZ_L}$ & $\frac{1}{128\pi}\left(\frac{m_\mu}{v^2}\right)^2|\Delta\kappa_\mu|^2\tan^2\beta$  & $\frac{2}{9}\tan^2\beta$\\
\hline 
 $\sigma_{\mu^+\mu^-\rightarrow AZ_L}$   & $\frac{1}{128\pi}\left(\frac{m_\mu}{v^2}\right)^2|\Delta\kappa_\mu|^2\tan^2\beta$  & $\frac{2}{9}\tan^2\beta$\\
\hline
 $\sigma_{\mu^+\mu^-\rightarrow Z_LZ_L}$ & $\frac{1}{256\pi}\left(\frac{m_\mu}{v^2}\right)^2|\Delta\kappa_\mu|^2$  & $\frac{1}{9}$\\
\hline
 $\sigma_{\mu^+\mu^-\rightarrow H^+W_L^-/W_L^+H^-}$ &  $\frac{1}{128\pi}\left(\frac{m_\mu}{v^2}\right)^2|\Delta\kappa_\mu|^2\tan^2\beta$ & $\frac{2}{9}\tan^2\beta$\\
\hline
 $\sigma_{\mu^+\mu^-\rightarrow W_L^+W_L^-}$  & $\frac{1}{128\pi}\left(\frac{m_\mu}{v^2}\right)^2|\Delta\kappa_\mu|^2$  &  $\frac{2}{9}$\\
 \end{tabular}
\end{ruledtabular}
\label{table:G_di-boson}
\end{table*}

\begin{table*}[h!]
\caption{Cross sections for tri-boson productions involving longitudinal gauge bosons.}
\begin{ruledtabular}
\begin{tabular}{ccc}
  & In terms of $\Delta\kappa_\mu$ & \;\;\;\;\;\;In units of $\sigma_{\mu^+\mu^-\rightarrow hhh}$ \;\;\;\;\;\; \\
\hline
 $\sigma_{\mu^+\mu^-\rightarrow hhZ_L}$  & $\frac{s}{2^{14}\pi^3}\left(\frac{m_\mu}{v^3}\right)^2|\Delta\kappa_\mu|^2$  &  $\frac{1}{3}$\\
\hline
 $\sigma_{\mu^+\mu^-\rightarrow hHZ_L}$  & $\frac{s}{2^{13}\pi^3}\left(\frac{m_\mu}{v^3}\right)^2|\Delta\kappa_\mu|^2\tan^2\beta$  & $\frac{2}{3}\tan^2\beta$\\
\hline
 $\sigma_{\mu^+\mu^-\rightarrow hAZ_L}$ & $\frac{s}{2^{13}\pi^3}\left(\frac{m_\mu}{v^3}\right)^2|\Delta\kappa_\mu|^2\tan^2\beta$ & $\frac{2}{3}\tan^2\beta$\\
\hline
 $\sigma_{\mu^+\mu^-\rightarrow HHZ_L}$ & $\frac{s}{2^{14}\pi^3}\left(\frac{m_\mu}{v^3}\right)^2|\Delta\kappa_\mu|^2\tan^4\beta$  & $\frac{1}{3}\tan^4\beta$\\
\hline
 $\sigma_{\mu^+\mu^-\rightarrow HAZ_L}$  & $\frac{s}{2^{13}\pi^3}\left(\frac{m_\mu}{v^3}\right)^2|\Delta\kappa_\mu|^2\tan^4\beta$   &  $\frac{2}{3}\tan^4\beta$\\
\hline
 $\sigma_{\mu^+\mu^-\rightarrow AAZ_L}$ & $\frac{9s}{2^{14}\pi^3}\left(\frac{m_\mu}{v^3}\right)^2|\Delta\kappa_\mu|^2\tan^4\beta$  & $3\tan^4\beta$ \\
\hline
 $\sigma_{\mu^+\mu^-\rightarrow hZ_LZ_L}$ & $\frac{s}{2^{14}\pi^3}\left(\frac{m_\mu}{v^3}\right)^2|\Delta\kappa_\mu|^2$ & $\frac{1}{3}$\\
\hline
 $\sigma_{\mu^+\mu^-\rightarrow HZ_LZ_L}$  & $\frac{s}{2^{14}\pi^3}\left(\frac{m_\mu}{v^3}\right)^2|\Delta\kappa_\mu|^2\tan^2\beta$  & $\frac{1}{3}\tan^2\beta$ \\
\hline
 $\sigma_{\mu^+\mu^-\rightarrow AZ_LZ_L}$  & $\frac{9s}{2^{14}\pi^3}\left(\frac{m_\mu}{v^3}\right)^2|\Delta\kappa_\mu|^2\tan^2\beta$  & $3\tan^2\beta$\\
\hline
 $\sigma_{\mu^+\mu^-\rightarrow Z_LZ_LZ_L}$  &  $\frac{3s}{2^{14}\pi^3}\left(\frac{m_\mu}{v^3}\right)^2|\Delta\kappa_\mu|^2$  & $1$\\
\hline
  $\sigma_{\mu^+\mu^-\rightarrow hW_L^+W_L^-}$  &  $\frac{s}{2^{13}\pi^3}\left(\frac{m_\mu}{v^3}\right)^2|\Delta\kappa_\mu|^2$  & $\frac{2}{3}$\\
\hline
  $\sigma_{\mu^+\mu^-\rightarrow HW_L^+W_L^-}$ & $\frac{s}{2^{13}\pi^3}\left(\frac{m_\mu}{v^3}\right)^2|\Delta\kappa_\mu|^2\tan^2\beta$  &  $\frac{2}{3}\tan^2\beta$\\
\hline
 $\sigma_{\mu^+\mu^-\rightarrow AW_L^+W_L^-}$ & $\frac{s}{2^{13}\pi^3}\left(\frac{m_\mu}{v^3}\right)^2|\Delta\kappa_\mu|^2\tan^2\beta$  & $\frac{2}{3}\tan^2\beta$
 \\
\hline
 $\sigma_{\mu^+\mu^-\rightarrow Z_LW_L^+W_L^-}$  & $\frac{s}{2^{13}\pi^3}\left(\frac{m_\mu}{v^3}\right)^2|\Delta\kappa_\mu|^2$  & $\frac{2}{3}$\\
\hline
 $\sigma_{\mu^+\mu^-\rightarrow hH^+W_L^-/hW_L^+H^-}$  & $\frac{s}{2^{13}\pi^3}\left(\frac{m_\mu}{v^3}\right)^2|\Delta\kappa_\mu|^2\tan^2\beta$   & $\frac{2}{3}\tan^2\beta$  \\
\hline
 $\sigma_{\mu^+\mu^-\rightarrow HH^+W_L^-/HW_L^+H^-}$ & $\frac{s}{2^{13}\pi^3}\left(\frac{m_\mu}{v^3}\right)^2|\Delta\kappa_\mu|^2\tan^4\beta$   & $\frac{2}{3}\tan^4\beta$\\
\hline
 $\sigma_{\mu^+\mu^-\rightarrow AH^+W_L^-/AW_L^+H^-}$ & $\frac{s}{2^{13}\pi^3}\left(\frac{m_\mu}{v^3}\right)^2|\Delta\kappa_\mu|^2\tan^4\beta$ & $\frac{2}{3}\tan^4\beta$ \\
\hline
 $\sigma_{\mu^+\mu^-\rightarrow Z_LH^+H^-}$ & $\frac{s}{2^{13}\pi^3}\left(\frac{m_\mu}{v^3}\right)^2|\Delta\kappa_\mu|^2\tan^4\beta$   & $\frac{2}{3}\tan^4\beta$\\
\hline
 $\sigma_{\mu^+\mu^-\rightarrow Z_LH^+W_L^-/Z_LW_L^+H^-}$ & $\frac{s}{2^{13}\pi^3}\left(\frac{m_\mu}{v^3}\right)^2|\Delta\kappa_\mu|^2\tan^2\beta$   & $\frac{2}{3}\tan^2\beta$\\
\end{tabular}
\end{ruledtabular}
\label{table:G_tri-boson}
\end{table*}

\subsection{Other operators, the golden channels, and other Higgs sectors}

Results for other dimension-six mass operators in Eq.~(\ref{eq:eff_lagrangian_2HDM}),  $\mathcal{O}_{\mu H_u}^{(1)}$, $\mathcal{O}_{\mu H_u}^{(2)}$, and $\mathcal{O}_{\mu H_u}^{(3)}$, are summarized in Appendix~\ref{sec:c1c2c3}. The main differences are the $\tan \beta$ dependences of various processes. 

With the Higgs sector of the 2HDM type-II, there are also  more possible dimension-six operators with covariant derivatives than in  the SM case that can contribute to di-boson and tri-boson processes. For example, symmetries allow for $C_{R,H_d}(\overline{\mu}_R H_d^{\dagger}) i \slashed{D} (\mu_R H_d)$, $C_{R,H_u}(\overline{\mu}_R H_u) i \slashed{D} (\mu_R H_u^{\dagger})$, $C_{L,H_d}(\overline{l}_L H_d) i \slashed{D} (l_L H_d^{\dagger})$, or $C_{L,H_u}(\overline{l}_L \cdot H_u^{\dagger}) i \slashed{D} (l_L \cdot H_u)$. As in the SM case, operators with derivatives acting on the muon fields are reduced via equations of motion to the mass operators $\mathcal{O}_{\mu H_d}$, $\mathcal{O}_{\mu H_u}^{(1)}$, or $\mathcal{O}_{\mu H_u}^{(2)}$. Similarly, operators with  derivatives acting on the Higgs doublets in  symmetric combinations can be written as the mass operators above.  The antisymmetric combinations result in the following six independent $(LL)$ and $(RR)$ operators: $C_{H_d l}^{(1)} (H_d^{\dagger}i \overleftrightarrow{D}_{\mu} H_d) \left(\bar{l}_L\gamma^\mu l_L\right)$, $C_{H_d l}^{(3)} (H_d^{\dagger}i \overleftrightarrow{D}_{\mu}^a H_d) \left(\bar{l}_L \tau^a \gamma^\mu l_L\right)$, $C_{H_u l}^{(1)} (H_u^{\dagger}i \overleftrightarrow{D}_{\mu} H_u)\left(\bar{l}_L\gamma^\mu l_L\right)$, $C_{H_u l}^{(3)} (H_u^{\dagger}i \overleftrightarrow{D}_{\mu}^a H_u)\left(\bar{l}_L \tau^a \gamma^\mu l_L\right)$, $C_{H_d \mu} (H_d^{\dagger}i \overleftrightarrow{D}_{\mu} H_d)\left(\bar{\mu}_R\gamma^\mu \mu_R\right)$, and $C_{H_u \mu} (H_u^{\dagger}i \overleftrightarrow{D}_{\mu} H_u)\left(\bar{\mu}_R\gamma^\mu \mu_R\right)$, which do not contribute to the muon mass or the Yukawa coupling. 

However, the operators with covariant derivatives  contribute to $\mu^+ \mu^- \rightarrow H^\pm W_L^\mp$, $H^+H^-$, $Ah$, $AH$, $Z_L H$, $ZH$, and $W^\pm H^\mp$  di-boson processes, in addition to those identified in the SM case. Similarly, these operators contribute to $\mu^+ \mu^- \rightarrow  ZW_L^\pm H^\mp$, $ZH^\pm H^\mp$, $ZHH$, $ZhH$, $ZAZ_L$, $ZAA$, $ W^\pm W_L^\mp A$, $W^\pm W_L^\mp H$, $W^\pm H^\mp Z_L$, $W^\pm H^\mp h$, $W^\pm H^\mp A$, and $W^\pm H^\mp H$ tri-boson processes, in addition to those identified in the SM case. Furthermore, the dipole operators contribute to $\mu^+\mu^- \to H Z$,  $AZ$, $H^\pm W^\mp$, $HW^+W^-$, $ZH^\pm W^\mp$, and $AW^+W^-$ processes, in addition to those identified in the SM case. All other di-Higgs processes in Table~\ref{table:di-Higgs} are not affected, namely $\mu^+\mu^-\rightarrow HH$, $AA$, and $hH$, in addition to $hh$ already identified in the SM case. Furthermore, all tri-Higgs final states in Table~\ref{table:tri-Higgs} are not affected. Among di-boson and tri-boson processes involving Goldstone bosons in Tables~\ref{table:G_di-boson} and \ref{table:G_tri-boson} the only unaffected processes are $\mu^+\mu^-\rightarrow HZ_LZ_L$, $HAZ_L$, and $hAZ_L$ in addition to $hZ_LZ_L$ already identified in the SM case.

The 2HDM type-II backgrounds for all the identified di-Higgs processes, $\mu^+\mu^-\rightarrow hh$, $HH$, $AA$, and $hH$, are negligible (proportional to the muon Yukawa coupling). 
Among the tri-Higgs final states not affected by other dimension-six operators, only $hhh$, $HHH$, $hhH$, $hAA$, $hHH$, and $HAA$ have negligible backgrounds.\footnote{Note however that if masses of $H$ and $A$ are close, some of the processes with negligible backgrounds might not be distinguishable from those with large backgrounds. For example, $\mu^+\mu^-\rightarrow HH$ might not be distinguishable from $\mu^+\mu^-\rightarrow HA$ or $\mu^+\mu^-\rightarrow HHH$ might not be distinguishable from $\mu^+\mu^-\rightarrow AAA$.}  Finally, among the identified tri-boson final states involving Goldstone bosons, only $hAZ_L$ and $HZ_LZ_L$ have negligible backgrounds (in the alignment limit). However, we should note that the non-negligible 2HDM background might not be the limiting factor for a process to be a sensitive probe of a modified muon Yukawa coupling, as can be seen from Fig.~\ref{fig:HA_and_HpHm}. 

The discussion of dimension-eight operators closely follows the discussion in the SM case.

\section{Conclusions}
\label{sec:conclusions} 

We studied multi-Higgs  boson signals which in general accompany a modification of the muon Yukawa coupling independently of the scale and other details of new physics. As long as the dominant effect of new physics on the muon Yukawa coupling is captured by the dimension-six mass operator, the cross sections for $\mu^+ \mu^- \to hh$ and $\mu^+ \mu^- \to hhh$ are uniquely tied to the modification of  the muon Yukawa coupling. As a result of negligible SM backgrounds for these processes, these signals could provide the first evidence for new physics even before a deviation of the muon Yukawa coupling from the SM prediction is established by $h \to \mu^+\mu^-$. 

Even a low energy muon collider would easily see clear signals associated with the opposite sign muon Yukawa coupling. For example, 191 di-Higgs and 30 tri-Higgs events are expected already at $\sqrt{s} = 3$ TeV. In addition,  the di-Higgs signal can be used to observe a deviation in  the muon Yukawa coupling at the 10\% level for $\sqrt{s} = 10$ TeV and at the 3.5\% level for $\sqrt{s} = 30$ TeV. The tri-Higgs signal leads to only a slightly better sensitivity at $\sqrt{s} = 10$ TeV, namely 7\%,  but would improve dramatically with increasing  $\sqrt{s}$, reaching 0.8\% at $\sqrt{s} = 30$ TeV (and 0.07\% at $\sqrt{s} = 100$ TeV).  

We further argued that if mass operators of higher dimensions also contribute  significantly to the muon Yukawa coupling, signals with more Higgs bosons in final states are expected and could be even stronger than $hh$ or $hhh$ (as an example, we showed predictions for final states with four and five Higgs bosons resulting from the dimension-eight mass operator). In such a case, the cross section of an individual process might not be directly linked to the modification of the muon Yukawa coupling, but by measuring all resulting multi-Higgs boson signals along with $h \rightarrow \mu^+ \mu^-$, the Wilson coefficients of all contributing operators including the sizes of their complex phases can be determined.

We also studied all processes involving Goldstone bosons originating from the same dimension-six mass operator. We argued that among the large number of such processes only $\mu^+ \mu^- \rightarrow hZ_LZ_L$ is directly related to a modification of the muon Yukawa coupling. All other final states can also originate from other dimension-six operators (operators with covariant derivatives and dipole operators) which are not related to the muon Yukawa coupling. However, the third identified unique signal of a modified muon Yukawa coupling, $\mu^+ \mu^- \rightarrow hZ_LZ_L$, has a 3 times smaller cross section than $hhh$ and the SM background for this final state is not negligible. We also noted, that sizable dimension-eight operators can  affect all three final states.  Measuring the relative strength of these signals can indicate whether other dimension-eight operators play a significant role.

We further extended the study to the  two Higgs doublet model type-II and showed that di-Higgs and tri-Higgs signals involving heavy Higgs bosons can be enhanced in the alignment limit by a factor of $(\tan \beta)^4$ and $(\tan \beta)^6$, respectively, which results in the potential sensitivity to a  modified muon Yukawa coupling at the $10^{-6}$ level already at  a $\sqrt{s} = 10 $ TeV muon collider. 
Considering only dimension-six operators, we identified $\mu^+ \mu^- \rightarrow HH$, $AA$, $hH$, all tri-Higgs final states in Table~\ref{table:tri-Higgs}, and $HZ_LZ_L$, $HAZ_L$, and $hAZ_L$  as possible additional unique signals of a modified muon Yukawa coupling that involve heavy Higgs bosons. The 2HDM background is more model dependent; however, as a result of possibly very large predicted cross sections, it might not play a significant role. Among the signals with the largest predicted cross sections and smallest backgrounds are $\mu^+\mu^-\rightarrow HH$ and $\mu^+\mu^-\rightarrow HHH$, which were the main focus of the paper. However, depending on the masses of heavy Higgs bosons and $\sqrt{s}$ of a muon collider, these processes might not be kinematically open or might be highly suppressed. In that case, $\mu^+ \mu^- \rightarrow hH$, $hHH$, or $hhH$ might  be the most sensitive probes.

The results could be applied to models with different Higgs sectors. If a new scalar $S$ results from a multiplet participating in electroweak symmetry breaking and entering the dimension-six mass operator, the effective couplings of the muon to $SS$, $SSS$, and mixed couplings involving both $S$ and $h$ are generated. In general, $SS$ and $SSS$ productions are expected to be related to $\Delta \kappa_\mu$. Depending on the details,  the production cross sections could be enhanced (or suppressed) by the fourth and sixth powers of the ratio of  mixing parameters. This motivates a broad exploration of pure new di-boson and tri-boson signatures.

\acknowledgments
We thank Tao Han, Wolfgang Kilian, Nils Kreher, Zhen Liu, Yang Ma, Jürgen Reuter, Tobias Striegl, and Keping Xie for useful discussions. TRIUMF receives federal funding via a contribution agreement with the National Research Council of Canada.

\appendix
\section{UNITARITY OF SCATTERING PROCESSES}
\label{sec:23scat}

In this appendix, we provide details regarding partial wave unitarity imposed on effective operators in our study. The operator $\mathcal{O}_{\mu H}$ generates $2 \to 2$ and $2 \to 3$ scattering of muons into Higgs final states. To begin, we are interested in providing a bound on Wilson coefficients through unitarity of the $S$-matrix. Three-body processes, such as $\overline{l}_L (p_1) \mu_R (p_2) \rightarrow H (p_3) (H^{\dagger} (p_4) H (p_5))$, are usually difficult to compute due to the number of free parameters integrated over the phase space. However, to simplify the calculation, we follow the procedure of \cite{DiLuzio:2016sur,Allwicher:2021jkr} by constructing an effective two-body final state in the following way. We define $\theta_4$ as the angle of $p_4$ in its center-of-mass (COM) frame with particle 5, $\vec{p}_4 + \vec{p}_5 = 0$. From here, in the COM frame of the three particles, $\vec{p}_3 + \vec{p}_4 + \vec{p}_5 = 0$, the angle $\theta_{45}$ defines the position of $\vec{p}_4 + \vec{p}_5$ with angular momentum $J_{45}$. Lastly, we define $s_1 = - (p_1 + p_2 - p_3)^2 = - (p_4 + p_5)^2$, which integrates over the $4-5$ system's invariant mass up to $s$. For massless states, the partial wave amplitude is
\begin{equation}
    \begin{split}
        a^{J}_{fi}(s, s_1) = & \frac{1}{\sqrt{\mathcal{S}}}\frac{\sqrt{s - s_1}}{256 \pi^2 \sqrt{s}} \left( \sum_{J_{45}} \frac{1}{2 J_{45} + 1}\right)^{-1/2} \\
        & \times \int_{-1}^{1} d (\cos \theta_4)  \int_{-1}^{1} d (\cos \theta_{45}) d_{\lambda, \lambda}^J (\theta_{45}) d_{\lambda + h_5, h_3 - h_4}^{J_{45}}(\theta_4) \mathcal{T}(s, s_1; \theta_4, \theta_{45}), 
    \end{split}
\end{equation}
where $d_{mn}^j(\theta)$ are the Wigner small $d$-matrices for two particle states with total angular momentum $j$ and $m,n$ are the helicity projections  between incoming and outgoing two-body systems. $h_i$ are the helicities of the $i$th particle, $\lambda = h_1 + h_2$ in the COM frame of particles 1 and 2 and $\mathcal{S}$ is the symmetry factor for indistinguishable final states. $\mathcal{T}$ is the matrix element of the $2 \rightarrow 3$ scattering amplitude in momentum space, $\mathcal{T}(s, s_1; \theta_4, \theta_{45}) = (2 \pi)^4 \delta^4(P_i - P_f) \langle f| T | i \rangle$ for $T$ being the interaction part of the $S$-matrix, $S = 1 + i T$. A final integration over $s_1$ is taken to complete the three-body phase space and yields the full partial wave,
\begin{equation}
    |a^{J}_{fi}(s)|^2 = \int_0^s ds_1 |a_{fi}^{J} (s, s_1)|^2.
\label{eq:fullpw}
\end{equation}
Unitarity of $S^{\dagger}S = 1$ relates the partial wave amplitudes as

\begin{equation}
    \frac{1}{2i} (a_{fi}^{J} - a_{if}^{J *}) = \sum_k a_{kf}^{J *} a_{ki}^J,
\end{equation}
for all intermediate $k$ states. In the limit of forward scattering, $f \rightarrow i$, the left-hand side reduces to $\textrm{Im}[a_{ii}^J]$ and the right-hand side is bounded from below by $|a_{ii}^{J}|^2$. This now implies $\textrm{Im} [a_{ii}^{J}] \geq (\textrm{Re} [a_{ii}^{J}])^2 + (\textrm{Im} [a_{ii}^{J}])^2$, which can be rearranged as $\textrm{Im} [a_{ii}^{J}](1 - \textrm{Im} [a_{ii}^{J}]) \geq (\textrm{Re} [a_{ii}^{J}])^2$ and is bounded by $1/4$ on the unit circle. Hence, the condition for unitarity translates as~\cite{DiLuzio:2016sur,Allwicher:2021jkr,Marciano:1989ns}
\begin{equation}
    |\textrm{Re} [a_{ii}^{J}]| \leq \frac{1}{2},
    \label{eq:bound}
\end{equation}
applied to the result of Eq.~(\ref{eq:fullpw}). The forward limit can be obtained by diagonalizing $a_{fi}^J$ including all relevant channels. Our bounds are obtained by applying equation Eq.~(\ref{eq:bound}) to the largest eigenvalue of $a_{fi}^J$. For the operator $\overline{l}_L \mu_R H (H^{\dagger} H)$ we can separate the doublet components into
$2 + 2\times 2 = 6$ scattering states attributing to the partial wave. In the basis of states defined as
$\{|(\overline{l}_L)_1 \mu_R \rangle, |(\overline{l}_L)_2 \mu_R \rangle, |H_1 (H_1^{\dagger} H_1) \rangle, |H_1 (H_2^{\dagger} H_2) \rangle, |H_2 (H_1^{\dagger} H_1) \rangle, |H_2 (H_2^{\dagger} H_2) \rangle \}$ where $(\overline{l}_L)_{i} \delta_{ij}$ is understood to contract with the first $H_j$ label, we additionally take the $J = 0 $ partial wave based on the helicities of the left- and right-handed fields, i.e. $h_1 = -1/2$ and $h_2 = + 1/2$ scattering into spinless scalars, which helps the phase space integration become trivial: $d_{00}^0 (\theta_4) = d_{00}^0 (\theta_{45})  = 1$. We find the partial wave to be
\begin{equation}
    a_{fi}^{J = 0} (s, s_1) = - \frac{\sqrt{s - s_1}}{64 \pi^2} C_{\mu H} \begin{pmatrix}
        0 & 0 & \sqrt{2} & 1 & 0 & 0 \\
        0 & 0 & 0 & 0 & 1 &\sqrt{2} \\
        \sqrt{2} & 0 & 0 & 0 & 0 & 0 \\
        1 & 0 & 0 & 0 & 0 & 0 \\
        0 & 1 & 0 & 0 & 0 & 0 \\
        0 & \sqrt{2} & 0 & 0 & 0 & 0
    \end{pmatrix},
\end{equation} whose largest eigenvalue of the matrix is $\sqrt{3}$. Integrating over the remaining variable $s_1$ and implementing the bound in Eq.~(\ref{eq:bound}), we find
\begin{equation}
    |C_{\mu H}| \leq \left( \frac{64 \pi^2}{\sqrt{6}} \right) \frac{1}{s} \rightarrow \left( \frac{64 \pi^2}{\sqrt{6}} \right) \frac{1}{\Lambda^2},
\label{eq:SM_3higgs_unitarity}
\end{equation}
where we require that the low energy theory preserves unitarity up to $\sqrt{s} = \Lambda$. 

Note that if we rather considered scattering of physical states after electroweak symmetry breaking, we can obtain bounds for all inelastic scattering cross sections for $2 \to k$ processes. For example, at the dimension-eight level, one now has access to $h^4$ and $h^5$ processes and generally, the operator $\overline{l}_L \mu_R H (H^{\dagger} H)^{n}$ for $n \geq 1$ (mass dimension $d = 2n + 4$) generates up to $h^{2n+1}$ final states, becoming highly inelastic. For general $2 \rightarrow k$ scattering, the inelastic cross section is bounded by $\sigma (2 \rightarrow k) \leq 4 \pi / s$, obtained from unitarity of the forward scattering amplitude~\cite{Maltoni:2001dc}. Applying this bound on the cross section for $\mu^+ \mu^- \rightarrow h^k$ with $k \leq 2n+1$ Higgses in the final state [Eq.~(\ref{eq:xsection_k})], we find
\begin{equation}
     \frac{s^{k-2}}{2^{4k-3}\pi^{2k-3}k!(k-1)!(k-2)!} \,\left| \lambda_{\mu\mu}^{h^k}\right|^2 \leq \frac{4 \pi}{s}.
\end{equation}
Thus, by using the definition of $\lambda_{\mu \mu}^{h^k}$ in Eq.~(\ref{eq:coupling_k}) and 
$\Delta \kappa_{\mu}$ in Eq.~(\ref{eq:general_dkappa}), we find a general unitarity bound on $|\Delta \kappa_{\mu}|$:
\begin{equation}
    |\Delta \kappa_{\mu}| \leq 2^{(5k + 1)/2} \pi^{k-1} n \left(\frac{(2n+1-k)!}{(2n+1)!} \right) \sqrt{k! (k-1)! (k-2)!} \left(\frac{\Lambda}{m_{\mu}} \right) \left( \frac{v}{\Lambda}\right)^k,
\end{equation}
which assumes the only contribution comes from the $n$th Wilson coefficient. When $n = 1$ for $k = 2$ and $k = 3$, we find, respectively,
\begin{equation}
    |\Delta \kappa_{\mu}| \leq \left(\frac{32 \pi}{3} \right) \frac{v^2}{m_{\mu} \Lambda}.
\end{equation}
and
\begin{equation}
    |\Delta \kappa_{\mu}| \leq \left(\frac{256 \pi^2}{\sqrt{3}} \right) \frac{v^3}{m_{\mu} \Lambda^2}.
\end{equation}
For $n = 2$ for $k = 4$ and 5, we find, respectively
\begin{equation}
        |\Delta \kappa_{\mu}| \leq \left(\frac{2^{11} \pi^3}{5} \right) \frac{v^4}{m_{\mu} \Lambda^3}
        \label{eq:unitarity_4h}    
\end{equation}
and
\begin{equation}
        |\Delta \kappa_{\mu}| \leq \left(2^{14} \pi^4 \sqrt{\frac{6}{5}} \right) \frac{v^5}{m_{\mu} \Lambda^4}.
\label{eq:unitarity_5h}        
\end{equation}

For the 2HDM-II-equivalent operator, $\overline{l}_L \mu_R H_d ( H_d^{\dagger} H_d)$, the expression in Eq.~(\ref{eq:bound}) is the same because of identical $SU(2)$ structure upon replacing $C_{\mu H} \rightarrow C_{\mu H_d}$. However, additional information constraining the parameter space of $C_{\mu H_d}$ can be exploited after EWSB. Particularly, for physical states, we can apply the inelastic cross section bound on $\mu^+ \mu^- \rightarrow h h h$ and $\mu^+ \mu^- \rightarrow H H H$ channels, revealing different $\tan \beta$ dependencies. We conclude that
\begin{equation}
    |C_{\mu H_d}| \leq \left( \frac{128 \pi^2}{\sqrt{3} |\sin^3 \alpha|} \right) \frac{1}{\Lambda^2} \rightarrow \left( \frac{128 \pi^2}{\sqrt{3} \cos^3 \beta} \right) \frac{1}{\Lambda^2}, 
\end{equation}
and
\begin{equation}
    |C_{\mu H_d}| \leq \left( \frac{128 \pi^2}{\sqrt{3} \cos^3 \alpha} \right) \frac{1}{\Lambda^2} \rightarrow \left( \frac{128 \pi^2}{\sqrt{3} \sin^3 \beta} \right) \frac{1}{\Lambda^2},
\end{equation}
in the alignment limit for $\mu^+ \mu^- \rightarrow hhh$ and $\mu^+ \mu^- \rightarrow HHH$, respectively, ignoring the heavy Higgs masses. Furthermore, one should apply the stronger of the two expressions depending on the domain of $\tan \beta$ considered. Similarly for the $2 \rightarrow 2$ processes, $\mu^+ \mu^- \rightarrow h h$ and $\mu^+ \mu^- \rightarrow H H$, we arrive at
\begin{equation}
    |C_{\mu H_d}| \leq \left(\frac{16 \pi}{3 \cos \beta \sin^2 \alpha} \right) \frac{1}{v \Lambda} \rightarrow \left(\frac{16\pi}{3 \cos^3 \beta} \right) \frac{1}{v \Lambda},
\end{equation}
\begin{equation}
    |C_{\mu H_d}| \leq \left(\frac{16 \pi}{3 \cos \beta \cos^2 \alpha} \right) \frac{1}{v \Lambda} \rightarrow \left(\frac{16\pi}{3 \cos \beta \sin^2 \beta} \right) \frac{1}{v \Lambda}.
\end{equation}

\section{OTHER DIMENSION-SIX MASS OPERATORS IN THE 2HDM TYPE-II}
\label{sec:c1c2c3}

Here, we derive the effects of other dimension-six mass operators in 2HDM type-II, namely $C_{\mu H_u}^{(1)} \overline{l}_L \mu_R H_d (H_u^{\dagger} H_u), C_{\mu H_u}^{(2)} \overline{l}_L \mu_R \cdot H_u^{\dagger} (H_d \cdot H_u)$, and $C_{\mu H_u}^{(3)} \overline{l}_L \mu_R \cdot H_u^{\dagger} (H_d^{\dagger} \cdot H_u^{\dagger})$, on the muon Yukawa coupling and summarize resulting effective di-boson and tri-boson couplings of the muon. 

In a similar way as in Eq.~(\ref{eq:mmu_CmuHd}), these operators, considered one at a time, generate additional contributions to the muon mass, 
\begin{equation}
	m_\mu = y_\mu v_d + C^{(i)}_{\mu H_u}v_dv_u^2,
\end{equation}
where $i=1,2$ or 3, and modify the muon Yukawa coupling by 
\begin{equation}
	\Delta\kappa_\mu^{(i)} = 2\, C^{(i)}_{\mu H_u}\, \frac{v_dv_u^2}{m_\mu}.
\end{equation}
If more operators are present simultaneously, their contributions should be added.
The resulting coupling constants describing interactions between the muon and 2HDM Higgs bosons are summarized in Tables~\ref{table:C1_couplings},~\ref{table:C2_couplings}, and \ref{table:C3_couplings} for $C^{(1)}_{\mu H_u}, C^{(2)}_{\mu H_u}$, and $C^{(3)}_{\mu H_u}$, respectively. As can be seen from the tables, for a given $\Delta\kappa_\mu$, the couplings are $\tan\beta$ suppressed compared to corresponding couplings in Table~\ref{table:couplings}, except for those with at most one $H$ or $A$: $\lambda^{hh}_{\mu\mu}$, $\lambda^{hH}_{\mu\mu}$, $\lambda^{hA}_{\mu\mu}$, $\lambda^{hhh}_{\mu\mu}$, $\lambda^{hhH}_{\mu\mu}$, and $\lambda^{hhA}_{\mu\mu}$. 

For the operator  $\mathcal{O}^{'}_{\mu H_u}$  that can be written as a linear combination of $\mathcal{O}^{(1)}_{\mu H_u}$ and $\mathcal{O}^{(2)}_{\mu H_u}$  [see Eq.~(\ref{eq:O'})], the resulting couplings can be obtained by the corresponding linear combinations of couplings in Tables~\ref{table:C1_couplings} and \ref{table:C2_couplings}. It is straightforward to see that all the couplings are zero except for $\lambda_{\mu\mu}^{H^+H^-}$ and $\lambda_{\mu\mu}^{hH^+H^-}$. Thus this operator can only contribute to  $\mu^+\mu^-\rightarrow H^+H^-$ and $\mu^+\mu^-\rightarrow hH^+H^-$. As already discussed in the main text,  this operator does not contribute to the muon mass, and thus its contributions to the processes above are not related to the modification of muon Yukawa coupling. 
\newpage

\begin{table*}[h!]
\caption{\label{tab:table100} Coupling constants describing interactions with the 2HDM Higgs bosons in Eq.~(\ref{eqn:couplings_2HDM}) resulting from $C^{(1)}_{\mu H_u}$.}
\begin{ruledtabular}
\begin{tabular}{cccc}
    & Alignment limit ($\alpha = \beta - \frac{\pi}{2})$ & In terms of $\Delta\kappa_\mu^{(1)}$\\
 \hline
 $\lambda_{\mu\mu}^{hh}$ & $3v\cos\beta\sin^2\beta\;C^{(1)}_{\mu H_u} $ & $\frac{3m_\mu}{2v^2}\Delta\kappa_\mu^{(1)}$\\
\hline  
 $\lambda_{\mu\mu}^{AA}$   & $v\cos^3\beta\;C^{(1)}_{\mu H_u}$ & $\frac{m_\mu}{2v^2}\Delta\kappa_\mu^{(1)}(\tan\beta)^{-2}$\\
\hline 
 $\lambda_{\mu\mu}^{HH}$  & $(v\cos^3\beta-2v\cos\beta\sin^2\beta)\;C^{(1)}_{\mu H_u}$ & $-\frac{m_\mu}{v^2}\Delta\kappa_\mu^{(1)}+\frac{m_\mu}{2v^2}\Delta\kappa_\mu^{(1)}(\tan\beta)^{-2}$\\
\hline
 $\lambda_{\mu\mu}^{hH}$   & $(v\sin^3\beta-2v\cos^2\beta\sin\beta)\;C^{(1)}_{\mu H_u}$ & $\frac{m_\mu}{2v^2}\Delta\kappa_\mu^{(1)}\tan\beta-\frac{m_\mu}{v^2}\Delta\kappa_\mu^{(1)}(\tan\beta)^{-1}$\\ 
\hline
 $\lambda_{\mu\mu}^{hA}$  & $-iv\sin^3\beta\;C^{(1)}_{\mu H_u}$ & $-i\frac{m_\mu}{2v^2}\Delta\kappa_\mu^{(1)}\tan\beta$\\
\hline
 $\lambda_{\mu\mu}^{HA}$  & $iv\sin^2\beta\cos\beta\;C^{(1)}_{\mu H_u} $ & $i\frac{m_\mu}{2v^2}\Delta\kappa_\mu^{(1)}$\\
\hline
 $\lambda_{\mu\mu}^{H^+H^-}$  & $v\cos^3\beta\;C^{(1)}_{\mu H_u}$ & $\frac{m_\mu}{2v^2}\Delta\kappa_\mu^{(1)}(\tan\beta)^{-2}$\\
\hline
 $\lambda_{\mu\mu}^{hhh}$  & $\frac{3}{\sqrt{2}}\cos\beta\sin^2\beta\;C^{(1)}_{\mu H_u}$ & $\frac{3m_\mu}{2\sqrt{2}v^3}\Delta\kappa_\mu^{(1)}$ \\
 \hline
 $\lambda_{\mu\mu}^{AAA}$  & $-i\frac{3}{\sqrt{2}}\cos^2\beta\sin\beta\;C^{(1)}_{\mu H_u}$ & $-i\frac{3m_\mu}{2\sqrt{2}v^3}\Delta\kappa_\mu^{(1)}(\tan\beta)^{-1}$\\
\hline
 $\lambda_{\mu\mu}^{HHH}$  & $\frac{3}{\sqrt{2}}\cos^2\beta\sin\beta\;C^{(1)}_{\mu H_u}$ & $\frac{3m_\mu}{2\sqrt{2}v^3}\Delta\kappa_\mu^{(1)}(\tan\beta)^{-1}$\\
\hline
 $\lambda_{\mu\mu}^{hhH}$  & $(-\sqrt{2}\cos^2\beta\sin\beta+\frac{1}{\sqrt{2}}\sin^3\beta)\;C^{(1)}_{\mu H_u}$ & $\frac{m_\mu}{2\sqrt{2}v^3}\Delta\kappa_\mu^{(1)} \tan\beta-\frac{m_\mu}{\sqrt{2}v^3}\Delta\kappa_\mu^{(1)}(\tan\beta)^{-1}$\\
\hline
 $\lambda_{\mu\mu}^{hhA}$  & $-i\frac{1}{\sqrt{2}}\sin^3\beta\;C^{(1)}_{\mu H_u}$ & $-i\frac{m_\mu}{2\sqrt{2}v^3}\Delta\kappa_\mu^{(1)}\tan\beta$\\ 
\hline
 $\lambda_{\mu\mu}^{hAA}$  & $\frac{1}{\sqrt{2}}\cos^3\beta\;C^{(1)}_{\mu H_u}$ & $\frac{m_\mu}{2\sqrt{2}v^3}\Delta\kappa_\mu^{(1)}(\tan\beta)^{-2}$ \\
\hline
 $\lambda_{\mu\mu}^{hHH}$ & $(\frac{1}{\sqrt{2}}\cos^3\beta-\sqrt{2}\cos\beta\sin^2\beta)\;C^{(1)}_{\mu H_u}$ & $-\frac{m_\mu}{\sqrt{2}v^3}\Delta\kappa_\mu^{(1)}+\frac{m_\mu}{2\sqrt{2}v^3}\Delta\kappa_\mu^{(1)}(\tan\beta)^{-2}$\\
\hline
 $\lambda_{\mu\mu}^{AHH}$  & $-i\frac{1}{\sqrt{2}}\cos^2\beta\sin\beta\;C^{(1)}_{\mu H_u}$ & $-i\frac{m_\mu}{2\sqrt{2}v^3}\Delta\kappa_\mu^{(1)}(\tan\beta)^{-1}$\\
\hline
 $\lambda_{\mu\mu}^{HAA}$  & $\frac{1}{\sqrt{2}}\cos^2\beta\sin\beta\;C^{(1)}_{\mu H_u}$ & $\frac{m_\mu}{2\sqrt{2}v^3}\Delta\kappa_\mu^{(1)}(\tan\beta)^{-1}$\\
\hline
 $\lambda_{\mu\mu}^{hH^+H^-}$  & $\frac{1}{\sqrt{2}}\cos^3\beta\;C^{(1)}_{\mu H_u}$ & $\frac{m_\mu}{2\sqrt{2}v^3}\Delta\kappa_\mu^{(1)}(\tan\beta)^{-2}$\\
\hline
 $\lambda_{\mu\mu}^{HH^+H^-}$ & $\frac{1}{\sqrt{2}}\cos^2\beta\sin\beta\;C^{(1)}_{\mu H_u}$ & $\frac{m_\mu}{2\sqrt{2}v^3}\Delta\kappa_\mu^{(1)}(\tan\beta)^{-1}$\\
\hline
 $\lambda_{\mu\mu}^{AH^+H^-}$  & $-i\frac{1}{\sqrt{2}}\cos^2\beta\sin\beta\;C^{(1)}_{\mu H_u}$ & $-i\frac{m_\mu}{2\sqrt{2}v^3}\Delta\kappa_\mu^{(1)}(\tan\beta)^{-1}$\\
\hline
 $\lambda_{\mu\mu}^{hHA}$   & $i\frac{1}{\sqrt{2}}\cos\beta\sin^2\beta\;C^{(1)}_{\mu H_u}$ & $i\frac{m_\mu}{2\sqrt{2}v^3}\Delta\kappa_\mu^{(1)}$
\end{tabular}
\end{ruledtabular}
\label{table:C1_couplings}
\end{table*}
\newpage

\begin{table*}[h!]
\caption{\label{tab:table101} Coupling constants describing interactions with the 2HDM Higgs bosons in Eq.~(\ref{eqn:couplings_2HDM}) resulting from $C^{(2)}_{\mu H_u}$.}
\begin{ruledtabular}
\begin{tabular}{cccc}
    & Alignment limit ($\alpha = \beta - \frac{\pi}{2})$ & In terms of $\Delta\kappa_\mu^{(2)}$\\
 \hline
 $\lambda_{\mu\mu}^{hh}$ & $3v\cos\beta\sin^2\beta\;C^{(2)}_{\mu H_u} $ & $\frac{3m_\mu}{2v^2}\Delta\kappa_\mu^{(2)}$\\
\hline  
 $\lambda_{\mu\mu}^{AA}$   & $v\cos^3\beta\;C^{(2)}_{\mu H_u}$ & $\frac{m_\mu}{2v^2}\Delta\kappa_\mu^{(2)}(\tan\beta)^{-2}$\\
\hline 
 $\lambda_{\mu\mu}^{HH}$  & $(v\cos^3\beta-2v\cos\beta\sin^2\beta)\;C^{(2)}_{\mu H_u}$ & $-\frac{m_\mu}{v^2}\Delta\kappa_\mu^{(2)}+\frac{m_\mu}{2v^2}\Delta\kappa_\mu^{(2)}(\tan\beta)^{-2}$\\
\hline
 $\lambda_{\mu\mu}^{hH}$   & $(v\sin^3\beta-2v\cos^2\beta\sin\beta)\;C^{(2)}_{\mu H_u}$ & $\frac{m_\mu}{2v^2}\Delta\kappa_\mu^{(2)}\tan\beta-\frac{m_\mu}{v^2}\Delta\kappa_\mu^{(2)}(\tan\beta)^{-1}$\\ 
\hline
 $\lambda_{\mu\mu}^{hA}$  & $-iv\sin^3\beta\;C^{(2)}_{\mu H_u}$ & $-i\frac{m_\mu}{2v^2}\Delta\kappa_\mu^{(2)}\tan\beta$\\
\hline
 $\lambda_{\mu\mu}^{HA}$  & $iv\sin^2\beta\cos\beta\;C^{(2)}_{\mu H_u} $ & $i\frac{m_\mu}{2v^2}\Delta\kappa_\mu^{(2)}$\\
\hline
 ${\lambda_{\mu\mu}^{H^+H^-}}$  & $-v\cos\beta\sin^2\beta\;C^{(2)}_{\mu H_u}$ & $-\frac{m_\mu}{2v^2}\Delta\kappa_\mu^{(2)}$\\
\hline
 $\lambda_{\mu\mu}^{hhh}$  & $\frac{3}{\sqrt{2}}\cos\beta\sin^2\beta\;C^{(2)}_{\mu H_u}$ & $\frac{3m_\mu}{2\sqrt{2}v^3}\Delta\kappa_\mu^{(2)}$ \\
 \hline
 $\lambda_{\mu\mu}^{AAA}$  & $-i\frac{3}{\sqrt{2}}\cos^2\beta\sin\beta\;C^{(2)}_{\mu H_u}$ & $-i\frac{3m_\mu}{2\sqrt{2}v^3}\Delta\kappa_\mu^{(2)}(\tan\beta)^{-1}$\\
\hline
 $\lambda_{\mu\mu}^{HHH}$  & $\frac{3}{\sqrt{2}}\cos^2\beta\sin\beta\;C^{(2)}_{\mu H_u}$ & $\frac{3m_\mu}{2\sqrt{2}v^3}\Delta\kappa_\mu^{(2)}(\tan\beta)^{-1}$\\
\hline
 $\lambda_{\mu\mu}^{hhH}$  & $(-\sqrt{2}\cos^2\beta\sin\beta+\frac{1}{\sqrt{2}}\sin^3\beta)\;C^{(2)}_{\mu H_u}$ & $\frac{m_\mu}{2\sqrt{2}v^3}\Delta\kappa_\mu^{(2)} \tan\beta-\frac{m_\mu}{\sqrt{2}v^3}\Delta\kappa_\mu^{(2)}(\tan\beta)^{-1}$\\
\hline
 $\lambda_{\mu\mu}^{hhA}$  & $-i\frac{1}{\sqrt{2}}\sin^3\beta\;C^{(2)}_{\mu H_u}$ & $-i\frac{m_\mu}{2\sqrt{2}v^3}\Delta\kappa_\mu^{(2)}\tan\beta$\\ 
\hline
 $\lambda_{\mu\mu}^{hAA}$  & $\frac{1}{\sqrt{2}}\cos^3\beta\;C^{(2)}_{\mu H_u}$ & $\frac{m_\mu}{2\sqrt{2}v^3}\Delta\kappa_\mu^{(2)}(\tan\beta)^{-2}$ \\
\hline
 $\lambda_{\mu\mu}^{hHH}$ & $(\frac{1}{\sqrt{2}}\cos^3\beta-\sqrt{2}\cos\beta\sin^2\beta)\;C^{(2)}_{\mu H_u}$ & $-\frac{m_\mu}{\sqrt{2}v^3}\Delta\kappa_\mu^{(2)}+\frac{m_\mu}{2\sqrt{2}v^3}\Delta\kappa_\mu^{(2)}(\tan\beta)^{-2}$\\
\hline
 $\lambda_{\mu\mu}^{AHH}$  & $-i\frac{1}{\sqrt{2}}\cos^2\beta\sin\beta\;C^{(2)}_{\mu H_u}$ & $-i\frac{m_\mu}{2\sqrt{2}v^3}\Delta\kappa_\mu^{(2)}(\tan\beta)^{-1}$\\
\hline
 $\lambda_{\mu\mu}^{HAA}$  & $\frac{1}{\sqrt{2}}\cos^2\beta\sin\beta\;C^{(2)}_{\mu H_u}$ & $\frac{m_\mu}{2\sqrt{2}v^3}\Delta\kappa_\mu^{(2)}(\tan\beta)^{-1}$\\
\hline
 ${\lambda_{\mu\mu}^{hH^+H^-}}$  & $-\frac{1}{\sqrt{2}}\cos\beta\sin^2\beta\;C^{(2)}_{\mu H_u}$ & $-\frac{m_\mu}{2\sqrt{2}v^3}\Delta\kappa_\mu^{(2)}$\\
\hline
 $\lambda_{\mu\mu}^{HH^+H^-}$ & $\frac{1}{\sqrt{2}}\cos^2\beta\sin\beta\;C^{(2)}_{\mu H_u}$ & $\frac{m_\mu}{2\sqrt{2}v^3}\Delta\kappa_\mu^{(2)}(\tan\beta)^{-1}$\\
\hline
 $\lambda_{\mu\mu}^{AH^+H^-}$  & $-i\frac{1}{\sqrt{2}}\cos^2\beta\sin\beta\;C^{(2)}_{\mu H_u}$ & $-i\frac{m_\mu}{2\sqrt{2}v^3}\Delta\kappa_\mu^{(2)}(\tan\beta)^{-1}$\\
\hline
 $\lambda_{\mu\mu}^{hHA}$   & $i\frac{1}{\sqrt{2}}\cos\beta\sin^2\beta\;C^{(2)}_{\mu H_u}$ & $i\frac{m_\mu}{2\sqrt{2}v^3}\Delta\kappa_\mu^{(2)}$
\end{tabular}
\end{ruledtabular}
\label{table:C2_couplings}
\end{table*}
\newpage

\begin{table*}[h!]
\caption{\label{tab:table102} Coupling constants describing interactions with the 2HDM Higgs bosons in Eq.~(\ref{eqn:couplings_2HDM}) resulting from $C^{(3)}_{\mu H_u}$.}
\begin{ruledtabular}
\begin{tabular}{cccc}
    & Alignment limit ($\alpha = \beta - \frac{\pi}{2})$ & In terms of $\Delta\kappa_\mu^{(3)}$\\
 \hline
 $\lambda_{\mu\mu}^{hh}$ & $3v\cos\beta\sin^2\beta\;C^{(3)}_{\mu H_u} $ & $\frac{3m_\mu}{2v^2}\Delta\kappa_\mu^{(3)}$\\
\hline  
 $\lambda_{\mu\mu}^{AA}$   & $(-2v\cos\beta\sin^2\beta-v\cos^3\beta\;)C^{(3)}_{\mu H_u}$ & $-\frac{m_\mu}{v^2}\Delta\kappa_\mu^{(3)}-\frac{m_\mu}{2v^2}\Delta\kappa_\mu^{(3)}(\tan\beta)^{-2}$\\
\hline 
 $\lambda_{\mu\mu}^{HH}$  & $(v\cos^3\beta-2v\cos\beta\sin^2\beta)\;C^{(3)}_{\mu H_u}$ & $-\frac{m_\mu}{v^2}\Delta\kappa_\mu^{(3)}+\frac{m_\mu}{2v^2}\Delta\kappa_\mu^{(3)}(\tan\beta)^{-2}$\\
\hline
 $\lambda_{\mu\mu}^{hH}$   & $(v\sin^3\beta-2v\cos^2\beta\sin\beta)\;C^{(3)}_{\mu H_u}$ & $\frac{m_\mu}{2v^2}\Delta\kappa_\mu^{(3)}\tan\beta-\frac{m_\mu}{v^2}\Delta\kappa_\mu^{(3)}(\tan\beta)^{-1}$\\ 
\hline
 $\lambda_{\mu\mu}^{hA}$  & $(iv\sin^3\beta+2iv\cos^2\beta\sin\beta\;)C^{(3)}_{\mu H_u}$ & $i\frac{m_\mu}{2v^2}\Delta\kappa_\mu^{(3)}\tan\beta +i\frac{m_\mu}{v^2}\Delta\kappa_\mu^{(3)}(\tan\beta)^{-1}$\\
\hline
 $\lambda_{\mu\mu}^{HA}$  & $-iv\cos^3\beta\;C^{(3)}_{\mu H_u}$ & $-i\frac{m_\mu}{2v^2}\Delta\kappa_\mu^{(3)}(\tan\beta)^{-2}$\\
\hline
 $\lambda_{\mu\mu}^{H^+H^-}$  & $-v\cos\beta\sin^2\beta\;C^{(3)}_{\mu H_u}$ &$-\frac{m_\mu}{2v^2}\Delta\kappa_\mu^{(3)}$\\
\hline
 $\lambda_{\mu\mu}^{hhh}$  & $\frac{3}{\sqrt{2}}\cos\beta\sin^2\beta\;C^{(3)}_{\mu H_u}$ & $\frac{3m_\mu}{2\sqrt{2}v^3}\Delta\kappa_\mu^{(3)}$ \\
 \hline
 $\lambda_{\mu\mu}^{AAA}$  & $-i\frac{3}{\sqrt{2}}\cos^2\beta\sin\beta\;C^{(3)}_{\mu H_u}$ & $-i\frac{3m_\mu}{2\sqrt{2}v^3}\Delta\kappa_\mu^{(3)}(\tan\beta)^{-1}$\\
\hline
 $\lambda_{\mu\mu}^{HHH}$  & $\frac{3}{\sqrt{2}}\cos^2\beta\sin\beta\;C^{(3)}_{\mu H_u}$ & $\frac{3m_\mu}{2\sqrt{2}v^3}\Delta\kappa_\mu^{(3)}(\tan\beta)^{-1}$\\
\hline
 $\lambda_{\mu\mu}^{hhH}$  & $(-\sqrt{2}\cos^2\beta\sin\beta+\frac{1}{\sqrt{2}}\sin^3\beta)\;C^{(3)}_{\mu H_u}$ & $\frac{m_\mu}{2\sqrt{2}v^3}\Delta\kappa_\mu^{(3)} \tan\beta-\frac{m_\mu}{\sqrt{2}v^3}\Delta\kappa_\mu^{(3)}(\tan\beta)^{-1}$\\
\hline
 $\lambda_{\mu\mu}^{hhA}$  & $(i\frac{1}{\sqrt{2}}\sin^3\beta+i\sqrt{2}\cos^2\beta\sin\beta\;)C^{(3)}_{\mu H_u}$ & $i\frac{m_\mu}{2\sqrt{2}v^3}\Delta\kappa_\mu^{(3)}\tan\beta+i\frac{m_\mu}{\sqrt{2}v^3}\Delta\kappa_\mu^{(3)}(\tan\beta)^{-1}$\\ 
\hline
 $\lambda_{\mu\mu}^{hAA}$  & $(-\sqrt{2}\cos\beta\sin^2\beta-\frac{1}{\sqrt{2}}\cos^3\beta\;)C^{(3)}_{\mu H_u}$ & $-\frac{m_\mu}{\sqrt{2}v^3}\Delta\kappa_\mu^{(3)}-\frac{m_\mu}{2\sqrt{2}v^3}\Delta\kappa_\mu^{(3)}(\tan\beta)^{-2}$\\
\hline
 $\lambda_{\mu\mu}^{hHH}$ & $(\frac{1}{\sqrt{2}}\cos^3\beta-\sqrt{2}\cos\beta\sin^2\beta)\;C^{(3)}_{\mu H_u}$ & $-\frac{m_\mu}{\sqrt{2}v^3}\Delta\kappa_\mu^{(3)}+\frac{m_\mu}{2\sqrt{2}v^3}\Delta\kappa_\mu^{(3)}(\tan\beta)^{-2}$\\
\hline
 $\lambda_{\mu\mu}^{AHH}$  & $-i\frac{1}{\sqrt{2}}\cos^2\beta\sin\beta\;C^{(3)}_{\mu H_u}$ & $-i\frac{m_\mu}{2\sqrt{2}v^3}\Delta\kappa_\mu^{(3)}(\tan\beta)^{-1}$\\
\hline
 $\lambda_{\mu\mu}^{HAA}$  & $\frac{1}{\sqrt{2}}\cos^2\beta\sin\beta\;C^{(3)}_{\mu H_u}$ & $\frac{m_\mu}{2\sqrt{2}v^3}\Delta\kappa_\mu^{(3)}(\tan\beta)^{-1}$\\
\hline
 $\lambda_{\mu\mu}^{hH^+H^-}$  & $-\frac{1}{\sqrt{2}}\cos\beta\sin^2\beta\;C^{(3)}_{\mu H_u}$ & $-\frac{m_\mu}{2\sqrt{2}v^3}\Delta\kappa_\mu^{(3)}$\\
\hline
 $\lambda_{\mu\mu}^{HH^+H^-}$ & $\frac{1}{\sqrt{2}}\cos^2\beta\sin\beta\;C^{(3)}_{\mu H_u}$ & $\frac{m_\mu}{2\sqrt{2}v^3}\Delta\kappa_\mu^{(3)}(\tan\beta)^{-1}$\\
\hline
 $\lambda_{\mu\mu}^{AH^+H^-}$  & $-i\frac{1}{\sqrt{2}}\cos^2\beta\sin\beta\;C^{(3)}_{\mu H_u}$ & $-i\frac{m_\mu}{2\sqrt{2}v^3}\Delta\kappa_\mu^{(3)}(\tan\beta)^{-1}$\\
\hline
 $\lambda_{\mu\mu}^{hHA}$   & $-\frac{i}{\sqrt{2}}\cos^3\beta\;C^{(3)}_{\mu H_u}$ & $-\frac{m_\mu}{2\sqrt{2}v^3}\Delta\kappa_\mu^{(3)}(\tan\beta)^{-2}$
\end{tabular}
\end{ruledtabular}
\label{table:C3_couplings}
\end{table*}
\newpage





\end{document}